\documentclass[draft,jgrga]{AGUTeX}

\usepackage{lineno}
\linenumbers*[1]

\usepackage{color}
\usepackage[final]{graphicx}

\authorrunninghead{KITE ET AL.}

\titlerunninghead{VALLES MARINERIS SNOWSTORMS}

\begin{document}

%% ------------------------------------------------------------------------ %%
%
%  TITLE
%
%% ------------------------------------------------------------------------ %%

\title{Chaos, storms, and climate on Mars}

%\title{Localized precipitation and channel formation on Mars. 2: Inverted streams near Juventae Chasma.}
%
% e.g., \title{Terrestrial ring current:
% Origin, formation, and decay $\alpha\beta\Gamma\Delta$}
% You may use \\ to break the title over several lines.

%% ------------------------------------------------------------------------ %%
%
%  AUTHORS AND AFFILIATIONS
%
%% ------------------------------------------------------------------------ %%

%Use \author{\altaffilmark{}} and \altaffiltext{}

% \altaffilmark will produce footnote;
% matching altaffiltext will appear at bottom of page.
% May use \\ to start a new line.

\authors{Edwin S. Kite, \altaffilmark{1,2}
 Scot Rafkin, \altaffilmark{3} Timothy I. Michaels, \altaffilmark{3}
William E. Dietrich, \altaffilmark{1}
and Michael Manga. \altaffilmark{1,2} }

\altaffiltext{1}{Earth and Planetary Science, University of California Berkeley, Berkeley, California, USA.}

\altaffiltext{2}{Center for Integrative Planetary Science, University of California Berkeley, Berkeley, California, USA.}

\altaffiltext{3}{Department of Space Studies, Southwest Research Institute, Boulder, Colorado, USA.}

%\altaffiltext{3}{Department of Space Sciences, University of
%Michigan, Ann Arbor, Michigan, USA.}

%\altaffiltext{4}{Division of Hydrologic Sciences, Desert Research
%Institute, Reno, Nevada, USA.}

%% ------------------------------------------------------------------------ %%
%
%  ABSTRACT
%
%% ------------------------------------------------------------------------ %%

% >> Do NOT include any \begin...\end commands within
% >> the body of the abstract.

\begin{abstract}
%(Assume present day topography)
Channel networks on the plateau adjacent to the Juventae outflow channel source region and chaos have the highest drainage densities reported on Mars. We model frozen precipitation on the Juventae plateau, finding that the trigger for forming these channel networks could have been ephemeral lakeshore precipitation, and that they do not require past temperatures higher than today. If short-lived and localized events explain some dendritic channel networks on Mars, this would weaken the link between dendritic valley networks and surface climate conditions that could sustain life. Our analysis uses Mars Regional Atmospheric Modeling System (MRAMS) simulations and HiRISE Digital Terrain Models. Following a suggestion from \citet{man08}, we model localized weather systems driven by water vapor release from ephemeral lakes during outflow channel formation. At Juventae Chasma, the interaction between lake-driven convergence, topography, and the regional wind field steers lake-induced precipitation to the southwest. Mean snowfall reaches a maximum of 0.9 mm/hr water equivalent (peak snowfall 1.7 mm/hr water equivalent) on the SW rim of the chasm. Radiative effects of the thick cloud cover raise mean plateau surface temperature by up to 18K locally. The key result is that the area of maximum modeled precipitation shows a striking correspondence to the mapped Juventae plateau channel networks. Three independent methods show this fit is unlikely to be due to chance. We use a snowpack energy balance model to show that if the snow has the albedo of dust (0.28), and for a solar luminosity of 0.8 ($\equiv$3.0 Gya), then if the atmospheric greenhouse effect is unchanged from today only 0.4\% of lake-induced precipitation events produce snowpack that undergoes melting (for a 6K increase in the atmospheric greenhouse effect, this rises to 21\%). However, warming from associated dense cloud cover would allow melting over a wider range of conditions. Complete melting of the snow from a single event is sufficient to move sand and gravel through the observed channel networks. At Echus Chasma, modeled precipitation maxima also correspond to mapped plateau channel networks. In these localized precipitation scenarios, global temperatures need not be higher than today, and the rest of the planet remains dry.

\end{abstract}

%% ------------------------------------------------------------------------ %%
%
%  BEGIN ARTICLE
%
%% ------------------------------------------------------------------------ %%

% The body of the article must start with a \begin{article} command
%
% \end{article} must follow the references section, before the figures
%  and tables.

\begin{article}

%% ------------------------------------------------------------------------ %%
%
%  TEXT
%
%% ------------------------------------------------------------------------ %%

\section{Introduction}
\noindent
The highlands of Mars show both erosional and depositional evidence for past fluvial flow, and geochemical and textural evidence at Meridiani indicates these features were formed by liquid water \citep{mal03, kraal08alu, mur09, gro06, grotzmclennan}. The distribution of fans and channels with elevation, together with their morphology and inferred discharge, argue against groundwater as the sole source, and demand precipitation \citep{car03,car10,hyn03}. Three models may explain these observations. (1) A globally prolonged climate interval that intermittently allowed surface runoff is the most straightforward interpretation, and one that can draw on knowledge of arid and polar Earth analogs \citep{hal07,bar09}. (2) Water vaporised by impacts would briefly warm the global atmosphere, and precipitation from these transient water vapor greenhouse atmospheres could cut some valleys where there is evidence of rapid discharge \citep{seg02,seg08,kra08,too10}. In addition, ancient channels are patchily distributed \citep{wil07,wei08,fas08b,hyn10,car00}, which suggests (3) localized precipitation as an alternative to transient, or prolonged, global wet conditions.

Localized precipitation is especially attractive as an explanation for channels exposed by erosion of layered deposits on the Valles Marineris plateau $\sim$ 3.0 Gya \citep{man04,wei08}, and also fans at Mojave Crater $\sim$ 10 Mya \citep{wil08}. That is because these channels and fans postdate the sharp decline in globally-averaged erosion rates, aqueous alteration, and channel formation near the Noachian - Hesperian boundary \citep{mur09,fas08,gol06}. The rarity of channels, fans, erosion, or aqueous minerals of similar age suggests that the Valles Marineris plateau channel networks and Mojave Crater fans do not record global episodes of surface runoff. Localized precipitation is an alternative, with vapor sourced from a transient event such as groundwater release during chaos terrain formation (at Valles Marineris) or partial melting of an ice-silicate mixture during impact (at Mojave Crater). This has previously been suggested, but never modeled \citep{man08,wil08}.

Here we report our study of the best-preserved of the Valles Marineris plateau channel networks, west of Juventae Chasma \citep{mal10,wei10}. The Juventae plateau networks include ``hillslope rills and low-order streams,'' leading \citet{mal10} to describe them as ``the best evidence yet found on Mars to indicate that rainfall and surface runoff occurred.'' In \S2-3, we summarize geologic observations: some constitute boundary conditions for the model, others are targets that the model must reproduce in order to be considered successful. In \S4 we analyse results from Mars Regional Atmospheric Modeling System (MRAMS, \citet{raf01,mic06}) simulations of chaos flood-effect precipitation that include detailed cloud microphysics (Appendix A). We find excellent agreement between the model-predicted precipitation locations and the previously-mapped area of layered deposits and inverted channel networks. In this paper, we concentrate on what controls the rate and location of precipitation. A companion paper, \citet{kit10} ( available at http://arxiv.org/abs/1012.5077) -- henceforth `Paper 1' -- describes idealized simulations of cloud formation and precipitation in a thin, cold atmosphere perturbed by a lake. Precipitation in our model falls as snow, but the geologic observations demand liquid water runoff. Snowpack melting on the Juventae plateau will occur under certain combinations of orbital parameters and snowpack physical properties. These are determined (\S5) by running a simple snowpack thermal model over all seasons, all plausible solar luminosities, and for the full range of orbital conditions sampled by the chaotic diffusion of Mars' orbital parameters \citep{las04}.
%On the assumption that chaos/outflow channel flood events are uncorrelated with orbital parameters,
We calculate the probability that snowpack melting will occur as a function of solar luminosity and additional greenhouse effect.
In \S6, we evaluate multiple working hypotheses for the mechanism of channel formation. We compare these hypotheses to data from a small number of key localities on the Juventae Plateau. We do not carry out a comprehensive geologic analysis.
%In \S6, we compare our proposed scenario with high-resolution data from key localities on the Juventae Plateau. We do not carry out a comprehensive geologic analysis. Instead we compare our model output with the observed thickness of the plateau layered deposits, maximum drainage density, and the hydrology of 2 adjacent catchments within the plateau channel network.
%A possible Earth analog is described: supraglacial melting on the Greenland ice sheet.
Finally we test the localized-precipitation hypothesis at a second site, Echus Chasma (\S7), and consider the implications of localized precipitation for global excursions to conditions warmer than the present-day on Hesperian Mars (\S8).

\section{Geologic constraints: Valles Marineris plateau channel networks}
\noindent
The Valles Marineris plateau layered deposits are distinct from layered deposits elsewhere on Mars \citep{wei08,mur09,wei10}. Dendritic channels with both positive and negative relief are commonly found in association with the plateau layered deposits \citep{mal03,edg05,wil05}, with preserved drainage densities as high as  15 km$^{-1}$ and commonly $>$1 km$^{-1}$. %(Figure \ref{DRAINAGEDENSITY}).
In contrast, most Mars light-toned layered deposits (LLD) have few or no channels visible from orbit. Channels at different levels within the plateau layered deposits crosscut one another \citep{wei10}. This requires that plateau channel formation was either interspersed with, or synchronous with, the depositional events that formed the layers, and suggests a common geologic scenario for channel formation and plateau layered deposit formation. Opal ($\pm$ hydroxylated ferric sulfate) has been reported in the plateau layered deposits \citep{mil08,bis09}, whereas most Mars LLD show sulfate ($\pm$ hematite $\pm$ phyllosilicates) \citep{mur09}.
%(\citet{led10} reinterpret the opal detection as a blend of phyllosilicates and sulfates.)
Plateau layered deposits show no evidence for regular bedding, while most LLD give the visual impression of quasi-periodic bedding, which has been statistically confirmed at many locations \citep{lew08,lew10}. Because the most likely pacemaker for quasi-periodic sedimentation is orbital forcing \citep{kui08,lew08}, this suggests the process that formed the plateau layered deposits was not sensitive to orbital forcing. This could be because stochastic processes controlled deposition, or because deposition timescales were much shorter than orbital timescales \citep{lew10}. The deposits are thin, no more than a few tens of meters total thickness \citep{wei10}, in contrast to the many km of sulfate-bearing deposits that accumulate within craters and canyons. Some show pitted surfaces that \citet{led10} relate to the ice-rich dissected mantle terrain of the northern midlatitudes. Stratigraphy constrains the opaline plateau layered deposits to be early Hesperian or younger, so they are among the youngest classes of aqueous minerals on Mars \citep{mur09}, and significantly postdate the maximum in valley formation during the Late Noachian -- Early Hesperian \citep{led10}. Taken together, these differences suggest the process which formed the Valles Marineris plateau layered deposits differed from that forming layered deposits elsewhere on Mars \citep{mur09}.

The plateau layered deposits and inverted channels formed over an extended interval of time. Tilted channels are present on downdropped fault blocks \citep{led10}, and some channels are truncated by chasm edges. This shows that some channels formed before backwasting of the chasm to its present day form. Crosscutting streams exposed at different levels within the plateau layered deposits confirm that multiple periods of runoff are recorded on the plateau \citep{wei10}. %In order for different generations of streams to crosscut one another, channel-forming events must have been interspersed with the layer-forming events \citep{wei10},

These observations raise several questions. What was the geologic context of plateau layered deposit formation? What was the source of the water for channel formation, and what permitted surface liquid water at this location? Does this require atmospheric pressures, temperatures, or water vapor loading different from contemporary Mars -- and if so, are the required changes global, regional or local? What were the mechanics of channel formation - for example, are these mechanically-eroded or thermally-eroded channels? Our focus in this paper is on the first three questions, although we do report some measurements relevant to channel formation mechanisms in \S 6.

%Where did the water come from? Where did the sediment come from? How did mineralization occur? Does this require atmospheric pressures, temperatures, or water vapor loading different from contemporary Mars? If so, are the required changes global, regional or local?

Channels are usually found on plateaux immediately adjacent to chasms, suggesting that location on a plateau immediately adjacent to a chasm provided a driver (limiting factor) for channel formation and/or preservation (Figure \ref{VALLESMAP}a). For example, the Echus, Juventae and Ganges plateau channel networks are all immediately West (downwind) of inferred paleolakes (Figure \ref{VALLESPLOT};\citet{col07,har08}). A relief-elevation plot confirms, independently, that all light-toned layered deposits are on the plateau near the edge of a chasm (Figure \ref{VALLESPLOT}). To form, a channel requires water supply sufficient to balance continuous losses to infiltration and evaporation, while generating runoff. Runoff must encounter sediment that is fine-grained enough to be mobilized. Inverted channels have additional requirements: to be exposed at the surface today, an inverted channel must be preferentially cemented or indurated, and then incompletely eroded. Therefore, there are five potential limiting factors for inverted channel formation: - heat, water, sediment, cementing fluids, or erosion. Location downwind of a chasm should logically supply at least one.

%Mobilizeable sediment in the form of sand dunes is found on the chasm floors (Figure VALLESPLOTc, \citet{hay06}), but not on the plateau, so sediment is probably not the limiting factor. Wind speeds peak on the chasm flanks, and are less on the plateau, although it is difficult . There is no evidence for volcanic constructs or vents in the area of the, although hydrothermal activity can occur without an obvious surface expression (for example, there are hot springs in Arkansas).

We hypothesize that the Juventae, Echus and Ganges plateau channel networks formed downwind of ephemeral chaos lakes, through precipitation from lake-effect storms (Paper 1, Figure \ref{HOWITWORKS}). The spatial association between chasm lakes and plateau channels/light-toned deposits can then be explained if rain or snow from nearby lakes is the limiting factor for forming plateau channels and light-toned deposits. This is analogous to the `snowbelts' that form downwind of the Great Lakes of N America from the cumulative precipitation of lake-effect storms (Figure \ref{VALLESMAP}b). Our hypothesis predicts that simulations of lake-effect storms in Valles Marineris should produce precipitation that is localized to the chasm rim and in the observed locations.
%(2) provides a water supply sufficient to mobilize sediment.
We test our hypothesis at Juventae Chasma, because it has the best-preserved plateau channels and the geologic evidence for a paleolake there is also compelling \citep{col07,har08}. Other possible scenarios for forming plateau channels and plateau layered deposits (Figure \ref{HOWITWORKS}) are discussed in \S 6.

\section{Geologic constraints: Boundary conditions at Juventae Chasma}
\noindent
Juventae Chasma has a spillway at $\approx$ +1 km elevation that is connected to the Maja Vallis outflow channel (\citet{cat06,col07}; Figure \ref{GEOLOGYCONSTRAINTS}). The floor of Juventae Chasma is below -4 km. Insofar as material was removed from the chasm by fluvial or debris-flow transport across the spillway, this suggests that a lake many km deep existed at least once in Juventae Chasma \citep{col07}. Evidence that multiple outbursts cut Maja Vallis also suggests that Juventae Chasma was flooded multiple times \citep{har09}.

The duration of groundwater outburst floods depends on the permeability of the source aquifer, which is unknown.
\citet{har08} calculate a duration of 0.1 -- 10$^5$ days per event, with $\leq$5 x 10$^3$ days per event preferred. The corresponding peak discharge rates are 3 x 10$^5$ -- $>$10$^8$ m$^3$s$^{-1}$. When discharge falls below the evaporation rate, icing-over of the lake surface is increasingly likely. For the 140 km-diameter cylindrical chaos modeled by \citet{har08}, and an evaporation rate of 2 mm/hr, the cryosphere fractures are frozen shut before the discharge falls below the evaporation rate. Alternatively, surface liquid water in a chaos terrain can be generated by failure of a dam confining an ice-covered lake, leading to mechanical disruption of the ice cover. For example, the 2002 stepwise collapse of Antarctica's Larsen B ice shelf took 3 weeks, with open water between collapsed ice blocks (Figure \ref{LARSENB}; \citet{sca03}).

Boundary conditions are as follows. We parameterize the dependence of lake surface roughness on wind speed following \citet{gar92}. We impose a lake temperature of 278.15K (5$^\circ$C), which is midway between dissipative throttling (Joule-Thompson) heating of aquifer water initially at 5km, and two-phase hydrostatic ascent \citep{gai03}. However, for the largest lake simulated (\texttt{juventae\_high}, Table 1), we set lake surface temperature to 273.15K (0$^\circ$C). This is because the high latent-energy flux associated with a relatively high temperature and a large lake area caused numerical instabilities in the model. We hold lake temperature steady. This is a reasonable assumption if the atmospheric response timescale (hours-days) is shorter than the timescale over which the chaos flood hydrograph changes, or if surface water continuously pours over the spillway, allowing the surface layer to be refreshed by warmer water from depth. Wind-dependent lake surface roughness is from equation 7.21 of \citet{pie02}. The inverted streams at Juventae required surface liquid water to form, but currently the low atmospheric pressure at plateau elevation (+2 km) makes surface liquid water unstable \citep{con10}. Therefore, we doubled initial and boundary pressure throughout our simulation. Doubling pressure
%increases atmospheric temperature by $<$1K,
halves the peak wind speed obtained by full release of Convectively Available Potential Energy (CAPE) (Paper 1). Therefore, it is conservative in terms of localized precipitation, because ascent of air parcels to cold altitudes will be less rapid (Paper 1). We use present-day topography in our model, apart from flooding to the lake level within Juventae Chasma. A minor exception is that we remove the sulfate-bearing \citep{bis09,cat06} light-toned layered deposits from within the chasm, which would otherwise form islands in our ephemeral lake. This is because we agree with the geologic interpretation that these deposits largely postdate chasm formation \citep{mur09b}, although there is disagreement on this point \citep{cat06}. The change has no effect on our best-fit model (\texttt{juventae\_high}) because, in our smoothed topography, the lake level in \texttt{juventae\_high} is higher than the summits of the sulfate-bearing light-toned layered deposits. We use present-day solar luminosity in our precipitation model (but not in our melting model; \S5). This is conservative in terms of localized precipitation, because lower atmospheric temperatures favor localized precipitation (Paper 1). More details of the model setup are provided in Appendix A.

%Plateau layered deposits and plateau channel networks are found exclusively on the SW rim of Juventae Chasma, and only within 50 km of the chasm edge (Figure \ref{GEOLOGYCONSTRAINTS}). This localisation is an important target for our model.

\section{Juventae mesoscale model: precipitation output}
\noindent
 This paper reports the results of model runs that vary lake temperature and lake level (Table 1), holding orbital elements and obliquity at present-day values and fixing the season at simulation start to southern Summer (Ls $\sim$ 270).  An MRAMS run simulating 7 days at Valles Marineris takes $\sim$0.5 years to complete (wall time), so we are limited in the number of parameters we can vary. In addition, we must rerun the General Circulation Model (GCM) that supplies the mesoscale boundary conditions if we wish to model a major change in pressure or dust loading. This limits the range of sensitivity tests we can practicably apply. Paper 1 describes a wider range of sensitivity tests.

\subsection{Analysis of best-fit model}
\noindent
{\it Lake-driven convergence.} The lake perturbation to the atmosphere is superimposed on the complex, topographically driven Valles Marineris mesoscale windfield \citep{rafmer,spigmeso}. Strong slope winds are seen at Juventae Chasma in the \texttt{juventae\_dry} run, as previously described \citep{rafmer,spigmeso}. The diurnal cycle of upslope daytime winds and downslope nighttime winds is broken by the lake, which drives low-level convergence strong enough to overcome the upslope daytime wind (Figure \ref{TEMPANDWIND}b). At altitudes $<$6 km, the convergence is toward a sheet-like time-averaged updraft running along the lake's long  axis (N-S), and concentrated in the southern half of the lake. Above 10 km, wind moves out from this updraft. East-directed outflow encounters the west-directed prevailing wind and slows, allowing more time for ice crystals to precipitate out before reaching the chasm edge.  West-directed outflow is much faster. Above 20 km, the windfield is increasingly dominated by radial flow away from a narrow, cylindrical updraft in the center of the lake. Lake storm effects on the background windfield are minor above 40km. The combined effect of a copious water vapor supply, the availability of dust for ice nucleation, water vapor and cloud radiative effects, and induced low-level convergence results in continuous modeled precipitation.

%[Structure as Paper 1, but with emphasis on precipitation since this is the observable].
\noindent
\emph{Radiative effects}: Mean temperature rises by 17K downwind of the lake, where icy scatterers have precipitated but the greenhouse effect of vapor remains (Figure \ref{TEMPANDWIND}c). There is a broad region of $>$5K warming. On average, downwelling longwave radiation increases by 60 Wm$^{-2}$ SW of the deep pit W of S Juventae. Ice cloud scattering partly compensates for this during the day, but maximum temperature also rises, by up to 8K, which is important for melting (\S 5). However, in the area of greatest modeled precipitation, maximum temperature is reduced by up to 3K because of the locally high atmospheric ice column abundance.

\noindent
\emph{Water vapor and ice column abundance}: The time-averaged water vapor column abundance (Figure \ref{WATERCOLUMN}a) shows that high vapor abundances are confined to the canyon. Ascent of vapor-laden parcels up canyon walls aids crystallization, and ice/total water column mass ratios increase from typically $<$50\% within Juventae Chasma to 60-70\% on the plateau. Most vapor is found at low elevations. The peak precipitable vapor column abundance is $\sim$0.27 cm and is located slightly SW of the lake's areal centroid. The peak in time-averaged ice column abundance (Figure \ref{WATERCOLUMN}b) is shifted 30km further WSW. This is the product of plume ascent timescales and the WSW-directed background wind speed (Paper 1). The falloff of ice column abundance with distance from this peak is almost symmetric. This is because the outward-directed pressure gradient at the top of the buoyant plume is much stronger than that driving the background wind field. The highest water ice column abundances away from the lake are above a promontory jutting into the lake on the SW edge of the chasm, and along the S edge of the chasm. These also correspond to the highest values of precipitation, as discussed below and shown in Figures \ref{CLOUDSTRUCTURE}, \ref{SNOWVSCHANNELS} and \ref{SNOWVSLAKELEVEL}.

\noindent
\emph{Rate and location of snowfall}.
The footprint of precipitation is displaced downwind of the centroid of the lake by a distance similar to the product of vapor lifetime (loss due to ice crystal growth and subsequent gravitational sedimentation) and characteristic wind velocity at cloud height (Figure \ref{SNOWVSLAKELEVEL}). Because the size of Juventae Chasma exceeds this distance, the peak in total precipitation (1.3 mm/hr)
%and maximum precipitation rate (2.9 mm/hr)
lies within the lake. We ignore this prediction because it cannot be geologically tested, and instead only describe results for precipitation on land.

Most snow falls close to the chasm edge (Figure \ref{SNOWVSCHANNELS}). Water-ice precipitation on the chasm flanks has a maximum on a promontory southwest of the lake center.
%r, and near the center of the south wall of Juventae Chasma.
Mean precipitation is $>$0.6 mm/hr only in a narrow belt $<$40 km from the chasm edge on the SW rim of the chasm. This area of high modeled precipitation corresponds to the mapped area of channels and layered deposits (Figure \ref{SNOWVSCHANNELS}).
%The ratio of peak precipitation to mean precipitation is 1.5-3 throughout the area of greatest precipitation.

The reduction in snowfall with distance from the chasm is rapid (Figure \ref{FALLOFF}). The decline is most rapid upwind (east) of the chasm. 200 km east of the chasm rim, peak snowfall and mean snowfall are both 1000 times less than at the chasm rim. The falloff is strongly modulated by topography: at a given distance from the lake, plateaux receive 10-100 times more snowfall than canyon floors.

Maximum precipitation on land (not shown) is 1.7 mm/hr. Maximum precipitation is between 1.5 and 2.0 times the mean rate in the area of greatest precipitation. This does not include transient high snowfall during model spin-up, immediately after aerosol microphysics are switched on.

The inflection in precipitation contours at the chasm edge (Figures \ref{SNOWVSLAKELEVEL} and \ref{SNOWVSCHANNELS}) is a projection effect of the steep chasm wall slopes: fall rates per unit column atmosphere decrease smoothly with distance from the lake, but at the chasm wall the snow from this column is spread over a larger surface area.

Because precipitation is localized and vapor lifetime is short, precipitation rates are comparable to evaporation rates.

%\emph{Plume structure}.
\noindent
\emph{Mass \& energy budgets:} The mean evaporation rate found from our mass balance ($\sim$2 mm/hr, Table 2) entails 1.4 kW/m$^2$ evaporative cooling. Is our assumption of constant lake surface temperature sustainable? 1.4 kW/m$^2$ cannot be supplied from a liquid lake by conduction. Thermal convection may transfer the required heat if lake temperature is above water's temperature of maximum density (277.13K) and

\begin{equation}
{q_{convect} = 0.05 \left( \frac{\rho g \alpha \Delta T}{\kappa \nu} \right)^{1/3} k  \Delta T > 1.4 kW/m^2}
\end{equation}

where $q_{convect}$ is convective heat flow in W/m$^2$, $\rho$ = 1000 kg/m$^3$ is water density, $g$ = 3.7 m/s$^2$ is Mars gravity, $\alpha$ = 2 x 10$^{-4}$ K$^{-1}$, $\kappa$ = 1.4 x 10$^{-7}$ m$^2$/s, $\nu$ = 10$^{-6}$ m$^2$/s, $k$ = 0.6 W/m/K is thermal conductivity, and $\Delta T$ is the temperature difference across the lake \citep{pos09}. With $\Delta T$ = 2.5 K, we obtain $q_{convect}$ = 1.8 kW/m$^2$ which can sustain the evaporation. An isolated, well-mixed lake of depth 5 km would cool only 0.006 K/sol through evaporative cooling, so the assumption of constant lake surface temperature is reasonable.

Evaporitic loss from a fully open lake, 1.3 x 10$^4$ m$^3$s$^{-1}$ (= 1.9 mm/hr) in our best-fitting model, is small compared to maximum Maja Valles outflow channel discharge (1.1 x 10$^8$ m$^3$ s$^{-1}$; \citet{kle05}).

\subsection{Sensitivity to flooding depth}
\noindent
The location of precipitation is sensitive to flooding depth, and flooding depths close to the spillway provide the best match to observations (Figure \ref{SNOWVSLAKELEVEL}). As lake level rises, the offset of maximum precipitation from the lake center changes. The main change on land as the flooding depth is moved closer to the spillway is that precipitation to the S of Juventae Chasma is reduced. This is because the area of convergence shifts NW and then N, tracking the centroid of lake area.
%In addition, the convergence corresponding to the lowest flooding depth (and smallest lake) creates a much weaker plume. This allows the diurnal
%In addition, the plateau slopes down from 2500m at the S end of Juventae Chasma to 1000m at the N end, so air has to ascend furthest at the S end of the canyon. This tends to favor ice crystallization and precipitation. As lake level rises, the amount of rise

More than 80\% of vapor released by the lake is trapped in or next to the lake as snow (Figure \ref{TERNARYFATE}). As plume intensity increases, so does the fraction of water vapor that snows out locally. Plume intensity - which we define using updraft velocity and cloud height - increases with increasing lake size (Paper 1) and with increasing lake temperature. Therefore, higher flooding depths within Juventae Chasma lead to more localized precipitation. Atmospheric water does increase with increasing lake area, but slowly (Table 2). Only \texttt{juventae\_low} and \texttt{juventae\_med} are directly comparable because of the lower lake temperature in \texttt{juventae\_high}.

%\subsection{Why are storms continuous?}
%\noindent
%Earth storms have a lifetime limited by hydrometeor drag. Unless strong shear horizontally separates rain-induced downdrafts from condensation-driven updrafts, hydrometeor drag shuts off the updrafts that supply vapor from low altitudes. In our simulations, hydrometeor loading has little obvious effect on the storms. Strong shear is present, but the low effectiveness of rain drag may also contribute. This can be parameterized as the ratio of the buoyancy change due to hydrometeor drag to the bouyancy change due to the latent heat released upon condensation of those hydrometeors:
%
%\begin{equation}
%{K_{rd} = \frac{\Delta B_{hm}}{\Delta B_{lh}}}
%\end{equation}
%
%where $K_{rd}$ is the rain drag effectiveness factor ...
%Earth rain has $K_{rd}$ = X and Mars snow has $K_{rd}$ has $K_{rd}$ = Y.

\subsection{Comparison between area of modeled \\ precipitation and area of observed channels}

\noindent
We use four independent metrics to quantify the agreement between geologic data and precipitation model that is qualitatively apparent in Figure \ref{SNOWVSCHANNELS}. Because (i) channels are exposed by incomplete erosion of the plateau layered deposits, (ii) the inverted channels are themselves layered, and (iii) channels are found wherever layered deposits are incompletely eroded on the Juventae plateau, we assume that areas containing light-toned layered deposits also contain channels.

\smallskip

\begin{list}{$\bullet$}
{ \setlength{\labelwidth}{2cm} }

\item \textit{Method of \citet{pie78}}. Following \citet{pie02}, let $A_4$ be the area of the inmost grid in the MRAMS domain, $P_{i,t}$ be the area of precipitation predicted by meteorological model $i$ above some threshold value $t$, and $G_j$ be the area of channels and light-toned layered deposits (collectively referred to as ``geology'', for brevity) mapped using criteria $j$. Then the skill of the model in predicting the ancient channels is

    \begin{equation}
    {skill = \frac{F_E}{F_M} = \frac{ (P_{i,t} \cap G_j) / G_j}{ P_{i,t} / A_4 } }
    \end{equation}

    where $F_E$ is the fraction of the geology in areas with model precipitation above the threshold, and $F_M$ is the fraction of the model domain over which the model precipitation is above the threshhold. The fraction of the model-predicted area that is occupied by geology, or coverage

    \begin{equation}
    {coverage = (P_{i,t} \cap G_j) / P_{i,t}}
    \end{equation}

    is a measure of the tendency of the model to overpredict the observed geology (where a value of 1 corresponds to no overprediction). Here, the null hypothesis is that the association between model and data is due to chance. $skill$ and $coverage$ are shown in Table 3. \texttt{juventae\_high} shows the greatest skill, for both $j$. We obtain $skill$ values $>$200 for \texttt{juventae\_high} and $t$ = 0.8 mm/hr. For these cases, the area of overlap between prediction and data is $>$200 times that expected by chance, showing that the agreement is almost certainly not due to chance.

\item \textit{Monte Carlo.} For this metric, we determine the fraction of randomly placed precipitation templates that provide a better fit to the geologic observations than the modeled precipitation. We take the predicted precipitation $P$ for each model $i$, including snow falling back into the lake (the precipitation `template'), and randomly translate it in latitude and longitude, wrapping around the boundaries of the inmost model grid ($\sim$1300 km  x $\sim$870 km). This creates a new modeled precipitation grid $Q_{i}$. A data grid $G_{j}$ is constructed by assigning 1 to areas containing mapped inverted channels or light-toned layered deposits and 0 otherwise. We then calculate the cross-correlation as

    \begin{equation}
    {X = G^{\prime}_{j} * Q^{\prime}_{i}}
    \end{equation}

    where the primes denote normalization (for each trial) by subtracting the nonlake mean and dividing by the nonlake standard deviation. The area of the present-day lake is masked out in all cases. We repeat this 10$^4$ times per model. Here, the null hypothesis is that the geology formed from a localized source of some kind, but that the location of this source was uncorrelated with present-day Juventae Chasma. The resulting p-values are 0.005 for \texttt{juventae\_high} (nominal $X$ = 0.476), 0.009 for \texttt{juventae\_med} (nominal $X$ = 0.365), and 0.058 for \texttt{juventae\_low} (nominal $X$ = 0.044). These are for the map of \citet{led10}. The null hypothesis is rejected at the 99.5\% level for for \texttt{juventae\_high}: if the observed channels and layered deposits formed from a localized source of some kind, it was almost certainly close to the present-day center of Juventae Chasma.
\item \textit{Azimuth.} We assign each land pixel to its nearest pixel on the perimeter of the lake. We then assign each perimeter pixel to its normalized distance along the circumference. Both the perimeter-matched geologic data and the perimeter-matched model output are then smoothed with a gaussian kernel of full width at half maximum equal to 5\% of lake circumference. The results are shown in Figure \ref{AZIMUTH}. All models produce a broader distribution of snowfall with azimuth and are also biased counterclockwise (from SW to S) with respect to the data. This bias is least severe for \texttt{juventae\_high}. \texttt{juventae\_high} also shows the best overall fit to the data: by cyclic translation of the smoothed and area-normalized precipitation pattern, we can find the percentage of perturbed patterns that would provide a better least-squared fit to the smoothed data than our actual model. For the \citet{led10}  (\citet{wei10}) geology, this percentage is 7\% (10\%) for \texttt{juventae\_high}, 12\% (16\%) for \texttt{juventae\_med}, and 36\% (39\%) for \texttt{juventae\_low}. (Note that the azimuthal distributions shown in Figure \ref{AZIMUTH} are maximum-normalized, not area-normalized).

\item \textit{Falloff of precipitation.} Figure \ref{FALLOFF} shows cumulative distribution functions (CDFs) of geology and precipitation. All geology is found within 60 km of the chasm edge, but the $e$-folding distance of the cumulative precipitation is significantly larger, 70-110 km. $\sim$95\% of snow is found within 250 km of the chasm edge.

\end{list}

\medskip
\noindent
All these metrics show that the area of precipitation is more extensive than the mapped area of channel networks. Extensive erosion has occurred and mapping efforts are incomplete, so this may be an artifact of incomplete preservation and high-resolution imaging. For example, there is a 25km-diameter ejecta blanket in our area of highest modeled precipitation that may be obscuring underlying channels. If precipitation is necessary to form or indurate plateau layered deposits, decline in the precipitation at 100-200 km from the chasm (Figure \ref{FALLOFF}) will create thin, or weakly-indurated deposits. These could easily be missed during mapping, removed by the wind, or both. It is possible that the relatively low horizontal resolution of our simulations is artificially broadening our modeled precipitation. Alternatively, there may be a threshhold precipitation level to produce layered deposits and channels. For example, sublimation losses may occur during the months between deposition and the beginning of the melt season. These sublimation losses are not included in our simulations, and will thin the snowpack everywhere while reducing the area of remaining snow. Sediment mobilization is very sensitive to small changes in runoff:

\begin{equation} Q_{sed} \propto (\tau_b - \tau_{crit})^{3/2} \end{equation}

where $Q_{sed}$ is sediment flux, $\tau_b \propto H$ is bed shear stress and is proportional to flow depth $H$, and $\tau_{crit}$ is critical bed shear stress. If snow availability is a limiting factor, this will focus channel incision on areas with high precipitation rates. %      If we make the tentative, but reasonable assumption that lithification requires some liquid water \citep{lew08}

%Discuss: which model is more reasonable?

%However, the model also predicts high precipitation on the S chasm rim where no channel networks, or layered deposits, have been observed. This mismatch may indicate that our model is in error. Alternatively, channel networks may have formed, but been eroded by backwasting of the chasm walls or by aeolian erosion (too much erosion). Backwasting of wallslopes continues - landslides with Amazonian CRA at Valles Marineris and the young CRA of the talus. Alternatively, channel networks may have formed, but be buried beneath a spectrally neutral caprock (too little erosion). Significant snowfall to the S occurs only during the night. During the day, precipitation occurs mostly to the SW. Therefore, our assumption of constant lake surface temperature may be to blame for this difference between model output and observation. If the lake surface cools (or freezes over) during the night, snowfall to the S would be reduced, and the discrepancy would disappear. Another possibility is that vapor release to the atmsphere occurred only during spillway dam failure, and that open water occurred mainly close to the spillway in the N of the chasm while the ice cover to the S remained nearly intact. This would bias precipitation Nwards, as required to match observations.

\section{Will snow melt?}
\noindent
We use a simple probabilistic melting model, described in Paper 1, to determine the likelihood of melting for the range of orbital conditions possible on pre-modern Mars. Our model is probabilistic because the true orbital parameters are not known for ages $\gg$10 Mya \citep{las04}. This model tracks energy flow in snowpack as a balance of radiation, conduction, and evaporative cooling. A constant downgoing longwave flux from the atmosphere of 55 W/m$^2$ is assumed. This is approximately the peak afternoon, equatorial, perihelion greenhouse effect in low-dust conditions on contemporary Mars \citep{lew99,dddmcd}, and a constant longwave flux at this value is found to correctly reproduce the effect of the atmosphere on annual peak temperature on contemporary Mars (Paper 1). The probability of melting Mars snowpack depends on surface albedo, snowpack thickness, latitude, solar luminosity, orbital conditions, and the atmospheric state. The Juventae channels are at 4$^\circ$S, and all our calculations are for the equator. We assume a pressure of 1220 Pa, and so we ignore transfer of sensible heat from the surface to the atmosphere. \citet{hec02} shows that sensible heat losses are more than 10 times smaller than evaporative cooling for this low pressure. As in Paper 1, we assume 70\% humidity, which is optimistic in terms of melting because it reduces evaporative cooling by $\approx$31\% relative to the zero-humidity case.

%(Michael comment: May be some irrelevant stuff).
%
%[Get dust content and compare to \citet{clo87} tables. Nominal content is about 10 ppm which is too small for albedos of 0.3 - 0.4 which are required for melting.]
%[My grains precipitate as Mars Snow Class II in the sense of \citet{clo87} table 1.]
%[Marsh \& Woo]

The probability of melting depends on snow albedo, which is high for pure snow but much lower for realistic, dust-contaminated snow. \citet{war80} show that 1000 ppmw dust can reduce ice albedo from $>$0.9 to 0.3. In the words of \citet{lan05}, ``Water ice is very bright in the visible spectrum when clean, but even a small amount of dust contamination can reduce the albedo to values close to that of the dust itself if the dust grains are embedded in ice grains.'' \citet{clo87} shows that 1000 ppmw dust reduces snow albedo to 0.45-0.6 for grain sizes 400$\mu$m - 100$\mu$m, respectively. This is for precipitation grain sizes in our model; metamorphism will increase grain size and decrease albedo. The mean bolometric albedo of bright regions in Mars' North Polar Residual Cap is inferred to be 0.41 from energy balance \citep{kie76}. Near-infrared spectroscopy has identified seasonal water ice layers up to 0.2 mm thick on pole-facing slopes in the Mars low latitudes \citep{vin10}. Analysis of the spatial and seasonal dependence of these detections indicates that low-latitude surface water ice has albedo 0.3 -- 0.4 \citep{vin10}. Modelling of OMEGA data indicates that water-rich terrains in the South Polar Layered Deposits have albedo $\sim$ 0.3 -- 0.4 (Figure 7 in \citet{dou07}). Measurements of the gray ring component of Dark Dune Spots in Richardson Crater at 72$^\circ$S show it to be composed of seasonal water ice deposits with an albedo of 0.25 -- 0.30 \citep{ker11}. When melting starts, the albedo of dust-contaminated ice remains low because ``when snow melts, the impurities often tend to collect at the surface rather than washing away with the meltwater'' \citep{war84}, forming a lag. Water has a low albedo, so stream and melt pond albedo is lower than unmelted surface albedo. \citet{gar10} show that 2 ppmw soot can greatly reduce snow albedo. Soot is 200 $\times$ more optically effective than Earth crustal dust, and presumably more effective than Mars dust. We model albedos from 0.28 (the albedo of Mars' dust continents; \citet{mel00}) to 0.4.

The melting is also sensitive to orbital conditions. For example, at the equator, high eccentricity is more favorable for melting than is low eccentricity.

Our preferred melting scenario is melting of snowpack on the plateau during or shortly after the lake storm (although it is also possible that snowpack persists for multiple years, and a separate transient warming event occurs). We assume that the snowpack persists until the next melting season. Low-latitude sublimation rates modeled by GCM \citep{mad09} suggest this requires snowpack thicknesses $>$0.25 m w.e. Lakes with lifetime $>$10$^2$ hours generate this snowpack, given our modeled precipitation rates. Annual maximum temperatures are then the metric relevant to melting. With these assumptions, the melting probabilities for equatorial snowpack are those shown in Figure \ref{MELTPROB}. The left panel is for melting if the storm does not continue through the melt season, and the right panel is for melting if the storm does continue through the melt season.

%\medskip
%\begin{description}
%\item[Melting Option 1: (Preferred)] {\it

%\item[Melting Option 2: (Less favored)] {\it Snowpack persists for multiple years,\- and a transient warming event occurs.} Snowpack lifetime is set by sublimation rate. Because of the nonlinear dependence of saturation vapor pressure on temperature, this lifetime can be extended by formation of a dusty lag deposit. A longer lifetime would increase the probability that the snowpack experiences subsequent transient warming events. Any event that lowers the albedo or raises the basal temperature could generate melting. Chaos regions and outflow channel source regions are thought to have been triggered by volcanic activity (e.g., \citet{mck99}), although we have no direct evidence of this association at Juventae Chasma. Effusive eruptions could inject greenhouse gases to the regional atmosphere, shallow intrusions could melt the snow from below, and explosive eruptions produce ash that could darken the snow. Snow tends to darken as it metamorphoses, since larger grains reduce the number of scattering surfaces. This would help melting. Ejecta from small impacts could also locally melt the snowpack.
%\end{description}

%It is though that the observed expansion of Greenland's supraglacial melt zone arises from an out-of-equilibrium situation: human increase of atmospheric CO2 concentrations. We propose that the Juventae channnels arise from an out-of-equilibrium situation: rapid localized precipitation delivers ice to where it is not stable.

Once liquid water is flowing, several feedbacks can extend its lifetime. In Greenland, supraglacial channels with water depth $>$0.5m reduce the albedo to close to that of water ($\sim$ 0.05) \citep{lut06}. Viscous dissipation can balance evaporative cooling only if

%Dissolution of salts can lead to freezing-point depression. The Juventae plateau layered deposits contain an hydroxylated ferric sulfate which could be copiapite (Fe$^{2+}$Fe$_4^{3+}$[OH(SO$_4$)$_3$]$_2 \cdot$ 20 H$_2$0), hydronium jarosite ( (H$_3$O)Fe$^{3+}_3$(SO$_4$)$_2$(OH)$_6$ ), or szomolnokite( Fe$^{2+}$(SO$_4$)$\cdot$(H$_2$O)). If water flows over jarosite and dissolves some jarosite, complete freezing will not occur until X K (Y K for copiapite; Z K for szomolnokite).

\begin{equation}
S > \frac{eL}{\rho g u D}
\end{equation}
 where $S$ is river slope, $e$ $\sim$ 1 mm/hr is evaporation rate, $L$ is the latent heat of vaporization, $\rho$ = 1000 kg/m$^3$ is density, $g$ is Mars gravity, $u$ = $O$(1)m/s is stream velocity, and $D$ = $O$(1)m is stream depth \citep{clo94}. This condition is not satisfied for the Juventae plateau channel networks (which have slopes $<$1\%; Table 4), so the stream will be roofed over by ice. Insulation of streams by ice can greatly extend liquid water lifetime \citep{clo87}.

%[Is rate sufficient to mobilize sediment?]

For the purpose of generating runoff, there is an optimum thickness of porous snowpack that scales as the diurnal thermal skin depth. If the porous layer is deeper than the diurnal thermal skin depth, melt percolating from the surface will refreeze.

%\section{Comparison of sedimentology and \- stratigraphy \- to model results}

\section{Discussion of channel and layered deposit formation}

In this section, we evaluate formation mechanisms for the layered deposits and interbedded channel networks. First we review constraints, including measurements from our own DTMs. We consider a series of options for the channel-forming mechanism, and test them against these constraints. These different mechanisms all assume that water is sourced from ephemeral lakes as in \S2 - \S5 (bottom row in Figure \ref{HOWITWORKS}). Next we drop the assumption that water is sourced from ephemeral lakes, and describe entirely different geological scenarios (top row in Figure \ref{HOWITWORKS}). We do not find decisive evidence that allows us to choose between these channel forming mechanisms or geological scenarios, and so are left with multiple working hypotheses. We conclude the section with a list of future tests that could help decide among these multiple working hypotheses.

In agreement with previous work \citep{man08,wei08}, we interpret the networks of sinuous ridges and troughs to be fossil fluvial channels (Figure \ref{GEOLCONTEXT}). Juventae plateau channels have low to moderate sinuosity and we have only found 2-3 highly sinuous (meandering) channels. This should be compared to Gale-Aeolis-Zephyria, where most channels meander \citep{bur09}. Most channels run down the present-day 0.2$^\circ$ NNE regional slope \citep{man08}. Sinuous ridge width distribution appears bimodal. We suspect that the broader $O$(10$^2$m) sinuous ridges are inverted valleys. The sinuous ridges are probably not eskers. Single channels transition between positive relief and negative relief. The sinuous ridges show horizontal layering, are flat topped, locally highly sinuous and reveal no coarse grains at HiRISE scale. These attributes do not resemble Earth eskers, nor Mars eskers in southern Argyre Planitia \citep{ban09}.

Given that the channel networks are fossil fluvial channels, endmember channel-forming mechanisms are thermal erosion into ice, and mechanical erosion of sediment. These have corollary implications for the composition of the light-toned layered deposits: in the thermal erosion endmember case, the present-day plateau layered deposits contain significant relict ice. An example on Earth of this endmember case is the snowmelt-fed channel network that forms each summer on the Greenland ice sheet. In the mechanical erosion endmember case, the present-day plateau layered deposits are composed of (indurated) sediment grains. We seek to understand whether the observations favor one or the other endmember mechanism.

\subsection{Implications of constraints for channel-forming mechanisms}

\noindent
\underline{Many of the channels are preserved in inverted relief:}
%The Juventae channels are often inverted, but this does not exclude a snowmelt origin.
On Earth channels can become inverted through cementation of channel fill, armoring by coarse grains of the channel floor against erosion, or infilling of the channel by an erosionally resistant material such as lava \citep{wil09}. We do not know how the Juventae plateau channel networks became inverted, but careful study of similar channels elsewhere suggests increased cementation of the channel thread, followed by differential erosion, as the most likely cause \citep{wil09,bur10}. The inverted channels lack evidence (such as pitting) for sublimation . They are often horizontally or subhorizontally layered, and the layers have variations in tone. They have well-defined, smooth, flat tops. These observations suggest that the inverted channels are composed of sediment and are not ice-cored. Sediment fill is more consistent with the mechanical erosion endmember scenario than the thermal erosion endmember scenario.

However, inverted channels on Mars do occur in ice-dominated, supraglacial and proglacial settings.  Midlatitude Amazonian glacier-associated channels at Lyot Crater (HiRISE image ESP\_016339\_2225), Acheron Fossae (HiWish image ESP\_018178\_2165), and E of Reull Valles (HiWish image ESP\_020055\_1410) have become inverted \citep{fas10}. The Acheron inverted channel is $\leq$80 Ma based on the age of its source glacier \citep{fas10}. In each cases, the glacier-associated channels appear to have incised into ice, but to have been incompletely filled with debris and sediment during or after incision. As the surrounding ice retreats due to sublimation, the channel-filling sediment is left as a sinuous ridge. Therefore, the interpretation that the inverted channels are composed of (indurated or cemented) sediment fill does not rule out the thermal erosion endmember scenario.

%Because global Mars climate $\leq$80 Mya is thought to have been cold and dry, with runoff only on or next to steep slopes, the formation and inversion of channels is not by itself evidence for a warmer climate than today.

\vspace{0.03in}
\noindent
\underline{The layered deposits are tens of meters thick:} The present-day thickness of the plateau layered deposits close to the chasm edge is 43$\pm$11 m ($n$ = 13), measured from HiRISE DTMs. 33-39 layers are visible in HiRISE images of the thickest exposures (defining layers using laterally continuous changes in tone, slope or erosional morphology).

In the thermal erosion endmember scenario in Figure \ref{HOWITWORKS}, the layered deposits are primarily water ice. The deposit thickness of 44$\pm$13m then requires a minimum of 1400-2600 sols to accumulate at our peak precipitation of 0.9 mm/hr, ignoring sublimation losses. The peak discharge of Maja Valles is 1.06 x 10$^8$ m$^3$s$^{-1}$ \citep{kle05}. The volume of Juventae Chasma below the pour point is $\sim$7x10$^4$ km$^3$ (ignoring the volume of Maja Valles itself). If this missing volume is 70\% rock by volume and was exported over the spillway, then with a fluvial sediment concentration of 1\% by volume, the minimum cumulative time to carve Maja Valles is 740 sols. Peak discharge exceeds mean discharge, so Maja Valles operated for longer than this calculation suggests. In addition, an unfrozen lake may have remained in Juventae after the end of channel formation. Making the conservative assumption that there is only open water at the lake surface only when a catastrophic flood is occurring, the predicted deposit thickness is 28\%-53\% of the observed thickness. Because this is a lower limit, we do not think this discrepancy rules out the thermal erosion endmember scenario.

%They appear to have irregular thickness. While it is possible that additional layers exist below the limit of resolution, if we take 1m thickness as representative we obtain a single flood duration of $\sim$1000 hours. This is consistent with groundwater model predictions of the duration of outflow channel formation events: e.g. 400-21000 hours (\citet{and07}, their Table 1). However, if the flood is caused by ice dam collapse (the `slow-filling' model of \S4.3) then these groundwater models will not yield the correct flood duration. In summary, the thermal-erosion endmember scenario is consistent with mass balance.

The thickness of the deposits is harder to explain in the mechanical erosion endmember scenario (Figure \ref{HOWITWORKS}), in which the present-day plateau deposits are (indurated) sediments. %The localized-precipitation hypothesis does not explain the source of the sediments in the mechanical erosion endmember scenario.
Enough atmospherically-transported material (sand, dust, and volcaniclastic material) must be brought in from other regions of a primarily dry Mars to account for observed thicknesses. For realistic cumulative outflow channel activity durations, not enough atmospheric dust flows into the storm zone {\it during the storm events} to precipitate as ice nuclei and account for observed thicknesses. It is not possible for the layered deposits to predate the channel-forming events, because channels at different levels within the deposit crosscut each other. Instead, the sediment would have to accumulate on the plateau between storm events as sand, dust or volcaniclastic materials, or (less likely) be supplied from the chasm floor during the early stages of outflow events by buoyant plumes. In this case, the lake storms would be responsible for the channels, but would not be the  source of the material making up the layers. However, fluid released during snow melting could be responsible for the induration or cementation of these materials, and thus their long-term preservation (Figure \ref{HOWITWORKS}).

\vspace{0.03in}
\noindent
\underline{Drainage density is up to 15 km$^{-1}$:}
%The plateau channel networks are preserved both in negative relief and as inverted channels. They are exposed by aeolian erosion of the plateau layered deposits.
Drainage density (units m$^{-1}$) is the sum of the channel lengths within a region of interest, divided by the area of that region. Impact ejecta locally armors the layered deposits against aeolian stripping, forming pedestal craters.  The highest drainage density that we have observed,  15 km$^{-1}$, is in the retreating margin of one of these ejecta blankets (Figure \ref{GEOLCONTEXT} shows context, Figure \ref{DRAINAGEDENSITY}a shows detail). Since erosion by the wind preferentially removes finer channels, it is possible that this maximum preserved value is representative of the original drainage density.

The drainage density observation is entirely consistent with the thermal erosion endmember channel-forming scenario (Figure \ref{HOWITWORKS}): supraglacial channels on the Greenland ice sheet today have comparable drainage densities (e.g., 29 km$^{-1}$ in the typical case shown in Figure \ref{DRAINAGEDENSITY}b).

The mechanical erosion endmember case is harder to reconcile with high drainage densities, given that our model predicts low rates of snowfall and snowmelt. On Earth, high drainage densities are found on steep topography (e.g., hillslopes in badlands) and in areas subject to rare, intense rainstorms. Neither of these factors is likely to have been present on the Juventae plateau, which has channels slopes typically $<$1$^\circ$. Several factors could mitigate the difficulty of producing high drainage densities from mechanical erosion driven by snowmelt on the Juventae plateau. Ground ice is stable at Juventae for obliquities $>$32$^\circ$ \citep{mel95}, which occur about $\frac{2}{3}$ of the time \citep{las04}. Using the same modelling approach that correctly predicted ground ice depth at the Phoenix landing site \citep{mel09b}, ground ice at Juventae is predicted to form an impermeable layer at $10^{-2}$m - $10^{-1}$m depth within the soil \citep{mel95}. A shallow impermeable layer would favor runoff over infiltration, so for a given afternoon melt rate, would produce higher peak discharges. These would lead to a higher drainage density than in the absence of a shallow impermeable layer. Assuming dry conditions between storm events, the sediment above any ice table is unlikely to have much inter-grain cohesion. This would make it easier for runoff and shallow groundwater flow to dislodge sediment grains from the channel head, lengthening the channel network and increasing drainage density.

High drainage density is also entirely consistent with rainfall. The difficulty of raising the atmosphere's temperature high enough to allow rainfall \citep{wordsworth,wordexoclimes} deters us from invoking rainfall so long as a snowmelt alternative is possible.

%Some light-toned material is found on the chasm wall \citep{wei10}, suggesting that light-toned material may have continued to accumulate after backwasting of the chasm to form present-day topography.

\vspace{0.03in}
\noindent
\underline{Drainage basin and channel dimensions and slopes:} Even if the channels were cut by thermal erosion, some of them must have been infilled by sediment in order to form inverted channels (Figure \ref{HOWITWORKS}). If the channels are formed in fine sediment by mechanical erosion, then sediment transport is required to form both the positive and the negative-relief channels. Therefore, in either endmember case, sediment must be transported through the observed channel network. The critical runoff $R$ needed to initiate sediment transport in a preexisting channel is \citep{per06}:

\begin{equation}
{R_{crit} = \frac{1}{A} \frac{w^2 \rho^{\prime} \tau^{*}_c D}{wS - 2 \rho^{\prime} \tau^{*}_c D } \left( \frac{8 \rho^{\prime} g \tau^{*}_c D }{ f} \right)^{ \frac{1}{2} }}
\end{equation}

where $R_{crit}$ is the critical runoff rate, $A$ is drainage area, $w$ is channel width, $\rho^{\prime} = (\rho_s / \rho_f ) - 1$ is the normalized density, $\rho_s$ is sediment density, $\rho_f$ is fluid density, $\tau^{*}_c \approx$ 0.05 is the critical Shields number, $D$ is sediment grain diameter, $S$ is slope, $g$ is Mars gravity, $f$ is the Darcy-Weisbach friction factor, and all units are mks. We assume basaltic sediment $\rho_s$ = 3000 kg/m$^3$ and water density $\rho_f$ = 1000 kg/m$^3$. In alluvial streams, $f$ is closely related to grainsize, but no similar relationship has been published for thermally-eroded channels. $A$, $w$, and $S$ are taken from our DTMs, as shown in Figure \ref{STRATDETAIL}, tabulated in Table 4, and described below. {\it Area:} Erosion has scoured the deposits between preserved inverted channels, removing the drainage basins that once sourced those channels. We are therefore forced to divide the area between channels by equidistance. Because there have been multiple episodes of crosscutting flow, some truncation of channels by subsequent generations of channel may have occurred, which may lead to a systematic underestimate of $A$ and corresponding overestimate of $R_{crit}$. {\it Width:} We measure channel widths using the distinctive, light-toned region at the top of the sinuous ridges visible in HiRISE images. DTMs confirm that this light-toned strip forms the summit of the much broader ridge, and often resolve a break-in slope near the top of the ridge approximately corresponding to the light-toned strip in the red-filter HiRISE images. By analogy with inverted channels near Green River, Utah \citep{wil07utah}, these observations suggest that the bright region whose width we are measuring corresponds to an erosionally resistant and vertically thin channel-fill deposit capping underlying weaker material. The channel widths we obtain are several meters for drainage areas $O$(1) km$^2$ (Table 4). These widths may correspond to indurated channel fill (or valley fill) sediments that are wider than was the instantaneous channel. {\it Slope:} We take an average slope along the exposed length of the channel. There is no visually obvious evidence of major postdepositional tilting of the plateau - the channels run approximately down the present-day slope. We do not understand why the channels frequently run parallel to each other before confluence.

 $R_{crit}$ for a catchment of area 1.6 km$^2$ and slope 0.8$^{\circ}$ feeding a channel of width 3.3m is shown in Figure \ref{DTMANALYSIS}. (The adjacent catchment of area 2.1 km$^2$ produced almost indistinguishable results). Both networks are from DTM1 (Appendix B).  We assume $R \approx M$, where $M$ is the melt rate. This is a reasonable approximation if background climate conditions are subfreezing and there is a shallow impermeable ice table, or if the substrate is covered with fine-grained material that has a low infiltration rate. An upper limit on $M$ is if all precipitation melts upon reaching the ground. Then the resulting runoff can mobilize coarse gravel (thick black line in Figure \ref{DTMANALYSIS}). However, a more realistic melt rate is 0.1 mm/hr. This is still capable of initiating the transport of coarse sand and fine gravel through the network (lower bound of gray envelope in Figure \ref{DTMANALYSIS}). The upper limit on the gray envelope in is an order-of-magnitude error intended to capture errors in the precipitation model, weather, and especially the width and area measurements discussed above. We conclude from the area covered by this gray envelope that if boulder-sized ($>$256 mm diameter) clasts have been transported through these Juventae channel networks, that would be strong evidence against localized precipitation. \citet{howpres} provides criteria for distinguishing fluvially-transported boulders from postdepositionally cemented blocks. Any clast that can be resolved by HiRISE is a boulder.

\vspace{0.03in}
\noindent
\underline{Pitted upper surfaces:} The plateau layered deposits have pitted upper surfaces. Pits are elongated approximately parallel to the present-day prevailing wind. \citet{led10} interpret some of these irregular depressions as sublimation pits. If correct, this would imply that the layered plateau deposits contain relict ice. It is unclear from HiRISE images (e.g., PSP\_008853\_1760) whether these are in fact sublimation features. Because Juventae is equatorial, the usual criterion employed for recognizing sublimation features in the midlatitudes -- north/south slope asymmetry -- cannot be used.

\subsection{Alternative interpretations and tests}
\noindent
In \S 2, we inferred that location adjacent to a chasm must supply one or more limiting factors for inverted channel formation, and that likely limiting factors included availability of snow or rain, sufficient heat for melting, availability of mobilizable sediments, cementing fluids, and incomplete erosion. In \S 3 - \S 5, we developed and tested the hypothesis that precipitation is limiting. The results are consistent with the hypothesis. What about heat, sediment, cementing fluids, and erosion? Can we rule any or all out?

Precipitation on the Valles Marineris plateau sourced from ephemeral chaos lakes is one parsimonious model compatible with the current data, but other scenarios are possible (Figure \ref{HOWITWORKS}). For example, we cannot currently rule out a scenario where the observed narrow, extended Valles Marineris cloud trails (0.4 micron effective diameter; \citet{cla10}) increase spectacularly in mean grain size and move closer to the canyon edge during different orbital conditions or dust loading, permitting regional precipitation. Near-surface water-ice morning fog occurs in Valles Marineris today \citep{moh09}, and OMEGA has detected surface water ice on the northern wall of Coprates Chasma \citep{vin10}. Net annual ice accumulation within the Valles Marineris occurs in GCMs at high and moderate obliquity \citep{mad09}, although this is largely due to the high thermal inertia of the chasm walls (J.B. Madeleine, via email). Taken together, this evidence suggests that Valles Marineris is a preferred site for equatorial ice precipitation, which is important at high obliquity \citep{mis03,for06}. In this scenario, snowpack is broadly distributed in the Valles Marineris region, but generally does not melt. This possibility is discussed by \citet{led10}. The limiting factor provided by the chaos regions is then airborne darkening agents (ash and/or debris) lofted by buoyant plumes (``dirty thunderstorms''; \citet{van10}), which lead to patchy melting or patchy preservation of a broadly distributed, preexisting snowpack. Alternatively, the ash and/or debris forms a cast of the channels, which form over a broad region but disappear elsewhere when the snowpack sublimes. A weakness of this scenario is that both the observed fog and surface ice, and the predicted high-obliquity net ice accumulation, are on the chasm floor and walls rather than the adjacent plateau. Figure \ref{VALLESPLOT} shows that opal-bearing layered deposits and channel networks are overwhelmingly found on the plateau, not the floor and walls.

Another alternative source of water is springs, before chasm opening had begun. Spillover of blister aufeis, or  groundwater, could source springs \citep{gai03,mur09b}. In this scenario, the layers and channel networks predate outflow channel formation. This would not explain the downwind preference for channel networks, but there is a 7\% possibility that this could be due to chance (\S4, Figure \ref{AZIMUTH}).

Among these three scenarios, our preference for the late formation model is tentative, but testable. First, a more complete study of the channel networks, using HiRISE DTMs where possible, and using channel width as a proxy for discharge (\citet{mon01} and references therein), would test the spring hypothesis and the precipitation hypotheses. If the channels were fed by a small number of point or line sources, channel widths should be constant between confluences (and channel sources should correlate to linear fractures, shear bands, or mineralogical anomalies). If precipitation or distributed groundwater flow fed the channels, channel widths should increase with contributing area as seen on Earth, and channel sources should not be correlated with fractures or mineralogical anomalies. Second, further tests of the localized-precipitation model at sites including Ganges Chasma, Cerberus Fossae and Mangala Fossae might uncover inconsistencies that would weaken the model's application at all sites including Juventae. We report initial results from one such test (at Echus) in \S 7.

Additional predictions are specific to the thermal-erosion endmember scenario (Figure \ref{HOWITWORKS}): failure of these predictions would not rule out the mechanical-erosion endmember scenario (Figure \ref{HOWITWORKS}). The thickness of the layered deposits should decrease away from the chasm edge in proportion to the mean precipitation rates predicted by localized precipitation models (e.g., Figure \ref{SNOWVSCHANNELS}). Finally, SHARAD observations can test if the light-toned layered deposits have a dielectric constant consistent with ice, as is required by the thermal-erosion endmember scenario. SHARAD has a free-space wavelength of 15m, so the light toned layered deposits are thick enough for this test.

We do not understand the cause of the opal and hydrated-ferric-sulfate mineralization of parts of the Juventae plateau layered deposits. In the K'au desert downwind of the active Kilauea caldera on Hawaii's Big Island, opal and jarosite depositional coatings accumulate with minimal liquid water \citep{sch06,che10}.  This Earth analog may not be directly applicable to Valles Marineris, because at Kilauea opal mineralization is not found more than a few kilometers from the volcanic vents, in contrast to the more extensive areas of opal mineralization on the Valles Marineris plateau \citep{che10}. Because of the quantities involved, the presence of inverted channels appears to set a more stringent lower limit on liquid water availability on the plateau than does the opal and hydrated-ferric-sulfate mineralization.

\section{Echus plateau}

\noindent
The greatest of the outflow channels is Kasei Valles \citep{wil00}. Kasei's source is lava-floored Echus Chasma. Beyond the 5km-high chasm wall, on the Hesperian plateau, dendritic channel networks are abundant (Figure \ref{ECHUSMANGOLD}a, \citet{cha09a,cha09b}\citep{man04,man08}. The floor of Echus Chasma is now 200m below a saddle that marks the start of the main channel. Regional topography suggests that the floor of Echus was once much deeper \citep{har08}.

Although no opal or hydroxylated sulfates have been reported on the Echus plateau \citep{mil08}, the erosional properties of the substrate for the Echus channel networks resembles the erosional properties of the Juventae plateau layered deposits. Material on the SE rim of Echus is relatively thinly layered, relatively light-toned in outcrop, and recessed from the chasm edge, suggesting a sedimentary or volcaniclastic origin (CTX image P14\_006586\_1800\_XN\_00N079W). It is cut by very broad valleys, which bottom out on more resistant basalt, suggesting it is less resistant to fluvial erosion (P02\_001839\_1806\_XN\_00N079W).

Our simulations of the Echus plateau channel networks, which are preserved in negative relief, show peak non-lake precipitation in the densest area of observed channels, at 0.7 mm/hr (Figure \ref{ECHUSMANGOLD}a). Because of the interaction of lake convergence, topography, and the regional windfield, precipitation is also predicted S and E of the chasm beyond the existing mapped area of channels. THEMIS and CTX show extensive channelization in these areas (dotted lines in Figure \ref{ECHUSMANGOLD}a), which die away on a length scale similar to the length scale of predicted precipitation (Figure \ref{ECHUSMANGOLD}b). Cursory inspection of CTX images further away from the channel edge does not show channelization. These results are consistent with localized precipitation.

The Echus plateau channel networks are deep and extensive. \citet{cha09b} gives 82 km$^3$ total erosion over an eroded outcrop of $\sim$ 30000 km$^2$, corresponding to an average of 2.7m material removed. With a melt season length of 30 sols and melting of 1 mm/day, this would require 90 years to remove if the material was primarily ice. However, if the material is non-ice with a fluvial sediment:water ratio of 100:1, 9000 years would be required to cut the observed channels.

Although the relatively incomplete geomorphic mapping of the Echus headwaters currently prevents more thorough hypothesis testing, this is a promising initial result that suggests our ability to match observations at Juventae with only localized precipitation can be reproduced elsewhere.

% (Michael notes:) Cerberus Fossae - could the discharge do something similar? another test? We are extending our analysis to localized vapor sources elsewhere on Mars.

\section{Implications for regional and global climate change}
\noindent
 The majority of water vapor released in our simulations is trapped by localized precipitation near the lake, and so is unavailable for broader climate change (Figure \ref{TERNARYFATE}). Because the remaining atmospheric vapor load is less than present-day global loads \citep{smi02}, we do not expect a global excursion to wet conditions (the MEGAOUTFLO hypothesis; \citet{bak91,bak01,qua05}) to result from water vapor release during chaos events (Fig \ref{TERNARYFATE}). There are four possible caveats. First, simultaneous triggering of chaos events in multiple chasms, or the broad area of outwash at the terminus of the outflow channels, would provide a larger interface for water vapor injection to the atmosphere than is simulated here. Supposing an evaporation rate of 2 mm/hr, for 5\% global water cover, and in the absence of precipitation, the atmosphere will contain 6 mbar water in 70 sols. Water vapor is a powerful greenhouse gas and 6 mbar of water vapor would cause noticeable global warming. But precipitation will occur, and it is likely that outwash freezes over in $<<$70 sols. In addition, higher-resolution images have shown inner channels within the outflow channels, indicating that much lower discharge rates, and multiple flooding events, incised the outflow channels \citep{wil00}. This does not support the hypothesis of large Late Hesperian and Amazonian seas, because at low discharges, water would freeze and perhaps sublime, redepositing at planetary cold traps. Second, noncondensible gases such as CO$_2$ or CH$_4$, stored in cryosphere clathrates (or deep aquifers) and outgassed during chaos events \citep{bar10}, could provide warming. Third, if chaos events are triggered by magma bodies laden with CO$_2$, SO$_2$, and halogens, then chaos events and transient greenhouse warming could occur simultaneously (e.g., \citet{joh08}). Fourth, if atmospheric pressure was higher (e.g. hundreds of mbar) this would suppress the buoyant instability that leads to localized precipitation (allowing more water vapor to escape to the background atmosphere; Paper 1).

%[This paragraph still needs some work].
Our model indicates that regional or global temperatures warmer than today are not required to explain the outflow-channel associated valley networks -- the melting probabilities for present-day solar luminosity are high (Figure \ref{MELTPROB}). With DTM drainage network and channel geometries, 0.1 mm/hr is approximately the required runoff for mobilization of fine gravel/granules, and 1.0 mm/hr will mobilize medium gravel (Figure \ref{DTMANALYSIS}). Assuming the albedo of dust, present-day solar luminosity, and the pdf of orbital parameters for 0.05 Gya, the probability of melting is 70\% and the exceedance probability for melt rates of 0.1 mm/hr is 30\% (although melt rates of 1.0 mm/hr are not found). These percentages are not precise, because of our crude treatment of the atmospheric contribution to surface temperature.

However, our calculations indicate a greatly reduced melting probability for reduced (3.0 Gya; Late Hesperian/Early Amazonian) solar luminosity, if the strength of the atmospheric greenhouse was no greater than today (Figure 
\ref{MELTPROB}). %Assuming the albedo of dust, the exceedance probability for melt rates of 0.1 mm/hr (1.0 mm/hr), is still lower - 15\% (0.01\%).
Taking into account the 6K increase in afternoon temperature due to the water vapor released by the lake increases the melting probability to 21\%, and the melting exceedance probability to 8\% for 0.1 mm/hr. However, this requires that the chaos lake starts in (or persists through) the melt season. These are still low probabilities, so why are channels found throughout the stratigraphic column? We suggest four possible explanations. (1) Melting is needed to produce water that indurates the sediment. Since induration increases resistance to aeolian erosion, the stratigraphic record on the Juventae plains is biased toward chaos storms that were accompanied by melting. (2) Runoff accompanied only a few layer-forming events, but the channels cut from the top to the bottom of the sedimentary stack. Meltwater percolated through the stack and cemented all layers. Deeply-incising, rather than superficial, channels are suggested by the observation of negative-relief channel continuous with channel-topped ridges (Figure \ref{STRATDETAIL}). If the 33-39 layers identifiable in the stratigraphic column correspond to 33-39 lake events, and modeling melt events as a Poisson process, the probability of melting occurring on the plateau at some point in the year following at least one chaos event is 12\% - 14\% for albedo = 0.28 for a greenhouse effect unchanged from today. This increases to $>$99.9\% if we assume a modest 6K increase in greenhouse forcing. (3) The chaos events were correlated with orbital conditions, with chaos events preferentially occurring during times that favored melt. (4) A greater past concentration of greenhouse gases compensated for the faint young Sun, so that the background climate state was warmer than in our calculations. This would lift the percentage of snowstorms that would deposit ice which would melt.

%Among the three northern Valles Marineris outflow channel source regions, Echus, Juventae, and Ganges, there is a positive correlation between outflow channel size and the extent of plateau channel networks. Echus has plateau valley networks covering 12000 km$^2$, suggestive evidence for ponding, and is the source of the largest outflow channel on Mars (Kasei), which has peak discharge 3.8 x 10$^8$ m$^3$ s$^{-1}$ \citep{har09,kle05}. Juventae has 2300 km,$^2$ of plateau channel networks and layered deposits, conclusive evidence for ponding, and feeds an outflow channel (Maja) with peak discharge 1.1 x 10$^8$ m$^3$ s$^{-1}$ \citep{led10,col07,kle05}. Ganges has 300 km$^2$ of plateau channel networks and layered deposits, and no evidence for ponding \citep{led10}. This positive correlation is consistent with localized precipitation forming each channel network.

Channel networks at SW Melas Chasma and Gale-Aeolis-Zephyria also formed in the Late Hesperian or Amazonian, but because there is no conclusive evidence for groundwater release at these locations it is difficult to make the case for localized precipitation. \citet{kit10b} propose a separate formation scenario for SW Melas Chasma and Gale-Aeolis-Zephyria which requires unusual orbital conditions favoring snowmelt, but also does not require a greenhouse effect stronger than the present day. Thermokarst at Ares Valles has also been argued to require transient global warm conditions during the Hesperian \citep{war10}.

The dendritic channels provide strong evidence for water runoff, so if the plateau layered deposits record global conditions, then they would be {\it prima facie} evidence that temperatures and pressures permitting surface liquid water runoff persisted (or were revived) well after the Late Noachian/Early Hesperian maximum in valley network formation. However, we do not believe that they required globally-moist conditions.

%[Orphan para] Groundwater release formed chaos terrain and outflow channels on Hesperian and Early Amazonian Mars. Early discharge estimates of 10$^9$ m$^3$s$^{-1}$ per outflow channel suggested large seas would have formed at the end of the channels.The greenhouse effect of water vapor released from these seas could have initiated brief global climate change - the MEGAOUTFLO hypothesis. Outflow-triggered global change could account for recently-discovered Hesperian and younger valley formation at many sites around Mars.

%Therefore, we may have to seek the geomorphological traces of chaos-induced weather systems closer to the chaos terrains themselves. [end orphan para].

\section{Conclusions}
\noindent
We conclude from this study that:-

(1) Material lofted from Juventae Chasma and transported W by the background wind field very likely contributed to channel formation on the Juventae plateau. This material could be water or sediment or both;

(2) Peak precipitation over land occurs to the SW of the chasm in mesoscale simulations of lake storms at Juventae Chasma. This corresponds to the location of the observed plateau channel networks. The location is sensitive to lake surface elevation;

(3) Peak precipitation of 1.7 mm/hr water equivalent (w.e.) and mean rates of 0.9 mm/hr w.e. occur at the location of the observed plateau channel networks. These rates are sensitive to lake surface elevation and lake temperature;

(4) Using present-day orbital conditions, 3.0 Gya solar luminosity, and the albedo of dust leads to $O$(10\%) probability of peak snowpack melting rate exceeding 0.1 mm/hr, per event bed. This includes the greenhouse effect of vapor from the storm (or, equivalently, a modest 6K increase in the background greenhouse effect). Neglecting infiltration and evaporation losses, this discharge is sufficient to move sand and fine gravel through the observed channels. The maximum observed drainage densities are consistent with snowmelt channels thermally eroded into ice;

(5) The minimum background atmospheric temperature to permit surface melting is sensitive to snowpack albedo, solar luminosity and orbital conditions;

(6) Juventae's plateau channel networks could have formed from localized precipitation, and do not require global climate change;

(7) The majority of water vapor released to the atmosphere during chaos flooding is trapped by localized precipitation at the chasm edge, and is not available to drive global climate change;

(8) Alternative explanations of channel formation that are compatible with stratigraphic constraints include patchy melting of broadly distributed equatorial snowpack, and spring discharge of groundwater on the plateau early in the formation of Juventae Chasma;

(9) Our model strongly predicts that plateau channel networks will not be found $>$ 250 km from a water vapor source. It also predicts that additional plateau channel networks and plateau layered deposits should be identified downwind of large, localized vapor sources elsewhere on Mars.

%%% End of body of article:

%%%%%%%%%%%%%%%%%%%%%%%%%%%%%%%%
%% Optional Appendix goes here
%
%%%%%%%%%%%%%%%%%
% Geophysical Research Letters only allows an appendix without a letter.
%% You can get this result with
%  \section*{Appendix}
%  or
%  \section*{Appendix: Title}
%%%%%%%%%%%%%%%%%
%
\appendix \section{Methods} %resets counters and redefines section heads

The fully compressible dynamical core of MRAMS is derived from the terrestrial RAMS code \citep{pie92}. A cloud microphysical scheme derived from CARMA was recently added to MRAMS by T.I. Michaels. This cloud microphysics scheme has been used to successfully reproduce observations of low-latitude clouds downwind of the Tharsis Montes and Olympus Mons \citep{mic06}. A recent description of MRAMS capabilities is \citet{mic08conf}.

We made minimal modifications to the MRAMS v.2.5-01 code to allow for surface liquid water.
%MRAMS is currently serial code and each single-CPU run lasted $\sim$0.5 years (wall time).
Liquid water microphysics is not included.
Water vapor thermodynamics are included in the energy equation, but water vapor is not included in the mass and momentum equations -- that is, we ignore pressure and virtual temperature effects of water vapor loading.

We do not permit dynamic dust lifting at the mesoscale.

Model vertical layer thickness varied from 2.3 km at altitude to 30 m near the ground. We used four grids with the outermost being hemispheric and a horizontal resolution of $\sim$8.9 km on the inmost grid. Output was sampled every 1/24 sol ($\approx$ 3699 s), or `Mars-hour.' We assume that this frequency, limited by available disk space, is enough to capture model behaviour -- for example, we refer to the warmest of 24 samples during a sol as the `day's maximum temperature.' The timestep varied between runs but was never more than 3.75s on the inmost grid.

%Ls $\sim$ 270$^\circ$ for the runs.
Atmospheric boundary conditions are from the NASA Ames MGCM \citep{hab93}, v2.1, and we use present-day orbital parameters.

\section{HiRISE stereo DTMs}
HiRISE DTMs were generated using SOCET SET following USGS recommended procedure \citep{usgs09}.

\emph{DTM1}:  Juventae1, PSP\_003223\_1755/PSP\_003724\_1755

\emph{DTM2}: Juventae2, PSP\_004423\_1755/PSP\_005412\_1755

\emph{DTM3}: Ganges1, PSP\_005161\_1720/ESP\_016237\_1720

All DTMs, together with the corresponding orthorectified HiRISE images, can be obtained for unrestricted further use from the lead author.
%Files are supplied as supplementary online data.
%Some artifacts exist -- refer to the \texttt{readme.txt} supplied with each image pair.

In addition, we made use of PDS released files for a Juventae plateau stereopair, PSP\_003434\_1755 / PSP\_003579\_1755.

% but doesn't print anything.
% After typing  \appendix
%
% \section{Here Is Appendix Title}
% will print
% Appendix A: Here Is Appendix Title
%
% \section*{Appendix}
% will print
% Appendix
%
% \section*{Appendix: Here Is Appendix Title}
% will print
% Appendix: Here Is Appendix Title
%
% For only 1 appendix \appendix \section{Appendix} is preferred.
% which will print
% Appendix A

%%%%%%%%%%%%%%%%%%%%%%%%%%%%%%%%%%%%%%%%%%%%%%%%%%%%%%%%%%%%%%%%
%
% Optional Glossary or Notation section, goes here
%
%%%%%%%%%%%%%%
% Glossary only allowed in Reviews of Geophysics
% \section*{Glossary}
% \paragraph{Term}
% Term Definition here
%
%%%%%%%%%%%%%%
% Notation -- End each entry with a period.
% \begin{notation}
% Term & definition.\\
% Second Term & second definition.
% \end{notation}
%%%%%%%%%%%%%%%%%%%%%%%%%%%%%%%%%%%%%%%%%%%%%%%%%%%%%%%%%%%%%%%%
%
%  ACKNOWLEDGMENTS

\begin{acknowledgments}
HiRISE stereo DTMs were produced on the SOCET SET workstation at the SETI Institute, generously made available by Cynthia Phillips. Ross Beyer, Annie Howington-Kraus, Audrie Fennema, and Sarah Mattson assisted with DTM generation. We thank Teresa Segura for advice on starting the project, and Inez Fung, Rebecca Williams, Jean-Baptiste Madeleine, Max Rudolph and Leif Karlstrom for useful discussions. We thank Janice Bishop, Keith Harrison, and Bob Grimm for commenting on the manuscript. Caleb Fassett drew our attention to inverted channels near Reull Vallis. This work made use of Bin Guan's scripts. We are grateful to the HiRISE team for maintaining a responsive public target request program (HiWish), that was useful for this work. We acknowledge support from Teragrid allocation TG-EAR100023, NASA Science Mission Directorate grant NNX08AN13G, NASA Science Mission Directorate grant NNX09AN18G, and NASA grants to SwRI which funded cloud microphysics capabilities.
\end{acknowledgments}

%% ------------------------------------------------------------------------ %%
%
%  REFERENCE LIST AND TEXT CITATIONS
%
% Either type in your references using
% \begin{thebibliography}{}
% \bibitem{}
% Text
% \end{thebibliography}
%
% Or,
%
% If you use BiBTeX for your References, please produce your .bbl
% file and copy the contents into your paper here.
%
% Follow these steps:
% 1. Run LaTeX on your LaTeX file.
%
% 2. Run BiBTeX on your LaTeX file.
%
% 3. Open the new .bbl file containing the reference list and
%   copy all the contents into your LaTeX file here.
%
% 4. Comment out the old \bibliographystyle and \bibliography commands.
%
% 5. Run LaTeX on your new file before submitting.
%
% AGU does not want a .bib or a .bbl file, but asks that you
% copy in the contents of your .bbl file here.

%Reference citation examples:

%...as shown by \textit{Kilby} [2008].
%...has been shown [\textit{Kilby et al.}, 2008].

%...as shown by \cite{jskilby}.
%...has been shown \citep{jskilbye}.

%% ------------------------------------------------------------------------ %%
%
%  END ARTICLE
%
%% ------------------------------------------------------------------------ %%

\end{article}

%% Enter Figures and Tables here:

% When submitting articles through the GEMS system:
% COMMENT OUT ANY COMMANDS THAT INCLUDE GRAPHICS.

% Figure captions go below this illustration; Table captions go above tables

\newpage
\pagebreak

 \begin{table}[ht]
\caption{List of runs with parameters.}
% title of Table
\centering
% used for centering table
{\small
\begin{tabular}{l l | r r r l}
% centered columns (4 columns)
\hline\hline
%inserts double horizontal lines
Run & Full name & Lake level (m) & Lake temp. (K) & Lake area (km$^2$) & Description\\ [0.5ex]
% inserts table
%heading
\hline
% inserts single horizontal line
\texttt{juventae\_dry}  & juventae\_Mar\_12\_2010 & -- & -- & -- & V.M. topography, no lake \\
% inserting body of the table
\texttt{juventae\_high}  & juventae\_May\_12\_2010 & + 0 m  & 273.15 & 25800 & At spillway\\
\texttt{juventae\_med} & juventae\_Mar\_13\_2010 & -1000 m & 278.15  & 19400 & Almost fills chasm \\
\texttt{juventae\_low}  & juventae\_Mar\_14\_2010 & -3000 m  & 278.15 & 6400 & In SE corner of chasm \\
\texttt{echus\_low}  & echus\_Apr\_5\_2010 & -900m & 278.15 &  3600 & Small equant lake\\
\texttt{echus\_high}  & echus\_Mar\_25\_2010 & -800m & 273.15 & 17100 & At spillway, elongated N.S.\\ [1ex]
%\texttt{m} & 45 & 300 & 556 \\
% [1ex] adds vertical space
\hline
%inserts single line
\end{tabular}}
\label{table:nonlin}
% is used to refer this table in the text
\end{table}

\begin{table}[ht]
\caption{Evaporation rate and vapor fate (end of sol 5). Units are Mt (10$^6$ metric tons). Italicized \textit{\texttt{dry}} runs
are subtracted from the runs below them. See also Figure \ref{TERNARYFATE}.}
% title of Table
\centering
% used for centering table
{\small
\begin{tabular}{l | r r r r r r r}
% centered columns (4 columns)
\hline\hline
%inserts double horizontal lines
Run & Evap. rate (mm/hr) & Water in atm. & Total atm.(\%) & Snow in lake & Snow beyond lake & Total snow (\%) \\ [0.5ex]
% inserts table
%heading
\hline
% inserts single horizontal line
\textit{\texttt{juventae\_dry}} &  \textit{0} & \textit{536} & \textit{100\%} & \textit{0} & \textit{0} & \textit{0\%}\\
\texttt{juventae\_low} & 2.47 & 304 & 18\% & 251 & 1166 & 82\% \\
\texttt{juventae\_med} & 2.82 & 536 & 10\% & 1356 & 3472 & 90\%\\
\texttt{juventae\_high} & 1.88 & 571 & 12\% & 1505 & 2590 & 88\% \\
% inserting body of the table
\hline
\textit{\texttt{juventae\_dry\_sol\_7}} & \textit{0} &  \textit{520} & \textit{100\%} & \textit{0}  & \textit{0} & \textit{0\%} \\
\texttt{juventae\_low\_sol\_7} & 2.54 & 403 & 15\% & 387 & 1887 & 85\% \\
\texttt{juventae\_med\_sol\_7} & 2.90 & 658 & 8\% & 2230 & 5393 & 92\% \\
% [1ex] adds vertical space
\hline
%inserts single line
\end{tabular}}
\label{table:nonlin}
% is used to refer this table in the text
\end{table}

 \begin{table}[ht]
\caption{Model skill using method of \citet{pie78}. The best-fitting model for each geological target is highlighted in bold. Italicized text corresponds to the original extent of layered deposits mapped by \citet{wei10}, and normal text is the value for \citet{led10}. 1.0 mm/hr is not shown because there is only 1 pixel with this value}
% title of Table
\centering
% used for centering table
{\small
\begin{tabular}{l | r r r r r | r r r r r}
% centered columns (4 columns)
\hline\hline
%inserts double horizontal lines
& & & Skill & & & & & Coverage & & \\
Precip. (mm/hr) &  0.02 & 0.2 & 0.4 & 0.6 & 0.8 & 0.02 & 0.2 & 0.4 & 0.6 & 0.8 \\ [0.5ex]
% inserts table
%heading
\hline
% inserts single horizontal line
% inserting body of the table
\texttt{juventae\_low} &  11    &     0     &    0   &   0 &  0  &  0.040     &    0        & 0     &    0    &     0      \\
& \textit{11}    &     \textit{0}     &    \textit{0}    &     \textit{0}       &  \textit{0}    & \textit{0.009}   &      \textit{0}       &  \textit{0}     &    \textit{0}    &     \textit{0}   \\
\texttt{juventae\_med} & 5 &  19 &  50   & 108  & 122   & 0.020  &  0.071 &   0.185  &  0.403 &   0.455      \\
& \textit{5}  & \textit{19}  & \textit{51} &  \textit{138}  & \textit{227} &     \textit{0.004} &   \textit{0.015}  &  \textit{0.041}  &  \textit{0.111} &   \textit{0.182}    \\
\texttt{juventae\_high} & 6 &  33 &  120 & 229 & \bf{269}   & 0.022  &  0.122   & 0.446  &  0.853 &   \textbf{1.000}       \\
&           \textit{6}   & \textit{33} &  \textit{136} & \textit{293} & \textbf{\emph{498}} &   \textit{0.005}  &  \textit{0.026}  &  \textit{0.109}  &  \textit{0.235}   & \textbf{\emph{0.400}}      \\ [1ex]
% [1ex] adds vertical space
\hline
%inserts single line
\end{tabular}}
\label{table:nonlin}
% is used to refer this table in the text
\end{table}

 \begin{table}[ht]
\caption{Measurements relevant to hydrology for 2 adjacent inverted channels (Figure \ref{STRATDETAIL}).}
% title of Table
\centering
% used for centering table
{\small
\begin{tabular}{l l l | r r }
% centered columns (4 columns)
\hline\hline
%inserts double horizontal lines
& & & Catchment 1 & Catchment 2  \\
% inserts table
%heading
\hline
% inserts single horizontal line
% inserting body of the table
$A$ &  area &     km$^2$   &  1.6 &  2.1       \\
$w$ & width &  m &  3.3$\pm$0.8 & 3.2$\pm$0.47 \\
$S$ & slope    &     rise:run     &   0.82$^\circ$ & 0.96$^\circ$     \\
% [1ex] adds vertical space
\hline
%inserts single line
\end{tabular}}
\label{table:nonlin}
% is used to refer this table in the text
\end{table}

\newpage
\pagebreak % ONE-COLUMN figure/table, including eps graphics

 \begin{figure}
 \noindent\includegraphics[width=40pc]{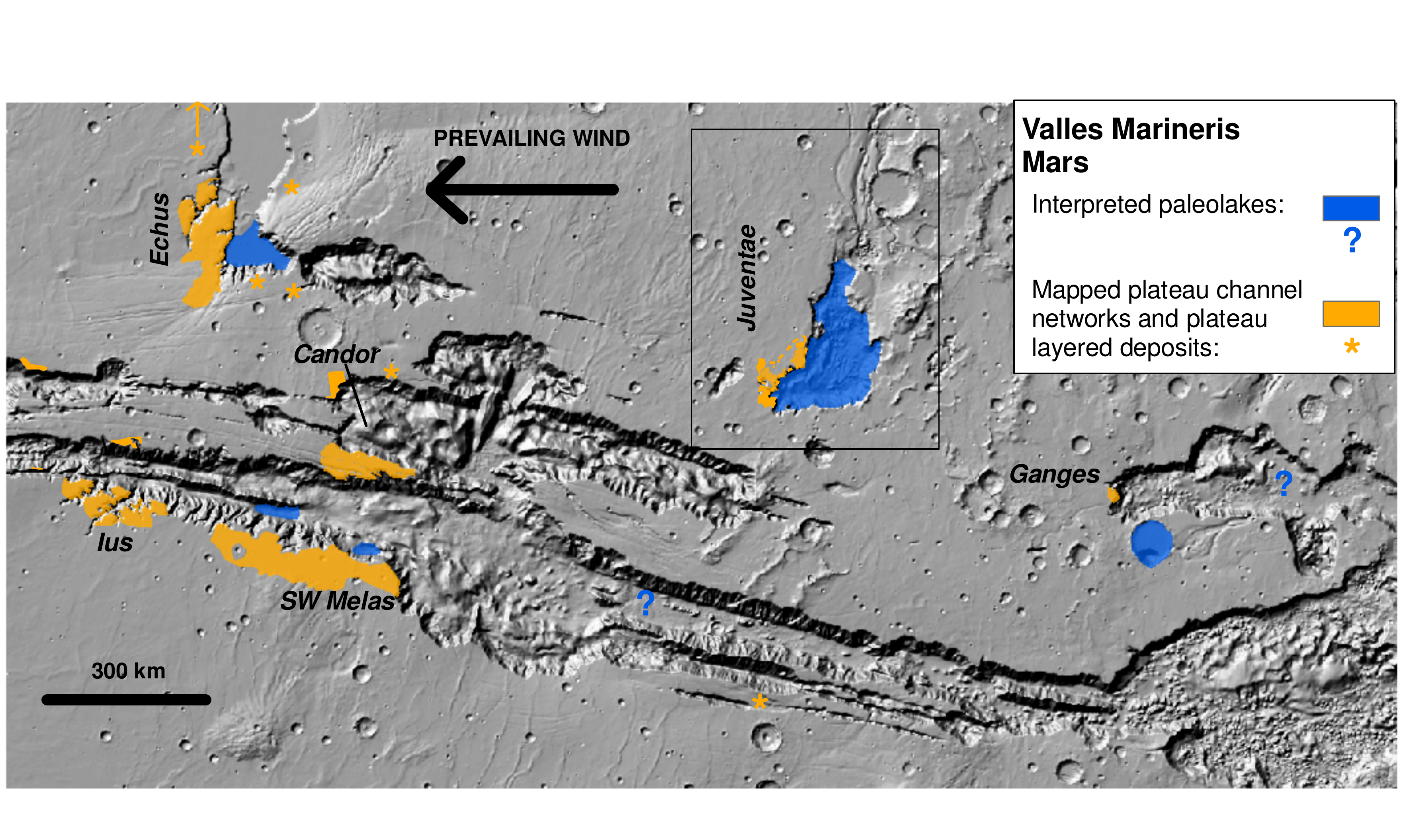}
 \noindent\includegraphics[width=40pc,clip=true,trim=0mm 0mm 16mm 0mm]{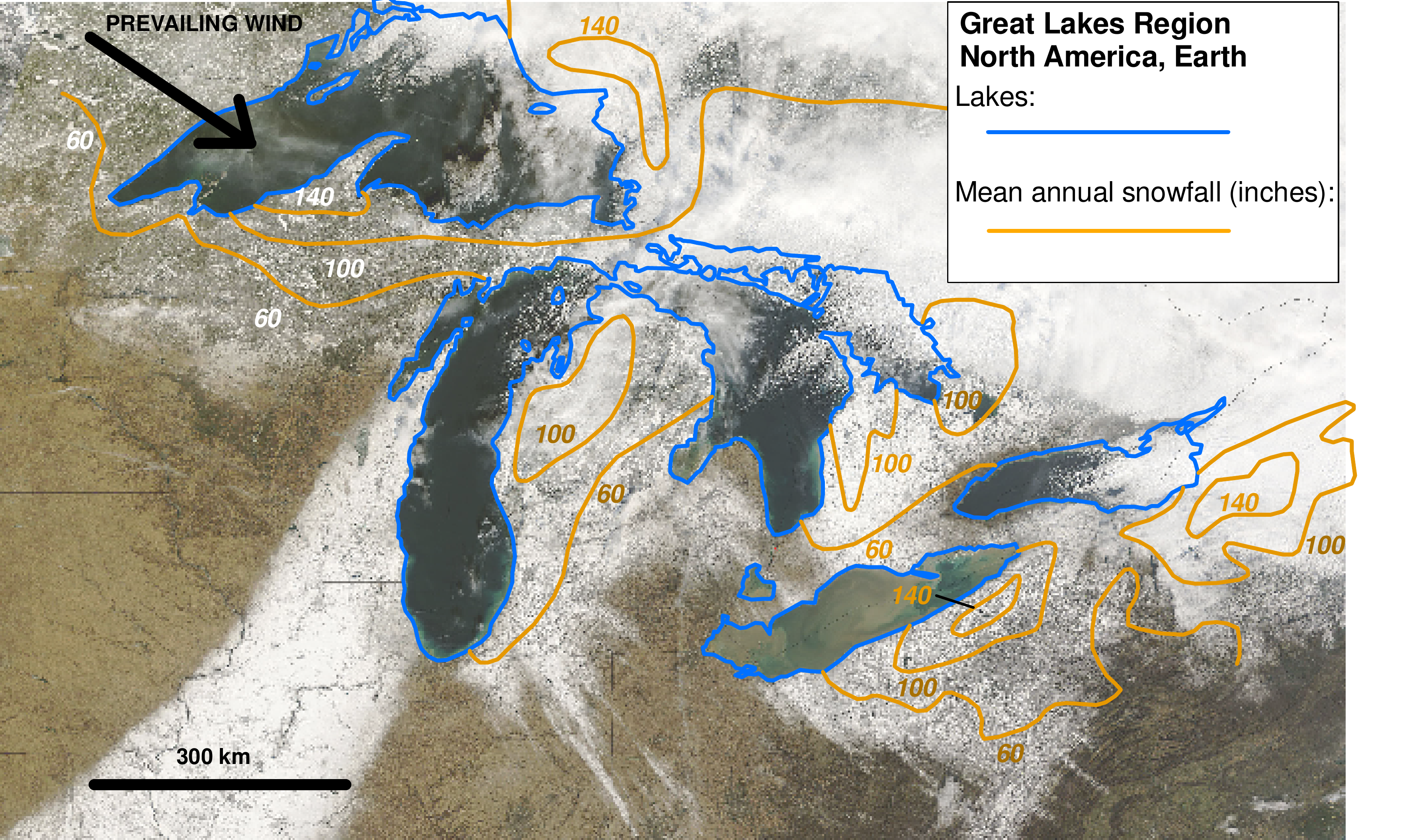}
 \end{figure}

 \newpage
 \pagebreak

 \begin{figure}
 % \noindent\includegraphics[width=20pc]{samplefigure.eps}
 \caption{\label{VALLESMAP} (Upper panel) Location of  Valles Marineris plateau layered deposits and plateau channel networks downwind of paleolakes. Reported plateau layered deposits and channel networks are shown by orange shading. Asterisks correspond to isolated, or incompletely mapped, occurrences. Reported paleolakes are shaded in blue: those where we consider the evidence to be less strong are shown by question marks. Box around Juventae corresponds to Figure \ref{GEOLOGYCONSTRAINTS}. Background is MOLA shaded relief. Sources:- \citet{wei10} (SW Melas, S Ius, S flank W Candor, Juventae and Ganges layered deposits); \citet{led10} (N Ius, N flank W Candor, N Tithonium and Juventae layered deposits) \citet{man08} (Echus networks); \citet{wil05} (Candor channel); \citet{har08} (questionable Candor lake); \citet{har09} (Juventae and Echus lakes); \citet{met09} (SW Melas lake); \citet{roa10} (Ius closed evaporitic basin); \citet{kom09} (Morella lake). The questionable Ganges lake is our own interpretation. (Lower panel) Location of snowbelts downwind of the Great Lakes, North America, Earth. Each local maximum in mean annual snowfall, as shown by orange contours, is downwind of a lake. Snowbelts form from the cumulative effect of lake-effect storms. Contours are from \citep{eic79}, as reproduced in \citet{mar10}. The prevailing wind direction is shown by the black arrow and the snow streaks in the background image. Background image was acquired 9 December 2006 by Terra/MODIS and shows the effects of a lake-effect storm on 7-8 December 2006 (image credit: NASA GSFC/Earth Observatory).}
 \end{figure}

\clearpage

 \begin{figure}
 \noindent\includegraphics[clip=true,trim=40mm 0mm 40mm 180mm]{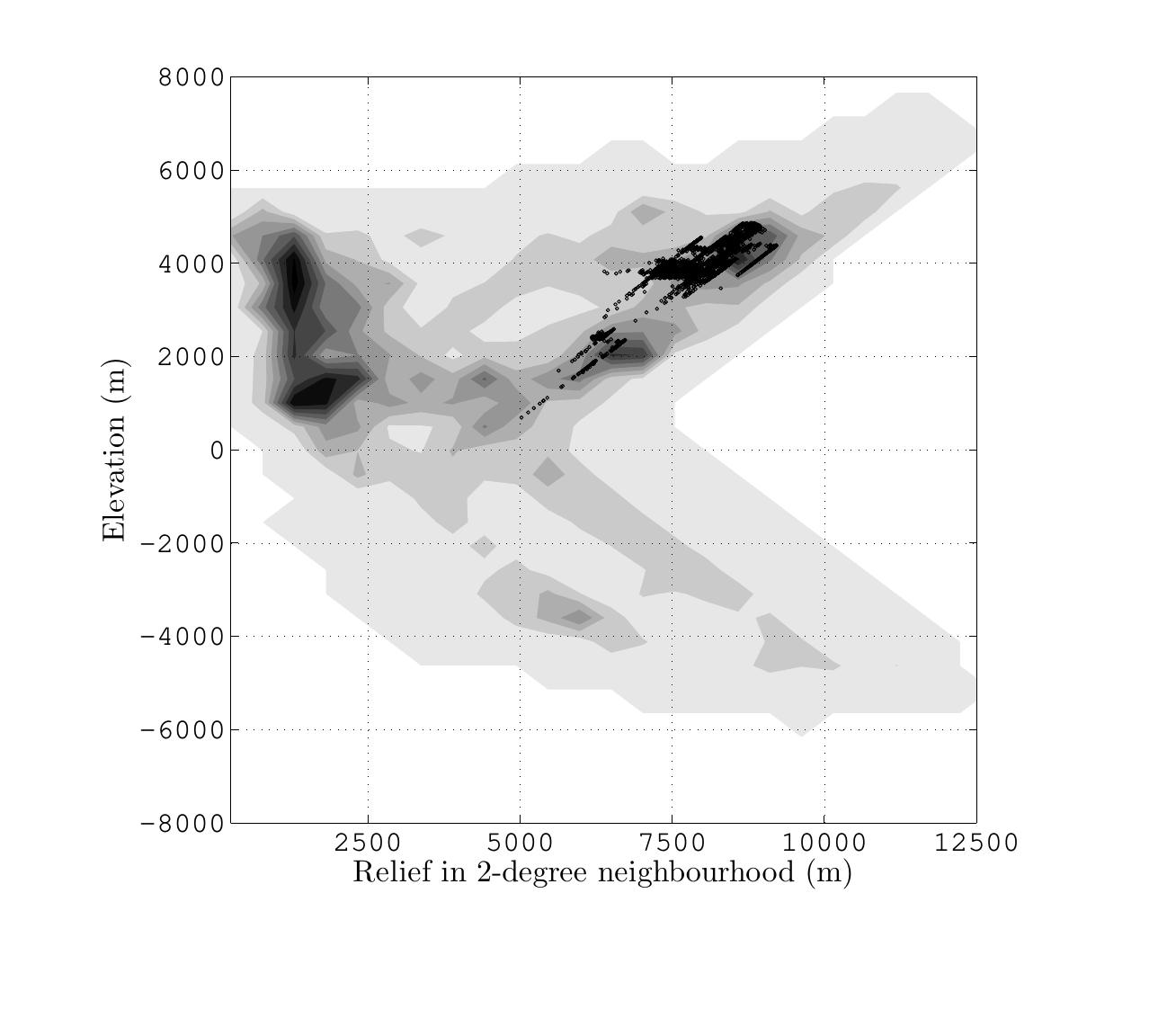}
 \caption{\label{VALLESPLOT} Topographic context of inverted channels. Black dots correspond to points mapped as `light-toned layered deposit' or `inverted channel' by \citet{wei10}. The association of inverted channels with light-toned layered deposits containing opal $\pm$ jarosite is only found near the rims of large canyons. Grayscale background is the probability distribution of all points, using a bin size of 500m. Relief at a point is defined as the maximum difference in elevation between that point and all other points in a neighbourhood with 2-degree radius. (Points with intermediate elevation cannot have high relief with this definition.)
 }
 \end{figure}

 \begin{figure}
 \noindent\includegraphics[width=180mm,clip=true,trim=400mm 300mm 200mm 150mm]{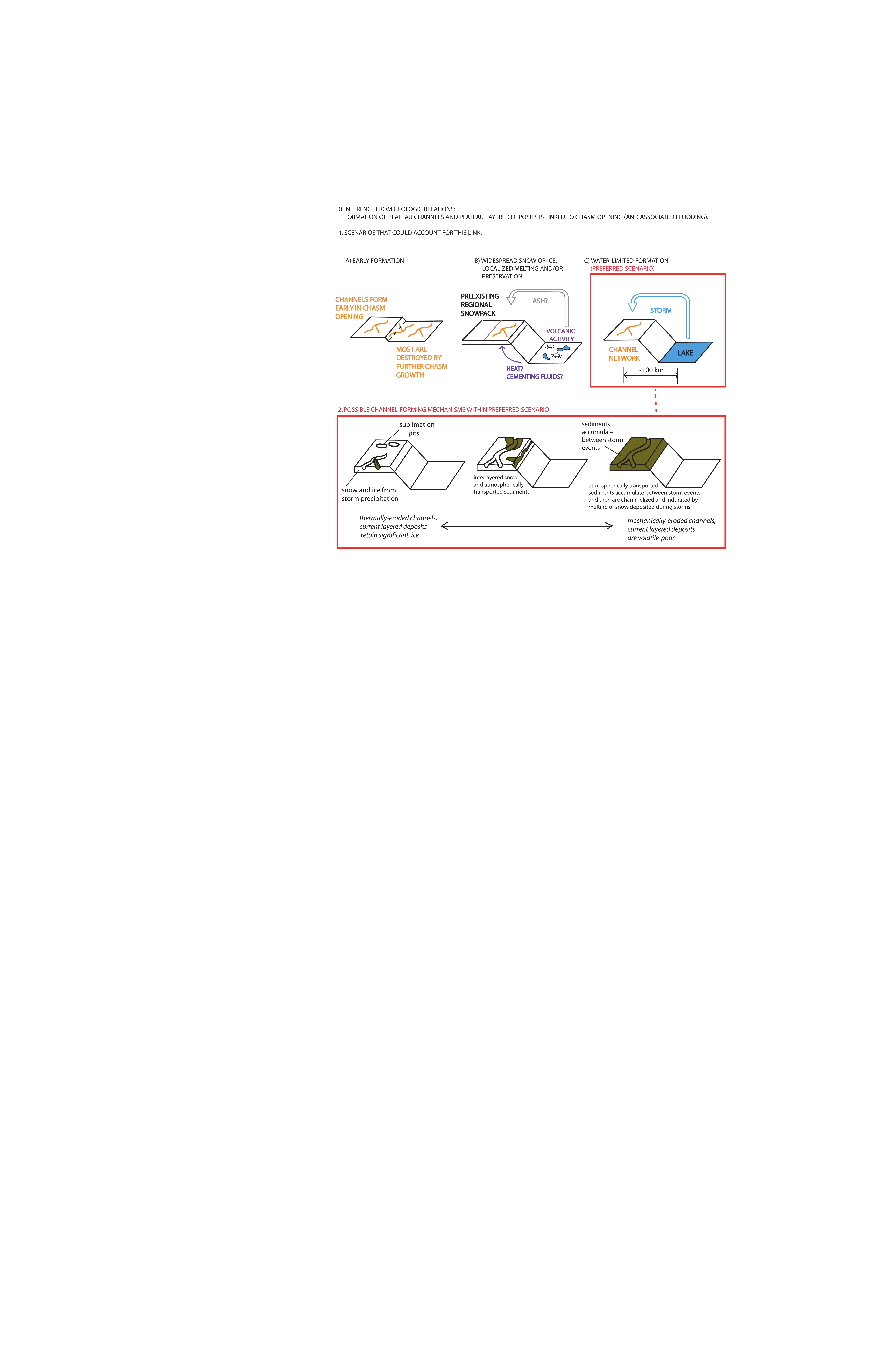}
% \noindent\includegraphics[width=180mm,clip=true,trim=0mm 0mm 0mm 0mm]{figureCHANNELFORMATIONloresbillrev.pdf}
 \end{figure}

   \newpage
  \clearpage

 \pagebreak

 \begin{figure}
 \noindent\includegraphics[width=180mm,clip=true,trim=130mm 0mm 0mm 100mm]{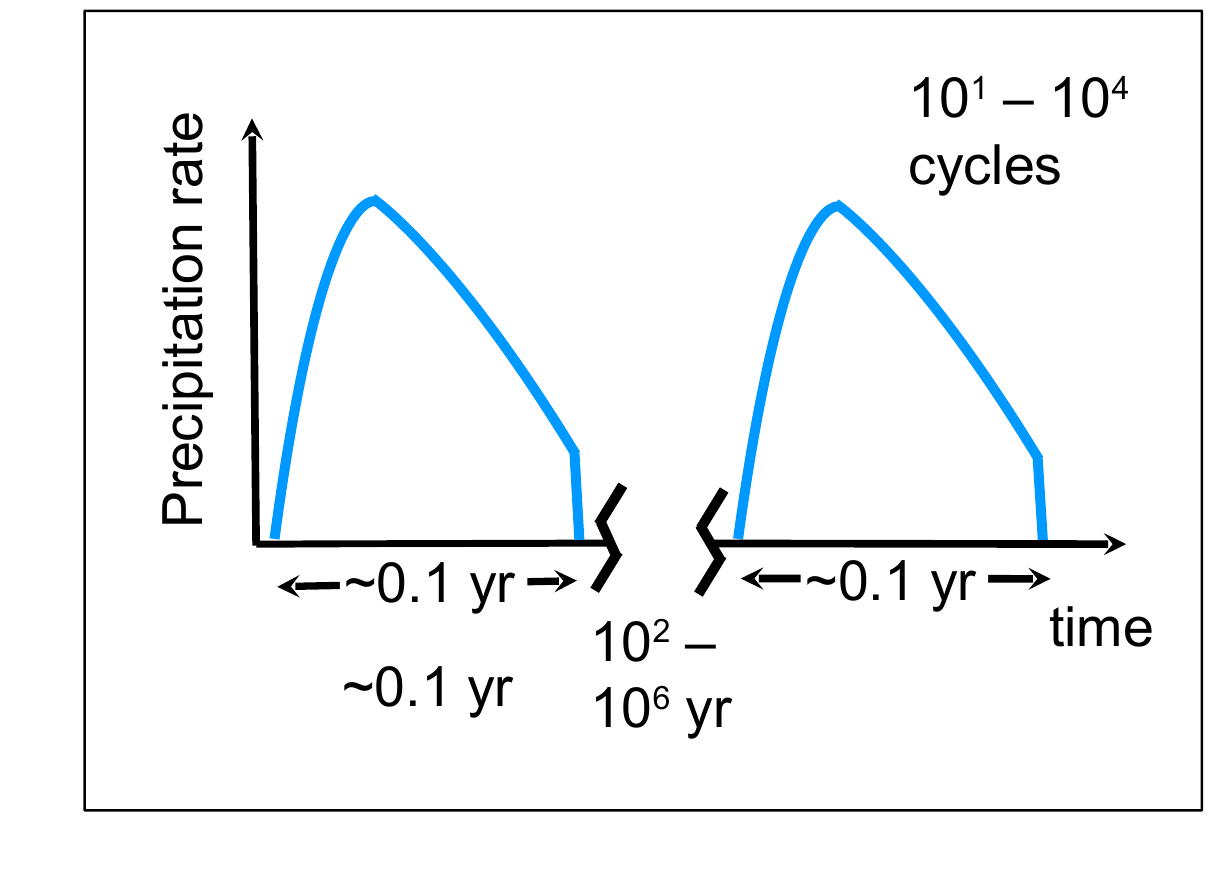}
 \caption{\label{HOWITWORKS} Top panel: Sketch of our hypothesis. Condensation of vapor released from short-lived lake in chasm creates a snowstorm. Precipitation on the plateau next to the chasm forms channel networks. Precipitation falling back into the lake leaves no geomorphic signature. Boxed inset shows the timescales implied by published models of chaos hydrology and the thermodynamics of lake freezing: each lake event lasts $\leq$1 year \citep{and07,har09}. Crosscutting channels, and discharge estimates from measurements of inner channels, require that many flood events occurred in each chaos chasm. Bottom panel: Estimates of repose interval between groundwater outflow events center on 10$^2$ - 10$^6$ years, and are considered to lengthen with time, but are sensitive to poorly-known crustal hydrologic properties \citep{and07,har09}.
 }
 \end{figure}

  \newpage
  \clearpage

 \pagebreak

 \begin{figure}
 \noindent\includegraphics[scale=0.7,clip=true,trim=50mm 5mm 70mm 0mm]{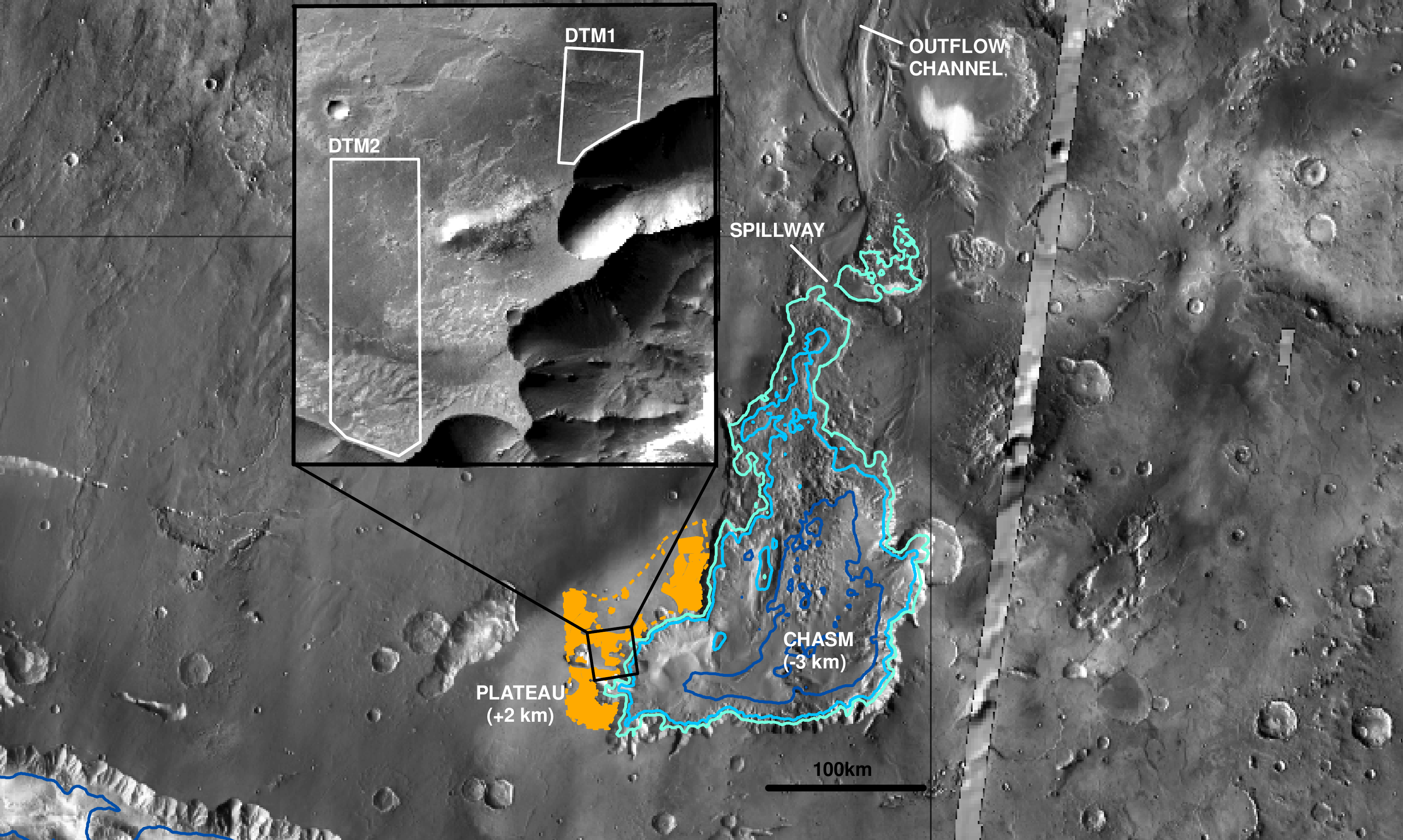}
 \end{figure}

\begin{figure}
 \caption{\label{GEOLOGYCONSTRAINTS} Site for our hypothesis test -- Juventae. Juventae Chasma is a 5 km deep, sharp-sided chaos chasm that sources Maja Valles (outflow channel to N). {\it Main image:} Orange shading corresponds to the area of plateau channel networks and plateau layered deposits mapped by \citet{led10}. Orange dotted line corresponds to pre-erosion extent estimated by \citet{led10} from outliers. Spillway to N of chasm indicates that flood level exceeded +1180m. We model flooding to depth -3000m (deep blue), -1000m (mid blue) and +0m (cyan). Main figure background is THEMIS VIS mosaic. {\it Inset image} shows the locations of 2 HiRISE DTMs (supplementary online data) that we constructed to characterize the hydraulic geometry and stratigraphic context of the inverted channel networks. Inset background is part of CTX image P18\_007983\_1751\_XN\_04S063W.}
 \end{figure}
\clearpage

 \newpage
 \pagebreak

\begin{figure}
\noindent\includegraphics[width=30pc,clip=true,trim=5mm 20mm 5mm 20mm]{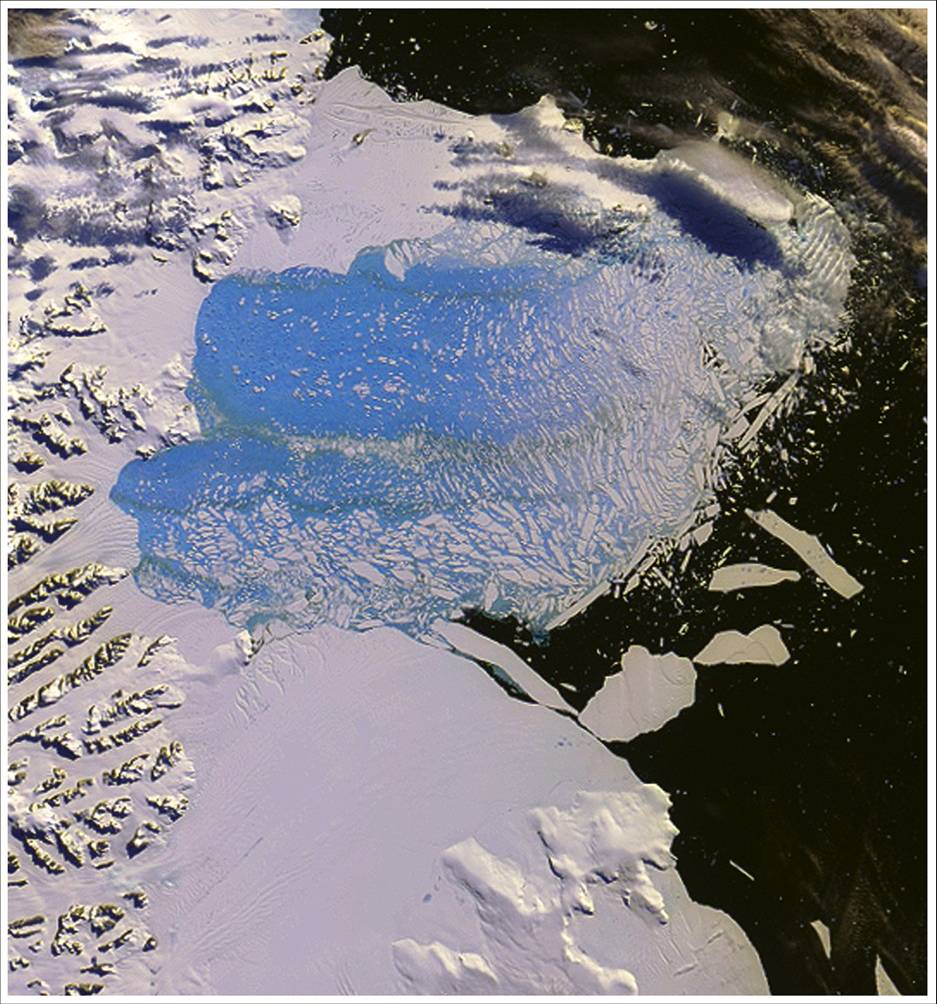}
\caption{\label{LARSENB} Disintegration of Antarctica's Larsen B ice shelf. Collapsed area is $\sim$3000 km$^2$. Acquired 7 March 2002 by Terra/MODIS. Image credit: NASA/GSFC.}
\end{figure}

\clearpage

  \newpage
 \pagebreak

 \begin{figure}
\noindent\includegraphics[clip=true,trim=0mm 20mm 0mm 0mm,width=80mm,height=80mm]{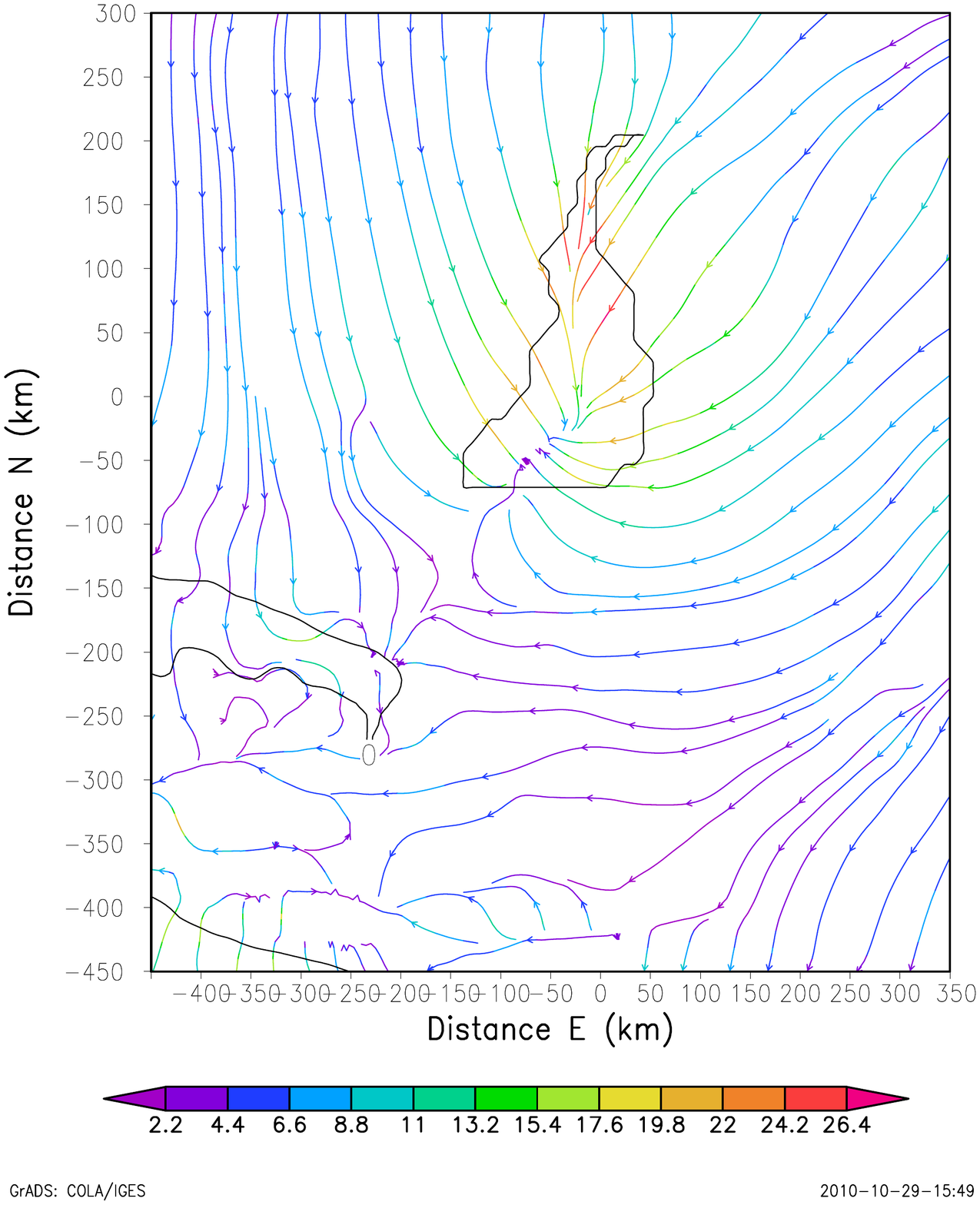}
\noindent\includegraphics[clip=true,trim=0mm 20mm 0mm 0mm,width=80mm,height=80mm]{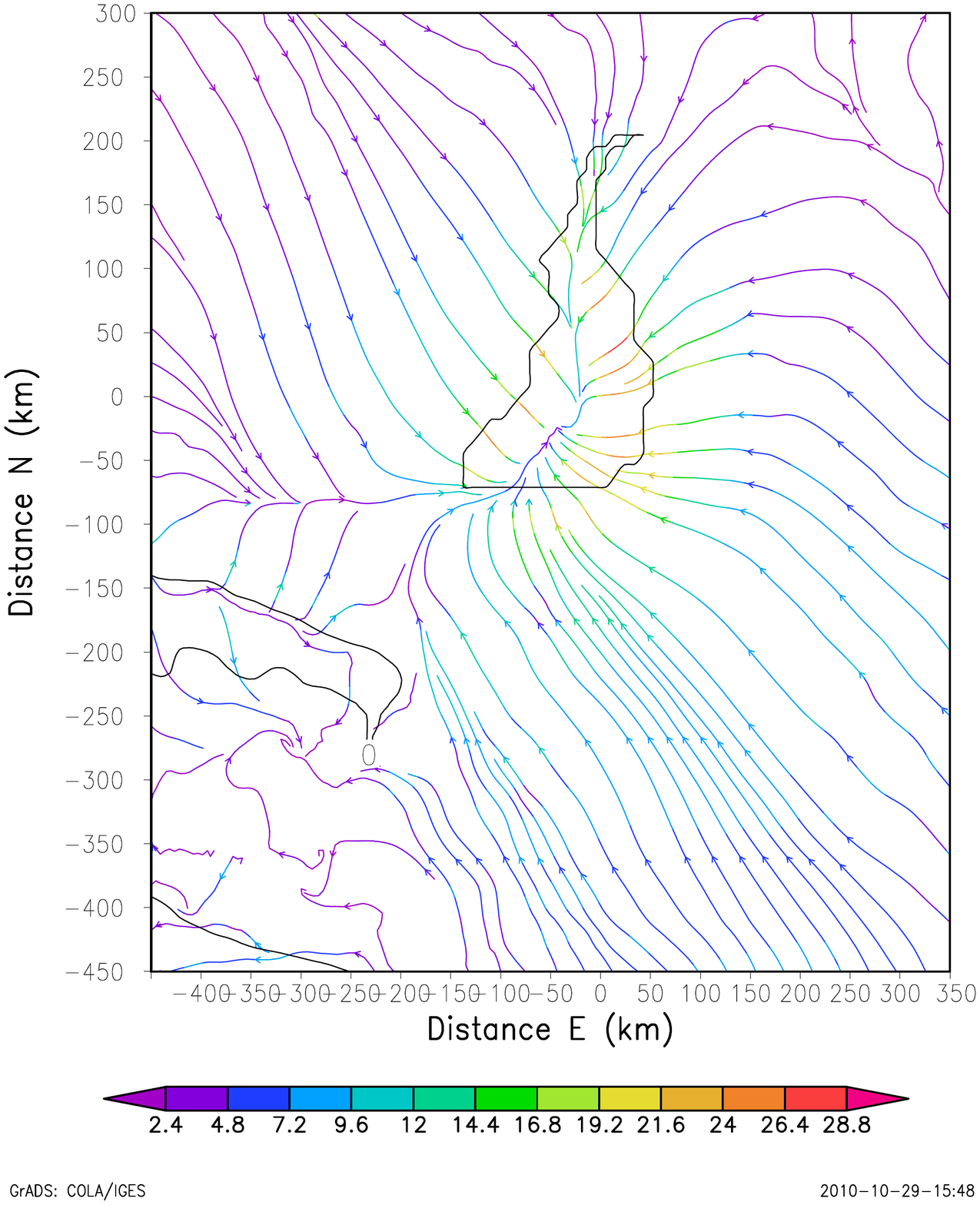}
\noindent\includegraphics[clip=true,trim=0mm 25mm 0mm 0mm,width=80mm,height=80mm]{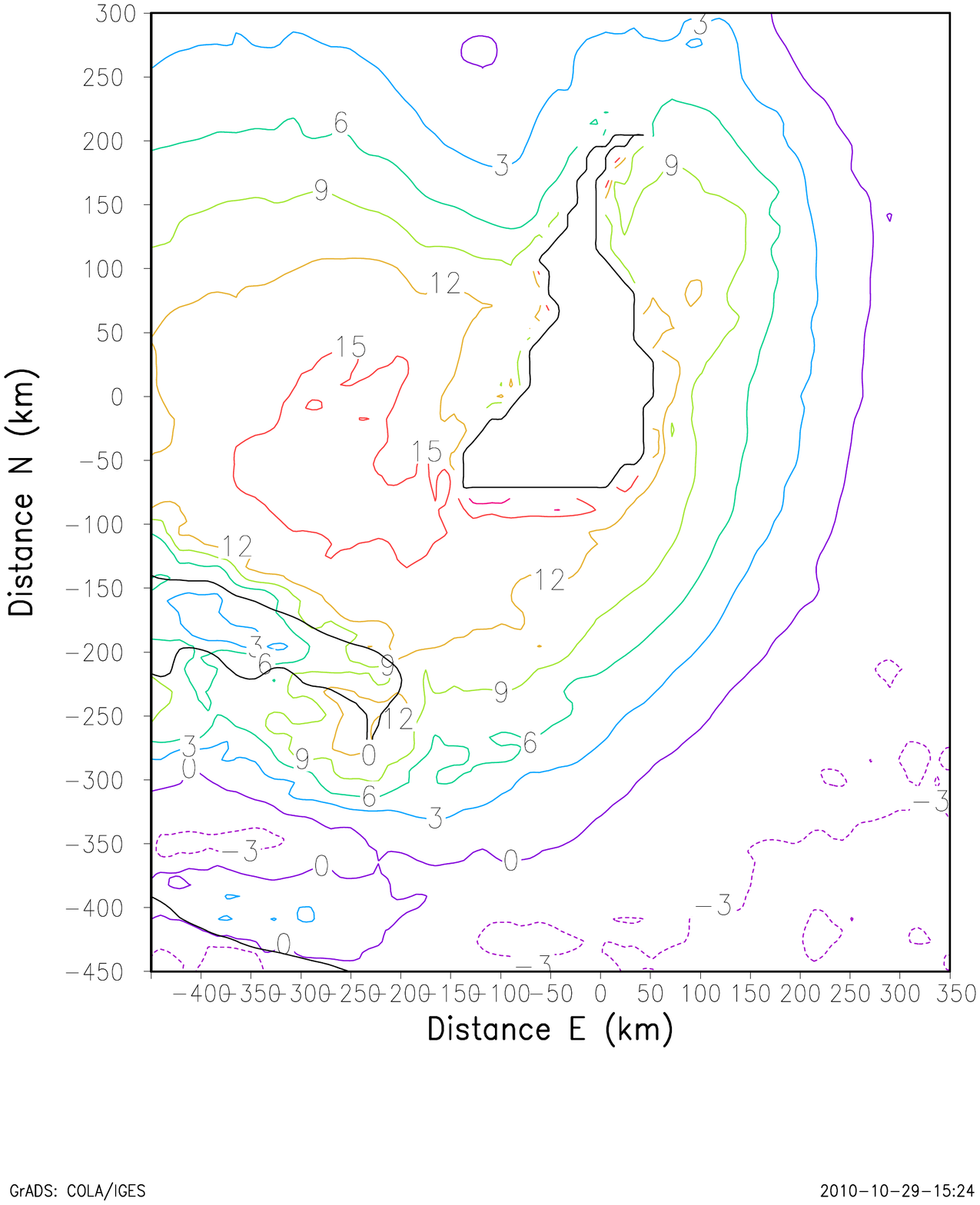}
%\noindent\includegraphics[clip=true,trim=0mm 20mm 0mm 0mm,width=80mm,height=80mm]{FigTEMPANDWINDc.pdf}
%\noindent\includegraphics[clip=true,trim=0mm 20mm 0mm 0mm,width=80mm,height=80mm]{FigTEMPANDWINDb.pdf}
%\noindent\includegraphics[clip=true,trim=0mm 7.5mm 0mm 0mm,width=80mm,height=80mm]{FigTEMPANDWINDa.pdf}
\end{figure}

\begin{figure}
 \caption{\label{TEMPANDWIND} Temperature and wind field at Juventae in our best-fit lake simulation (\texttt{juventae\_high}). 0m contour (black) defines canyons. (a) Net time-averaged wind at 13m elevation. Overall Easterly \& South-Easterly winds are reversed at Juventae Chasma because of the lake-driven circulation (b) shows change in wind field due to the lake (differencing dry and wet runs). Lake storm drives low-level convergence of up to 30 m s$^{-1}$. Magnitude of lake-driven circulation is comparable to magnitude of non-lake circulation. (c) Temperature difference due to lake. Temperature increases by up to 18K downwind of the lake. Axis tick labels correspond to distance from 3.91S 298.53E, near the lake center (Note: \texttt{juventae\_high}, average over 3 sols, from Mars-hours 49 to 121).}
 \end{figure}

 \newpage
 \pagebreak

 \begin{figure}
\noindent\includegraphics[clip=true,trim=0mm 20mm 0mm 0mm,width=80mm,height=80mm]{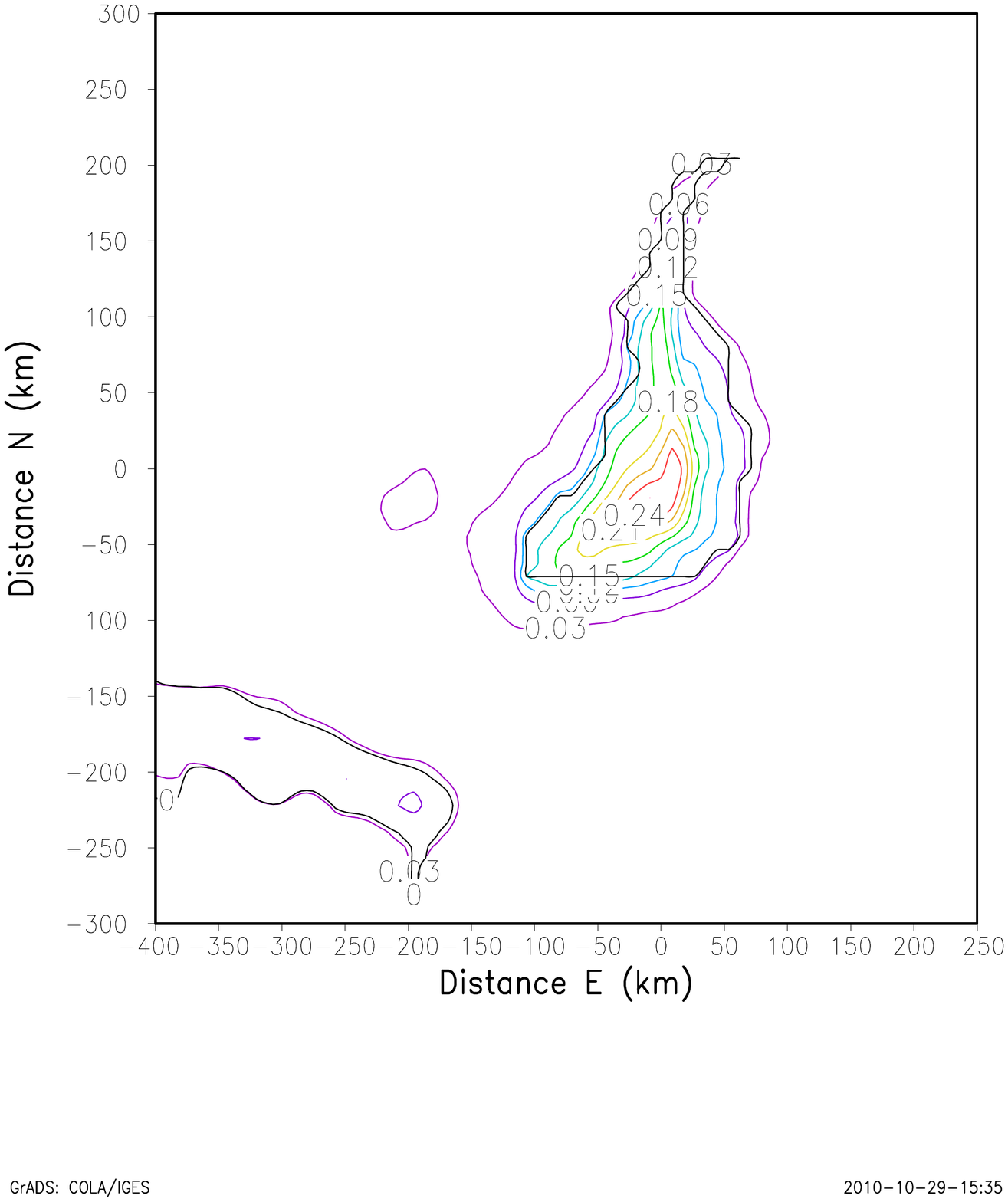}
\noindent\includegraphics[clip=true,trim=0mm 20mm 0mm 0mm,width=80mm,height=80mm]{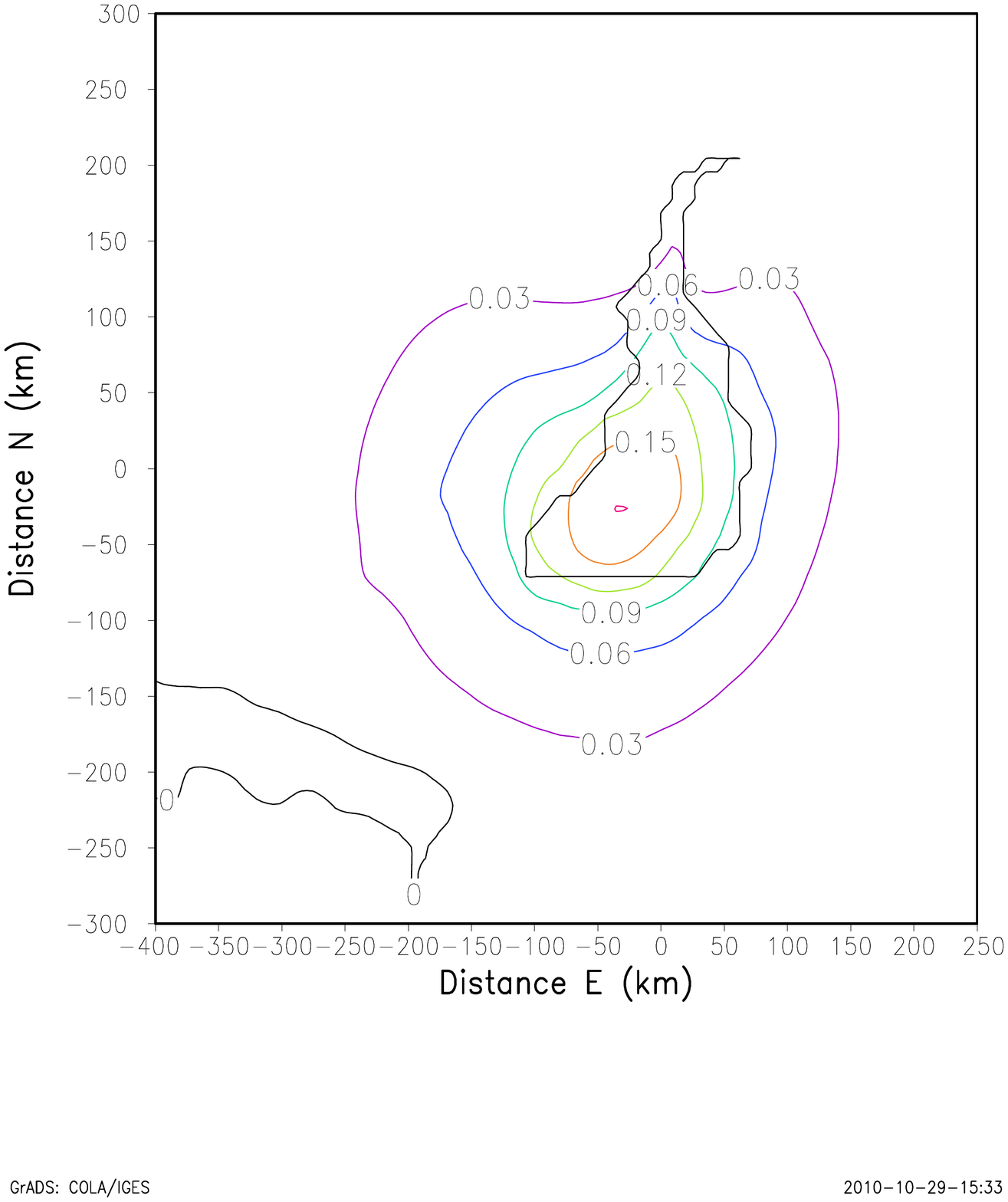}
%\noindent\includegraphics[clip=true,trim=0mm 20mm 0mm 0mm,width=80mm,height=80mm]{FigWATERCOLUMNa.pdf}
%\noindent\includegraphics[clip=true,trim=0mm 20mm 0mm 0mm,width=80mm,height=80mm]{FigWATERCOLUMNb.pdf}
 \caption{\label{WATERCOLUMN} Ice and vapor columns at Juventae Chasma. (a) Time-averaged precipitable vapor column abundance in cm water equivalent. (b) Time-averaged precipitable ice column abundance in cm water equivalent.
 (Note: \texttt{juventae\_high}, Average for 3 sols, from Mars-hours 49 to 121). }
 \end{figure}

  \newpage
 \pagebreak

 \begin{figure}
 \noindent\includegraphics[width=20pc,clip=true,trim=0mm 20mm 0mm 0mm]{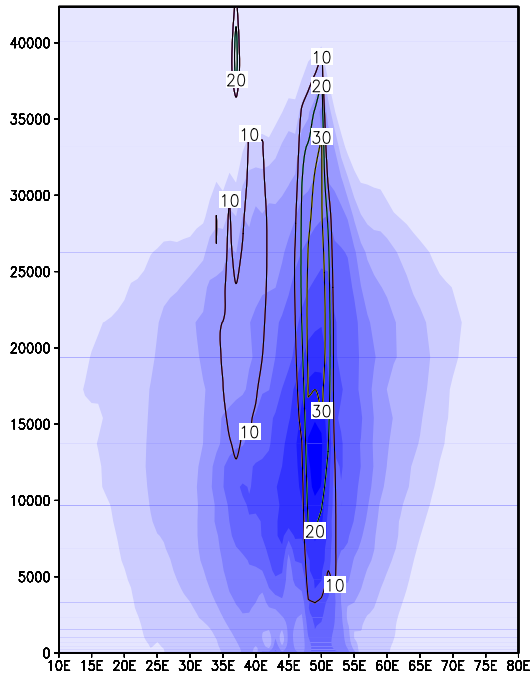}
% \noindent\includegraphics[width=20pc]{FigCLOUDSTRUCTURE.pdf}
 \caption{\label{CLOUDSTRUCTURE} E-W cross section through lake storm. Blue tint corresponds to increasing water ice fraction (interval 0.001, maximum value 0.009). Labelled contours correspond to bulk vertical velocity in m s$^{-1}$, which are comparable to the most intense supercell storms on Earth. The y-axis is vertical distance in m. The x-axis is horizontal distance: 10 units = 83 km. Lake extends from 41E to 55E on this scale.}
 \end{figure}

% \newpage
% \pagebreak
% \begin{figure}
% %\noindent\includegraphics[width=20pc,clip=true,trim=0mm 20mm 0mm 0mm]{FigSTORMSLICE.pdf}
%% \noindent\includegraphics[width=20pc]{FigCLOUDSTRUCTURE.pdf}
% \caption{\label{3DVIS} Possible 3D figure showing storm structure with 3D Visualizer. Show ice mixing ratio and maybe streamtube windfield also.}
% \end{figure}
% \newpage
% \pagebreak

 \begin{figure}
 \noindent\includegraphics[width=170mm,clip=true,trim=80mm 0mm 60mm 0mm]{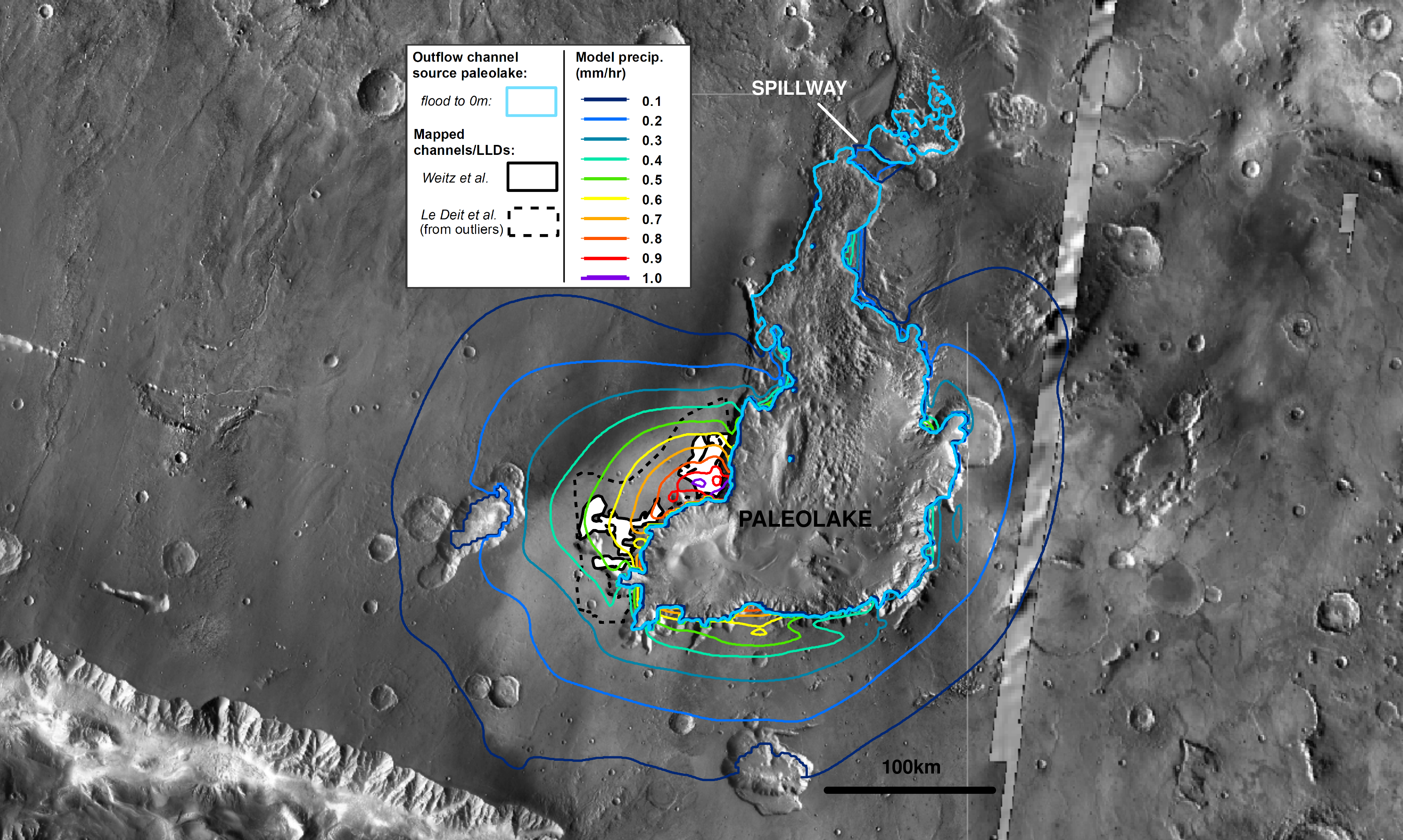}
 \end{figure}

\begin{figure}
 \caption{\label{SNOWVSCHANNELS} Modeled precipitation contours overlain on observed geology at Juventae. White shading with thick solid black outline corresponds to area of layered deposits and inverted channels reported by \citet{wei10}. The dashed black line corresponds to the pre-erosion area of layered deposits inferred from outlier buttes and pedestal craters by \citet{led10}. The thick cyan line defines the flooded area for this simulation (the -1000m contour). The colored lines are modeled time-averaged precipitation contours at intervals of 0.1 mm/hr water equivalent. Precipitation falling back into the lake is not shown. The spatial maximum in mean precipitation is $\approx$1.0 mm/hr. (Note: sol 5 average).}
 \end{figure}

\clearpage

 \begin{figure}
% \noindent\includegraphics[width=20pc]{samplefigure.eps}
\noindent\includegraphics[clip=true,trim=0mm 20mm 0mm 0mm,width=80mm,height=80mm]{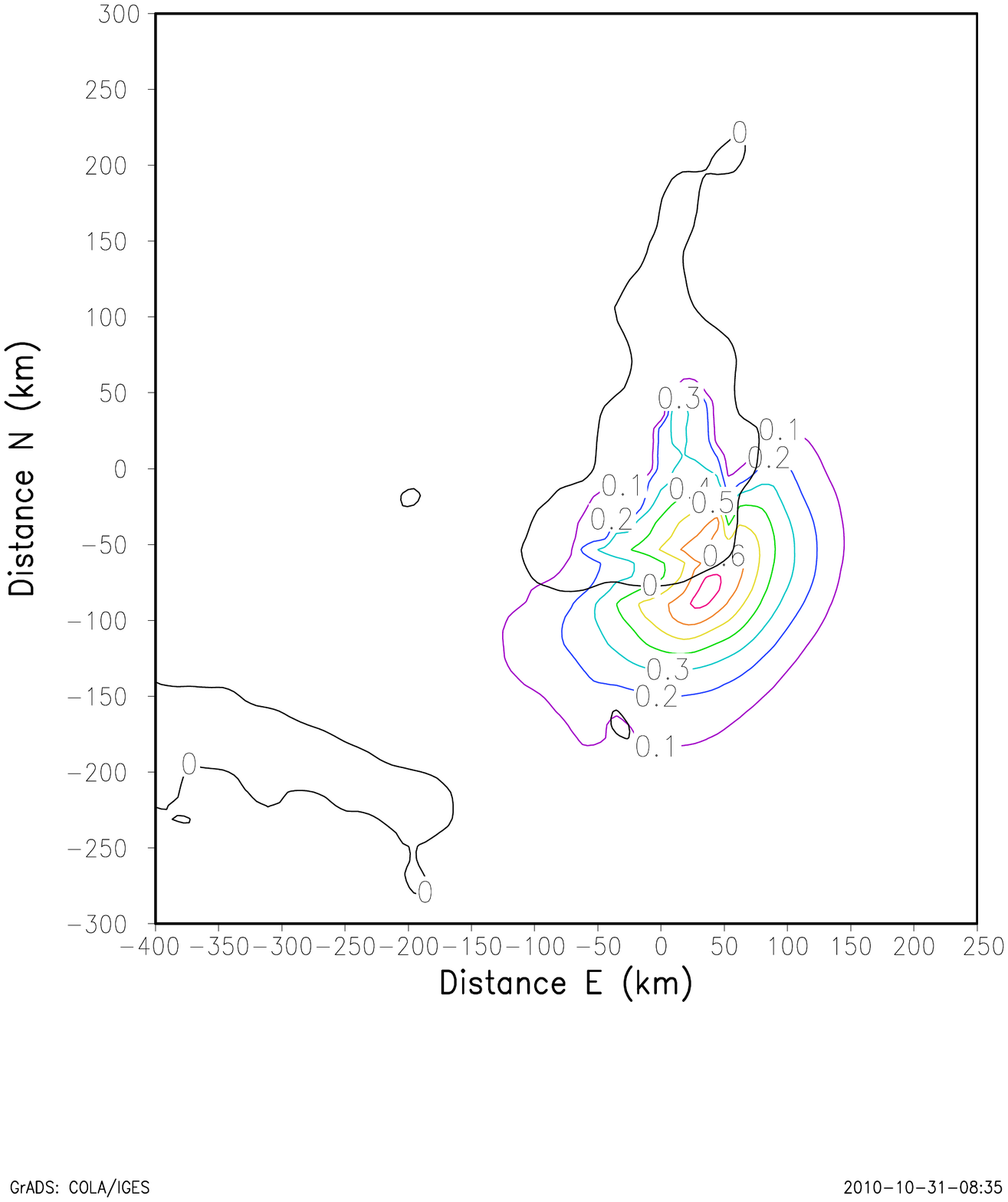}
\noindent\includegraphics[clip=true,trim=0mm 20mm 0mm 0mm,width=80mm,height=80mm]{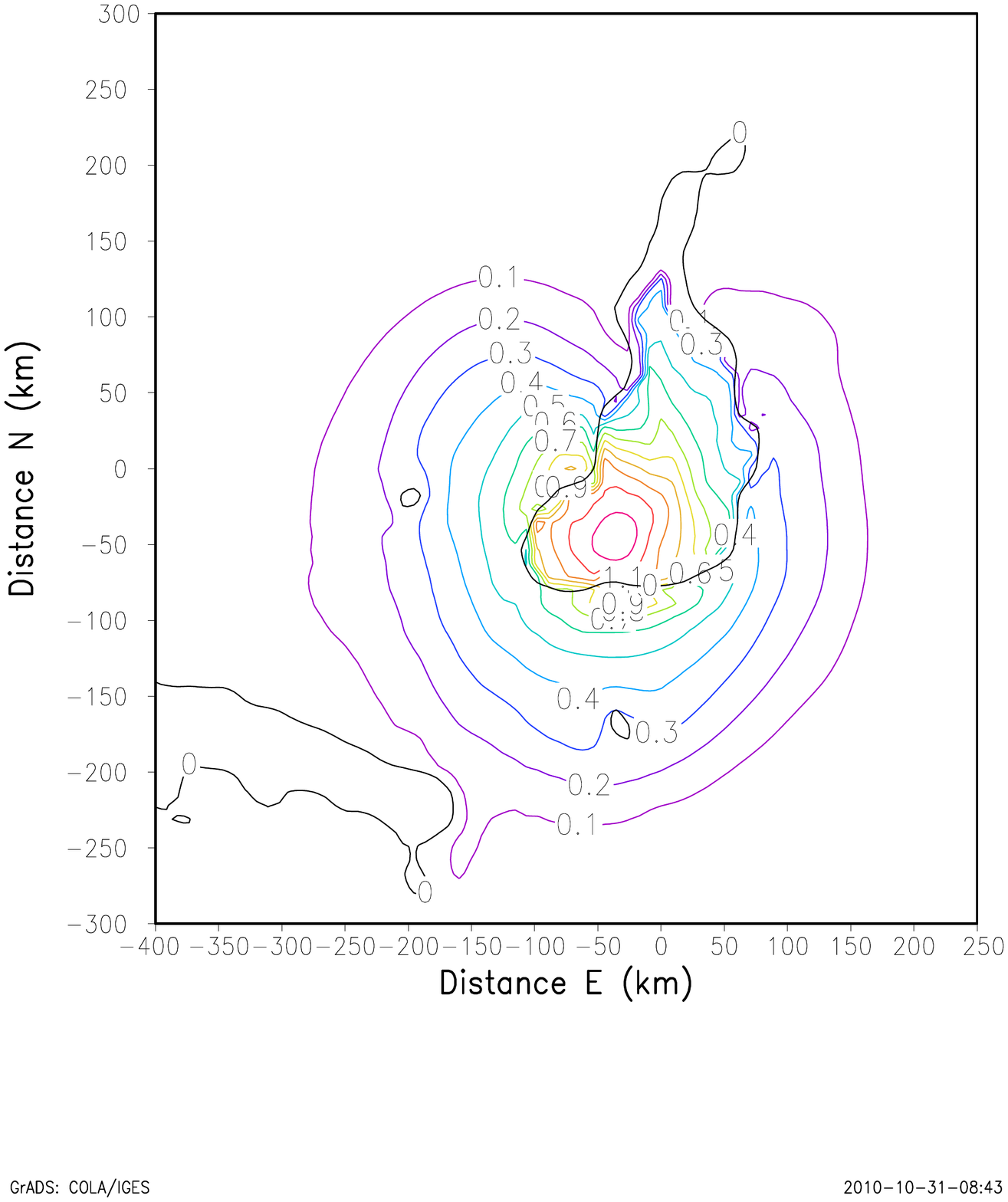}
\noindent\includegraphics[clip=true,trim=0mm 20mm 0mm 0mm,width=80mm,height=80mm]{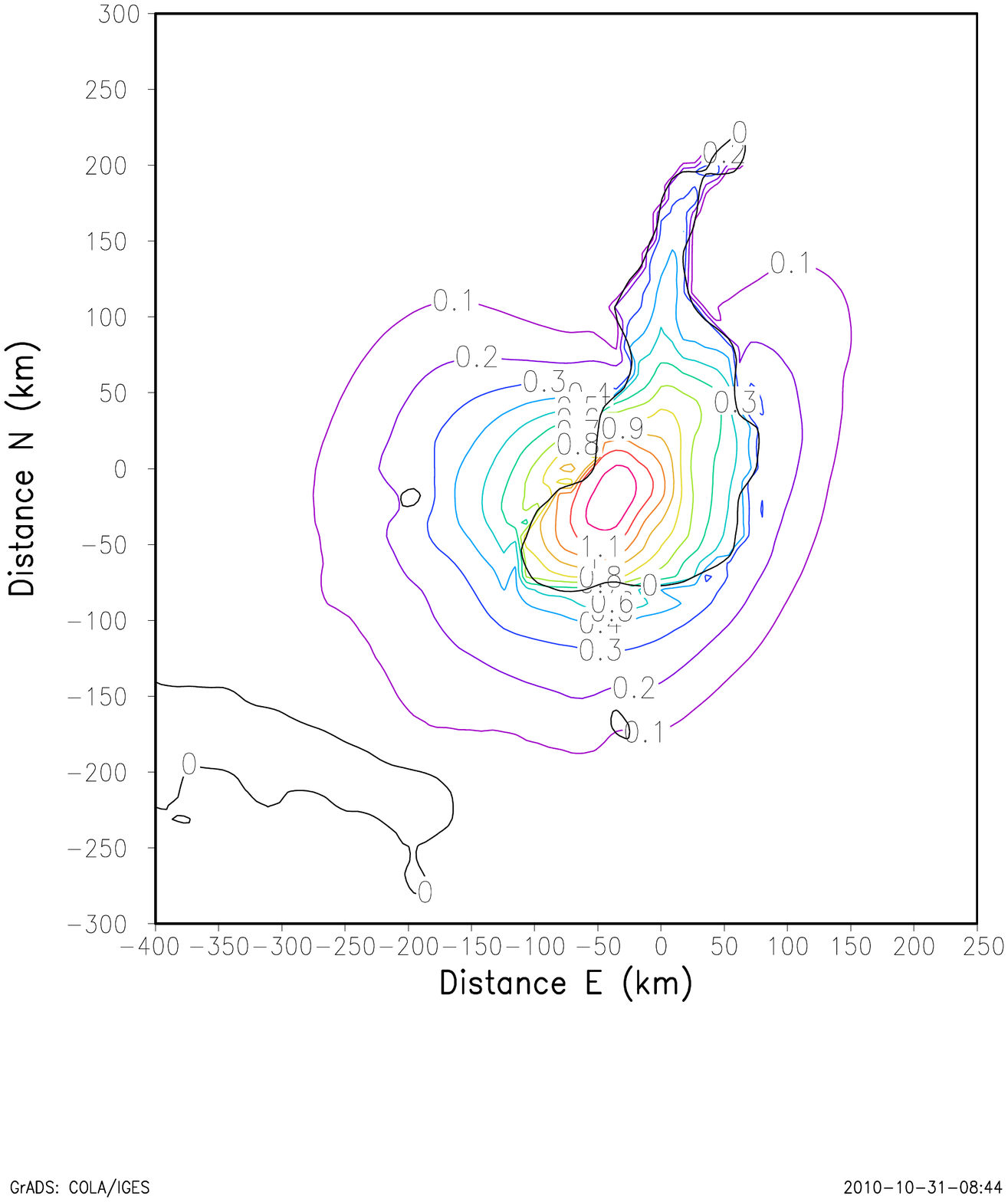}
 \caption{\label{SNOWVSLAKELEVEL} Sensitivity test comparing mean precipitation (mm/hr) for 3 different lake levels. a) -3000m (\texttt{juventae\_low}), b) -1000m (\texttt{juventae\_medium}), c) 0m (\texttt{juventae\_high}).}
 \end{figure}

\clearpage

  \newpage
 \pagebreak
\clearpage

 \begin{figure}
 \noindent\includegraphics[width=180mm,clip=true,trim=40mm 0mm 40mm 140mm]{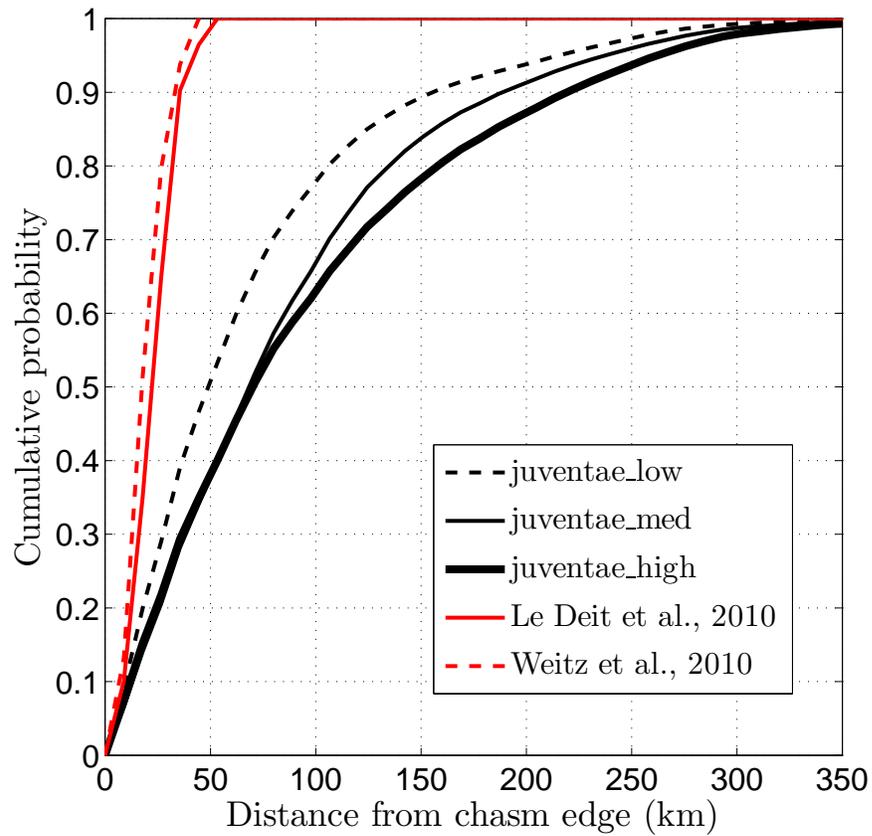}
 \caption{\label{FALLOFF} Comparison of falloff of precipitation with falloff of mapped area of channels and layered deposits. Because of an arbitrary mapping cutoff, layered deposits likely extend further than shown.}
 \end{figure}

  \newpage
 \pagebreak
\clearpage

 \begin{figure}
 \noindent\includegraphics[width=180mm,clip=true,trim=40mm 0mm 40mm 140mm]{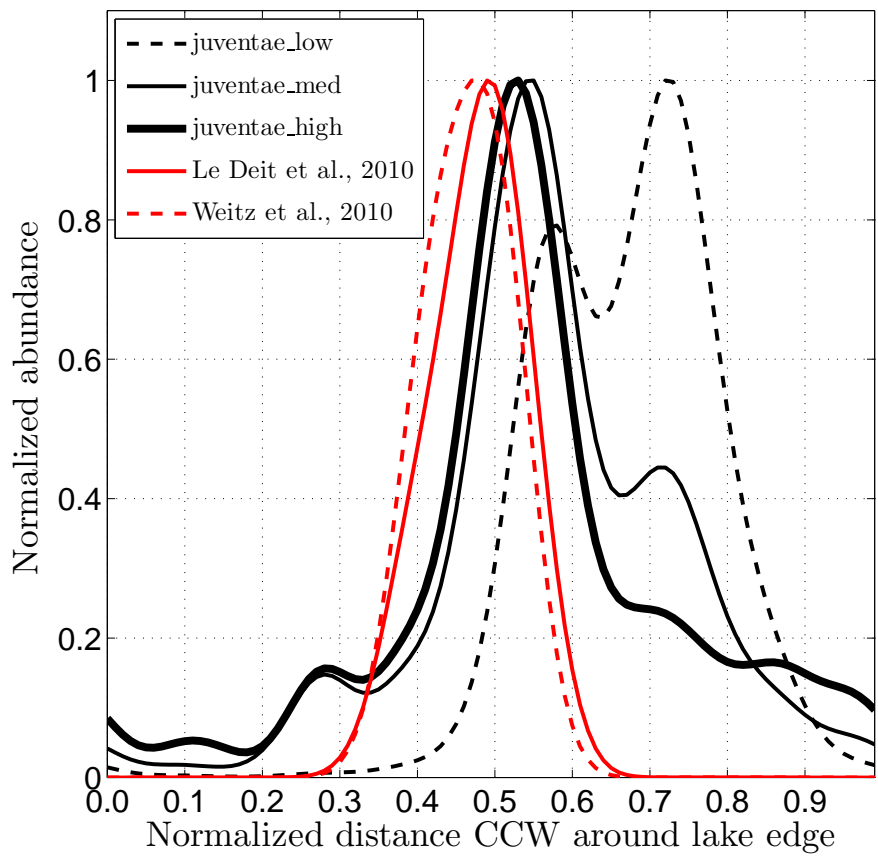}
 \caption{\label{AZIMUTH} Comparison of azimuth of precipitation with azimuth of mapped channels.}
 \end{figure}

\clearpage

 \begin{figure}
\noindent\includegraphics[clip=true,trim=20mm 50mm 20mm 50mm]{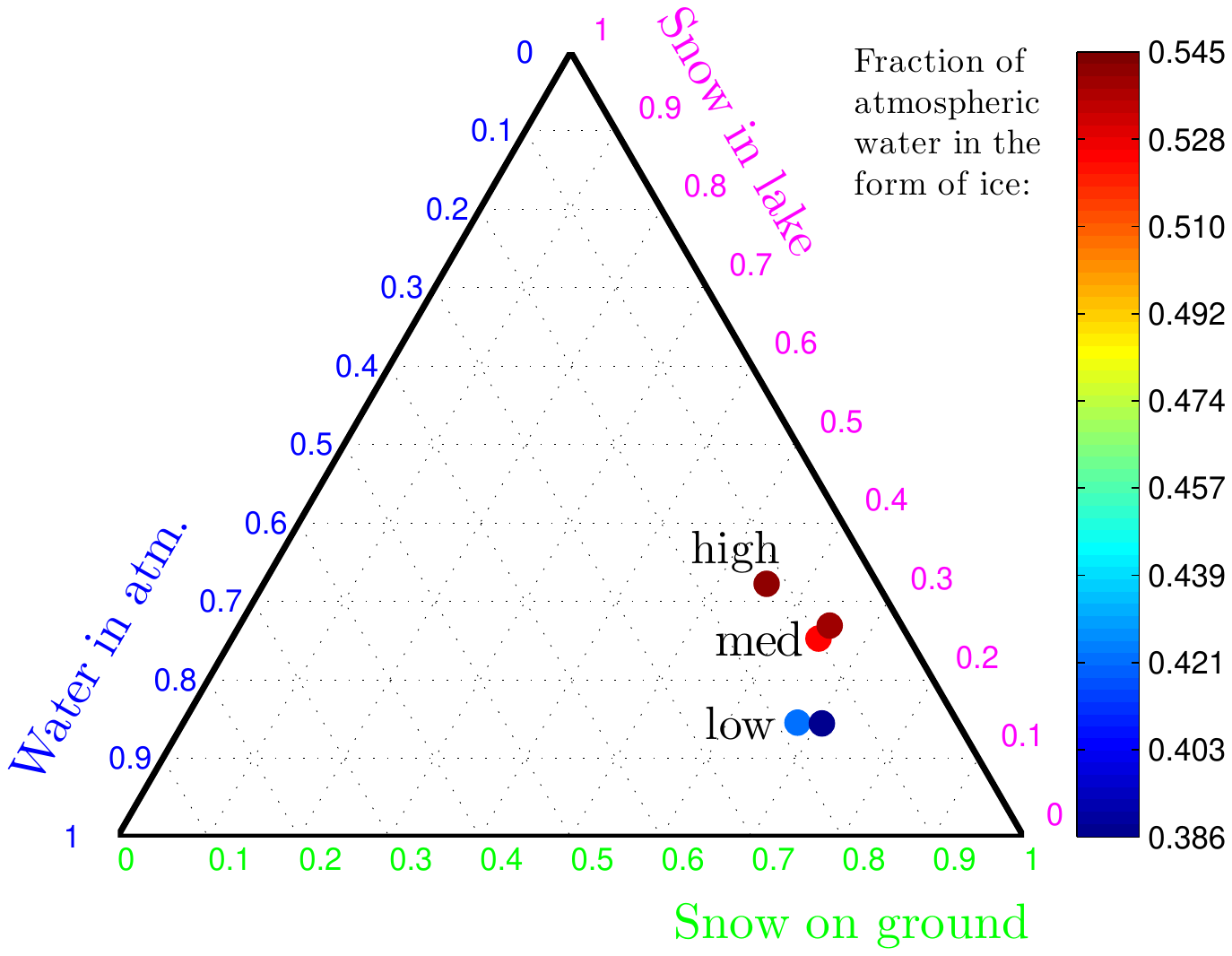}
\end{figure}

\begin{figure}
\caption{\label{TERNARYFATE} Fate of vapor released from the lake (precipitation efficiency). Colored dots corresponds to simulations' inventory of water (ice + vapor) after either 5 or 7 sols, after subtracting the inventory of a lake-free run. Color corresponds to fraction of atmospheric water in the ice phase: red is more ice-rich, blue is more vapor-rich. Only snow on ground can contribute to localized geomorphology. Water vapor in the atmosphere can contribute to regional and global climate change: greenhouse warming is increasingly likely as the mass of atmospheric water increases.}
 \end{figure}

\clearpage

\begin{figure}
\noindent\includegraphics[width=90mm,clip=true,trim=40mm 0mm 75mm 50mm]{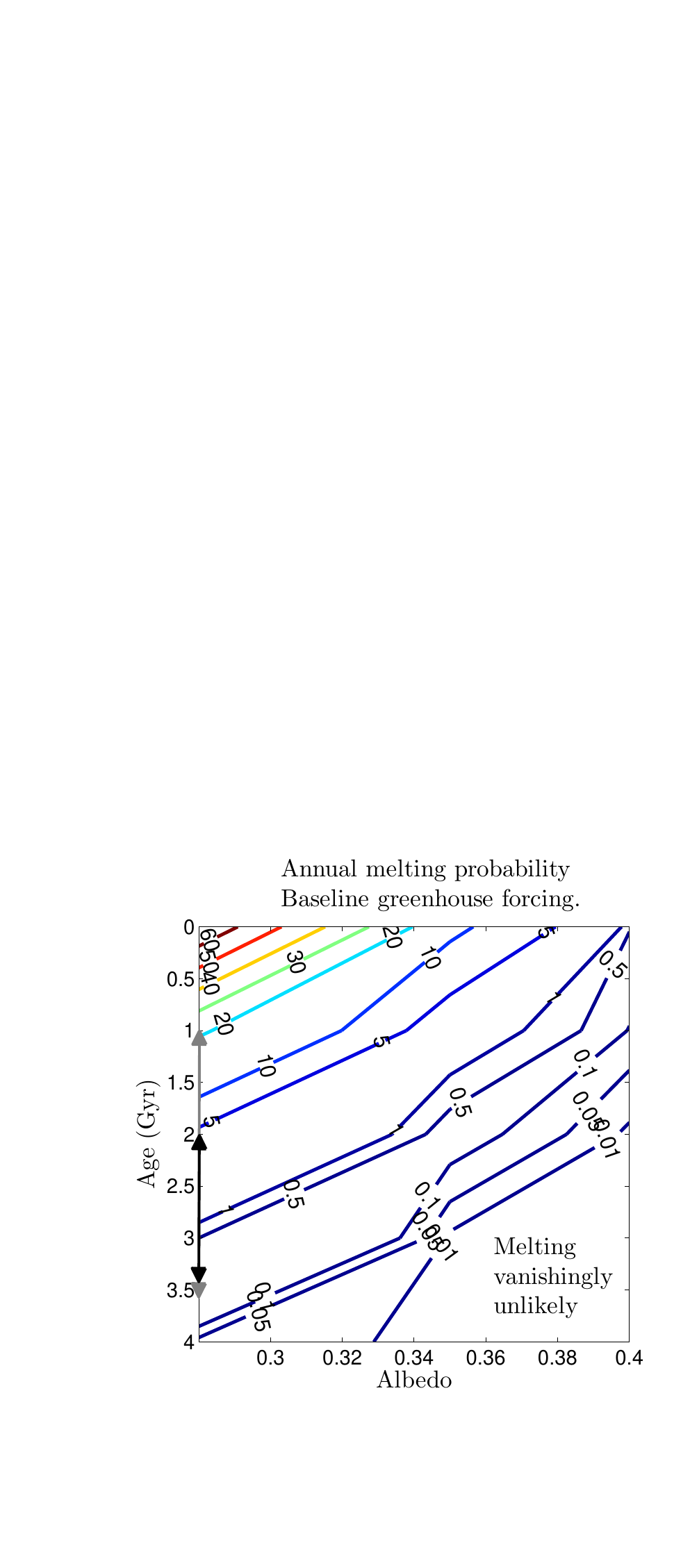}
\noindent\includegraphics[width=90mm,clip=true,trim=75mm 0mm 40mm 50mm]{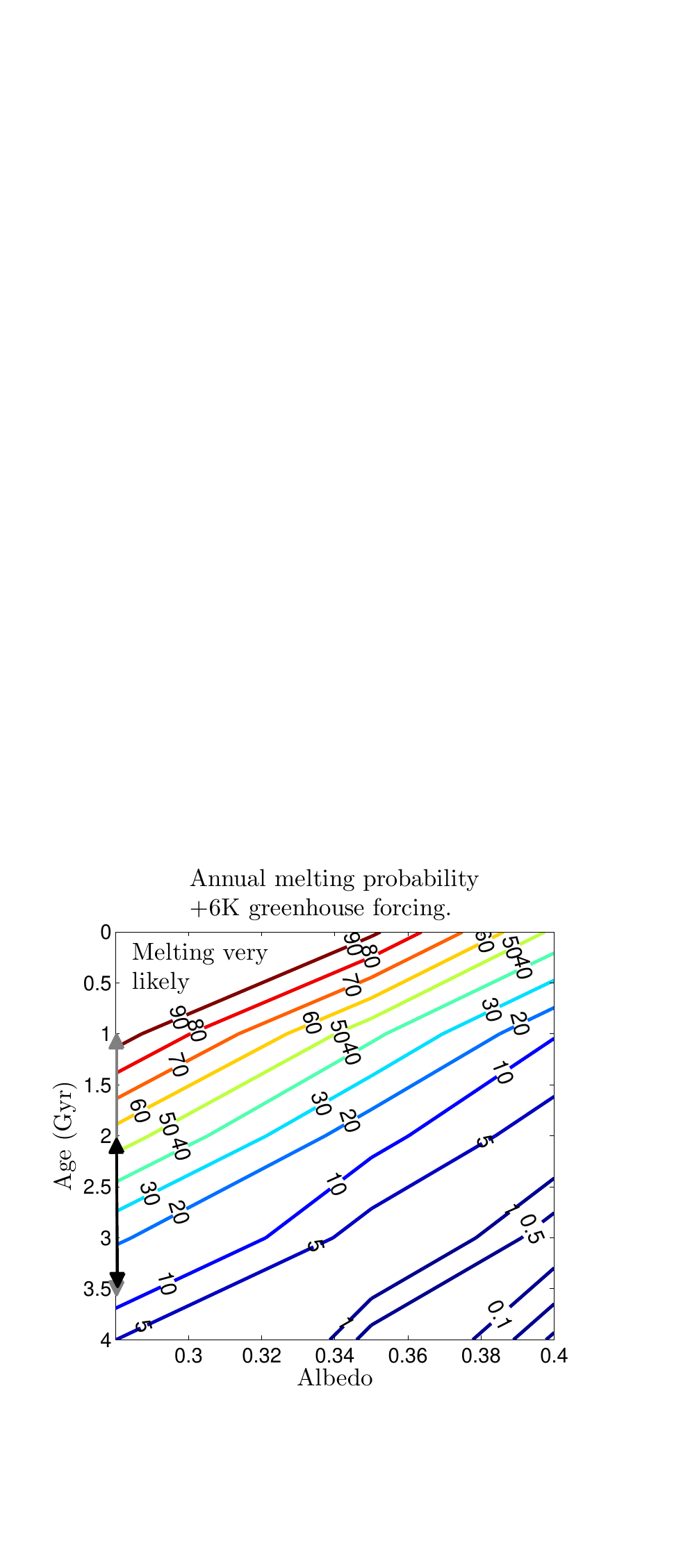}
\end{figure}

\begin{figure}
\caption{\label{MELTPROB} Equatorial melting probability (\%) for flatlying snowpack and solar luminosities appropriate to 0 to 4 Gya. Gray arrow on the y-axis corresponds to the range of ages for opaline layered deposits stated in the text of \citet{mur09}. Black arrow corresponds to the smallest range of stratigraphic ages that could accommodate all the plateau channel networks according to \citet{led10}. These ranges are large because of the uncertainties in mapping crater counts onto absolute ages \citep{harrecent}, and it is possible that all the plateau channel networks formed at roughly the same time.}
\end{figure}

\clearpage

\begin{figure}
\noindent\includegraphics[width=180mm,clip=true,trim=0mm 0mm 0mm 20mm]{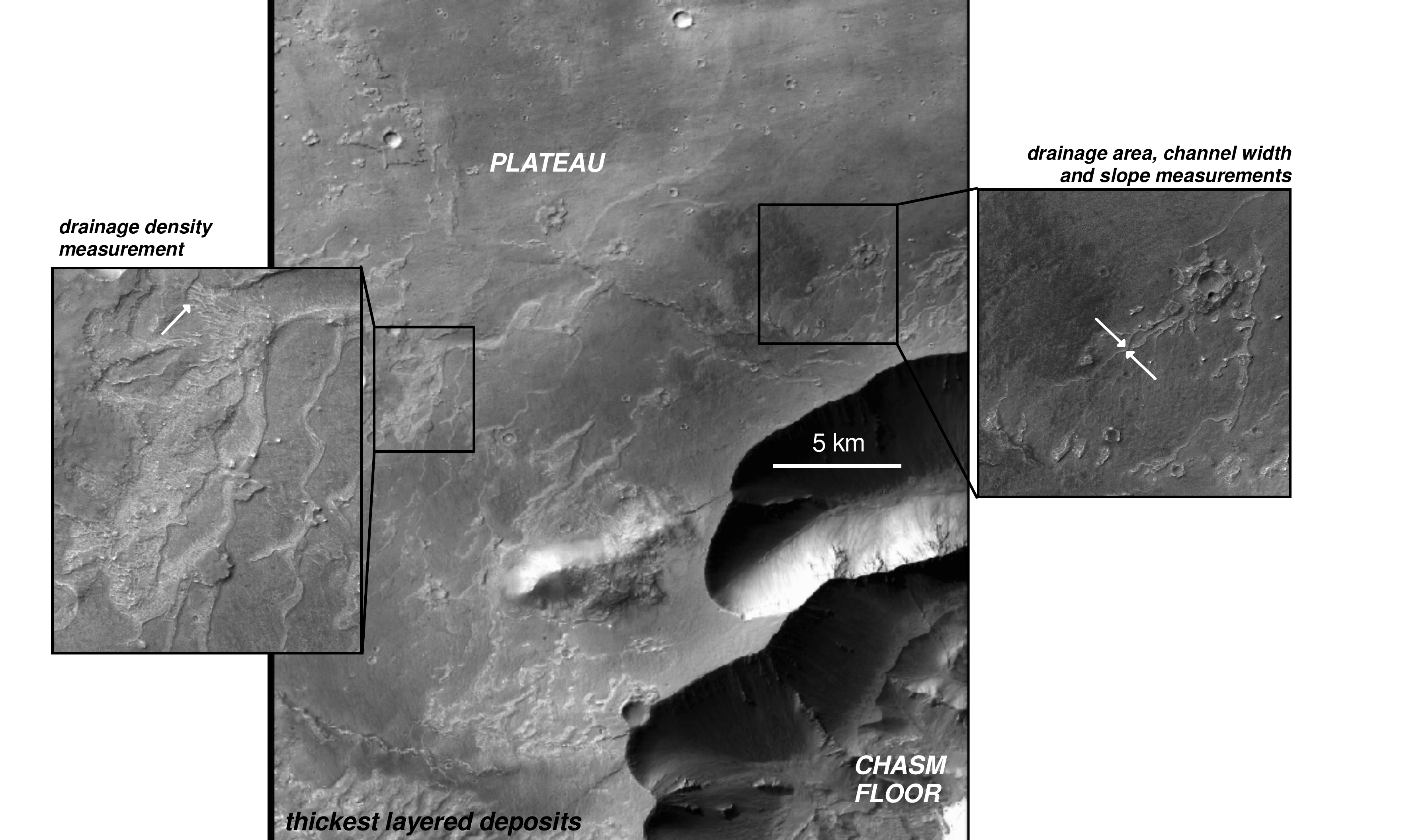}
\caption{\label{GEOLCONTEXT} Geological context of measurements of layered deposit thickness, channel drainage density, catchment area, channel slope and channel width. White arrows in zoomed-in panels show localities. Some thickness measurements come from an area just SW of this image. Background is part of CTX image P18\_007983\_1751\_XN\_04S063W.}
\end{figure}

 \begin{figure}
 \noindent\includegraphics[width=40pc,clip=true,trim=10mm 0mm 0mm 0mm]{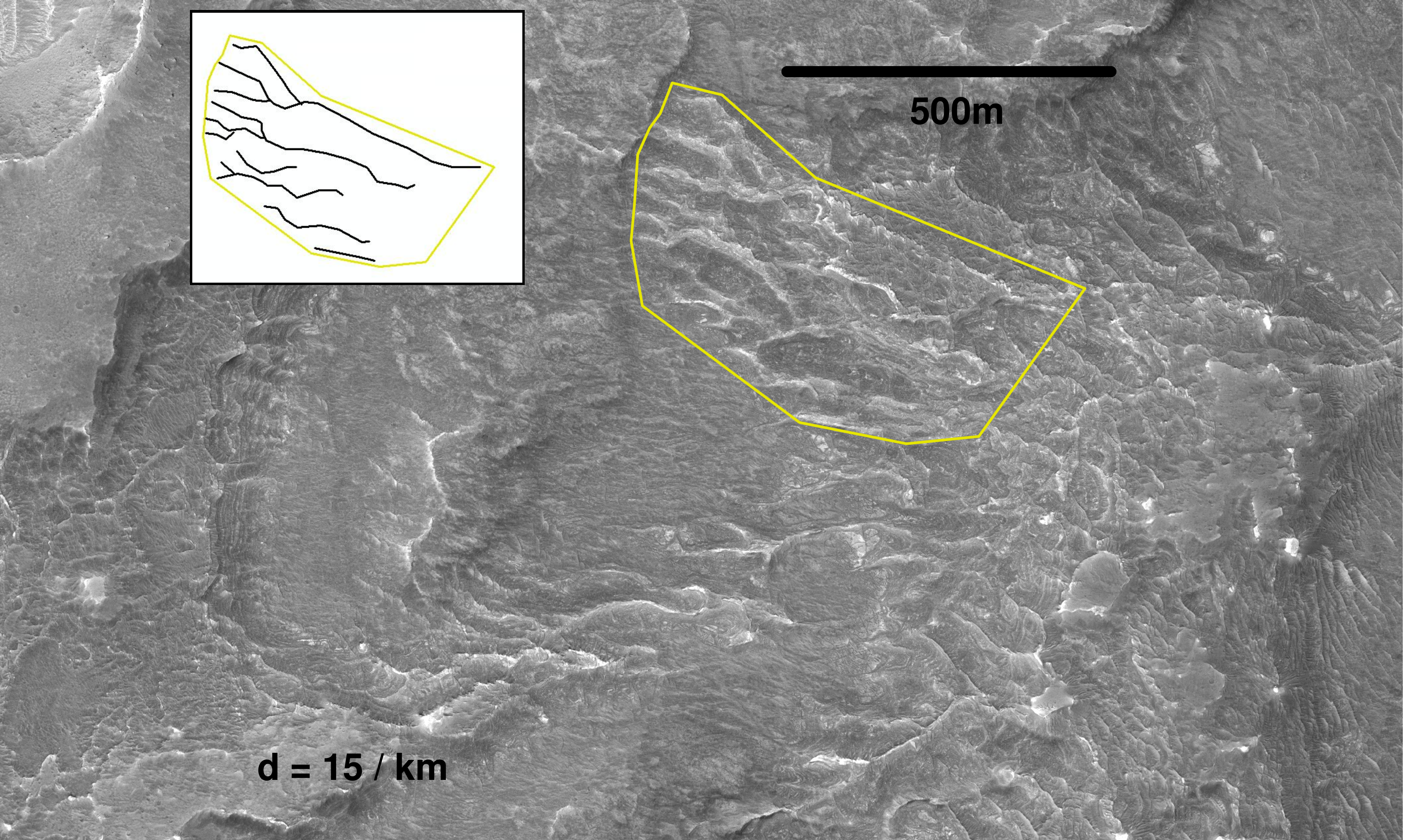}
 \noindent\includegraphics[width=40pc,clip=true,trim=0mm 0mm 0mm 0mm]{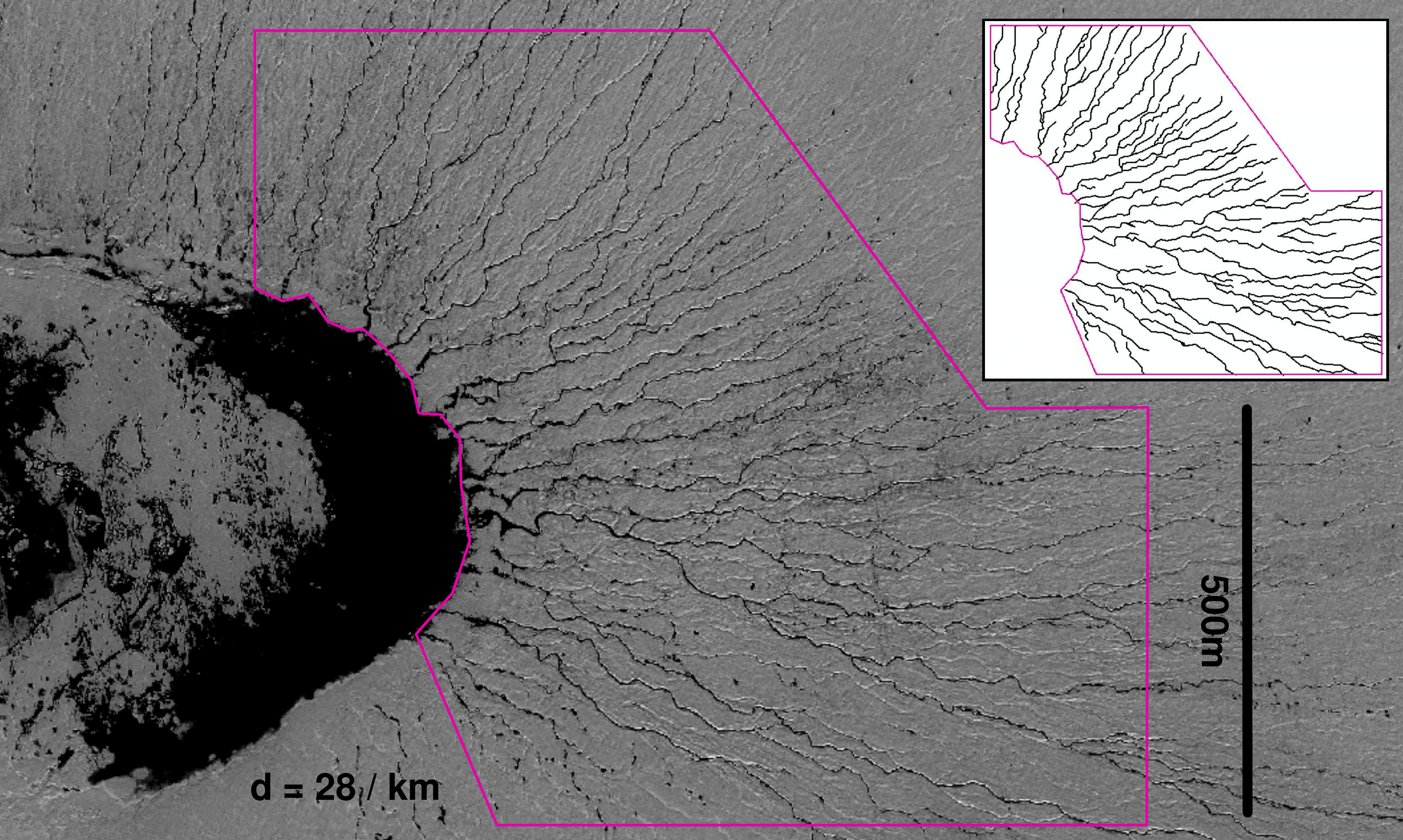}
 \end{figure}

\begin{figure}
% \noindent\includegraphics[width=20pc]{samplefigure.eps}
 \caption{\label{DRAINAGEDENSITY} Maximum drainage density, $d$, at Juventae ($d$ = 15 km$^{-1}$) is less than drainage density for snowmelt-carved systems on the Greenland ice plateau ($d$ = 29 km$^{-1}$). Polygons enclose the area for which the drainage density was measured, and the inset panels show channels which contributed to the count. {\it Upper panel:} PSP\_005346\_1755, Juventae Plateau. This locality was identified by \citet{mal10} as ``the best evidence yet found on Mars to indicate that rainfall and surface runoff occurred.'' {\it Lower panel:} IKONOS image of Greenland ice sheet, showing snowmelt-carved channels draining into a supraglacial lake. (Image courtesy Jason Box/OSU/Discovery Channel; \citet{box07}).
 {\it Details:} Greenland - length = 21.3 km, area = 0.77 km$^2$, giving $d$ = 28/km.
 Mars - length = 3.4 km, area = 0.22 km, giving $d$ = 15/km.}
 \end{figure}

\clearpage

\begin{figure}
\noindent\includegraphics[clip=true,trim=10mm 0mm 0mm 0mm, width=180mm]{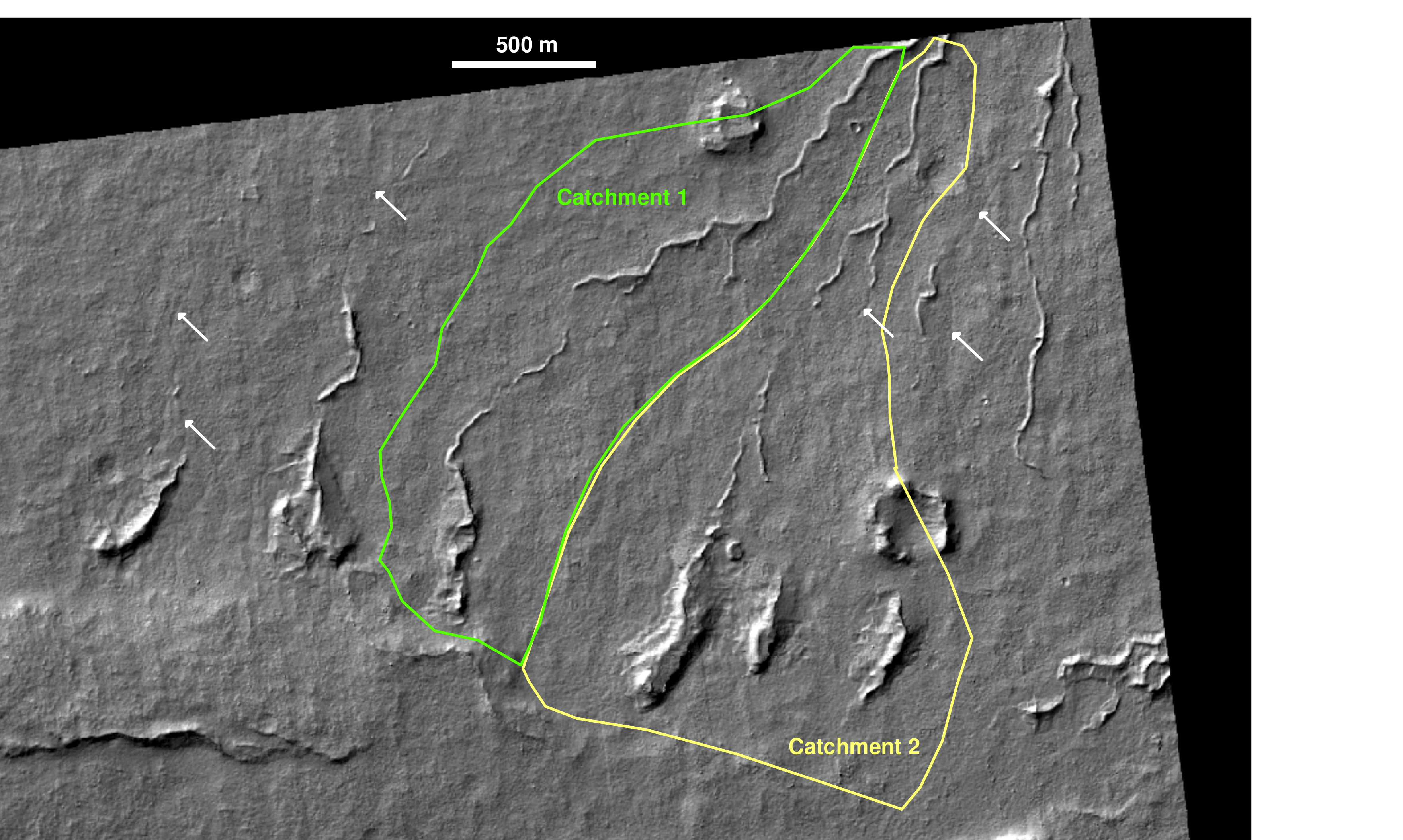}
\caption{\label{STRATDETAIL} Inverted channel systems on the Juventae plateau from which hydraulic parameters were measured. Shaded relief from HiRISE stereo DTM 1 (Figure \ref{GEOLOGYCONSTRAINTS}), illuminated from top left. North is up. Barely-visible striping parallel to the edge of the shaded relief raster is an artifact of stereo DTM generation.
Inverted channel heights are typically 2-5m. Shallow negative-relief channels (white arrows, depth $<$ 1m from DTM) join some inverted channel segments. The corresponding HiRISE image pair is PSP\_003223\_1755 and PSP\_003724\_1755.}
\end{figure}

%Rightmost channel rise:run 0.012, adjacent channel rise:run 0.009. Color scale runs from 2352m (white) to 2381m (red).

\begin{figure}
 \noindent\includegraphics[clip=true,trim=50mm 0mm 50mm 150mm, height=150mm]{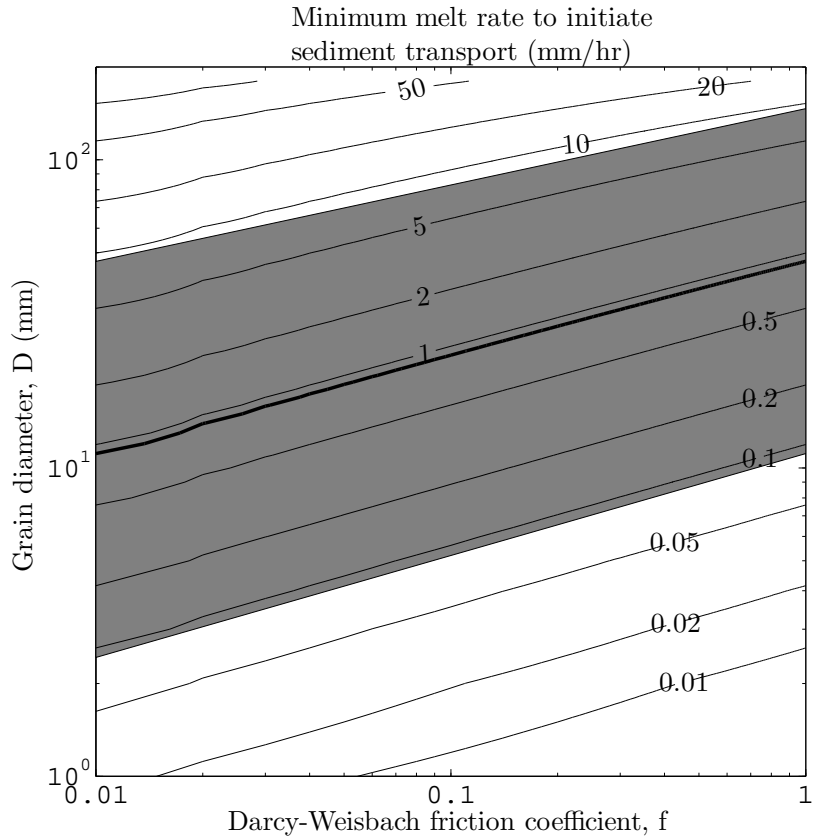}
 \caption{\label{DTMANALYSIS} Grain sizes that can be mobilized by the modeled precipitation. The central black line corresponds to the modeled snowfall rate, 0.9 mm/hr. Because we consider the modeled precipitation rate to be only an order-of-magnitude guide to the true precipitation rate, the gray envelope shows order-of-magnitude errors in both directions. The melting rate will be lower than the precipitation rate, and in our probabilistic discharge model the exceedance probability for a melt rate of 1.0 mm/hr is extremely low. Melt rates of 0.09 mm/hr (lower end of gray envelope) are more likely, and are sufficient to mobilize sand and gravel.}
 \end{figure}

%Exceedance probabilities for melt rate 0.1 mm/hr are 15\% (1.0\%, 0.03\%) for albedo 0.28 (0.35,0.4) and an unchanged greenhouse effect, and 46\% (8.2\%, 0.9\%) for a greenhouse effect 6K stronger than today. All calculations are for solar luminosity appropriate to 3.0 Gya.}

\begin{figure}
 \noindent\includegraphics[width=170mm,clip=true,trim=15mm 5mm 35mm 10mm]{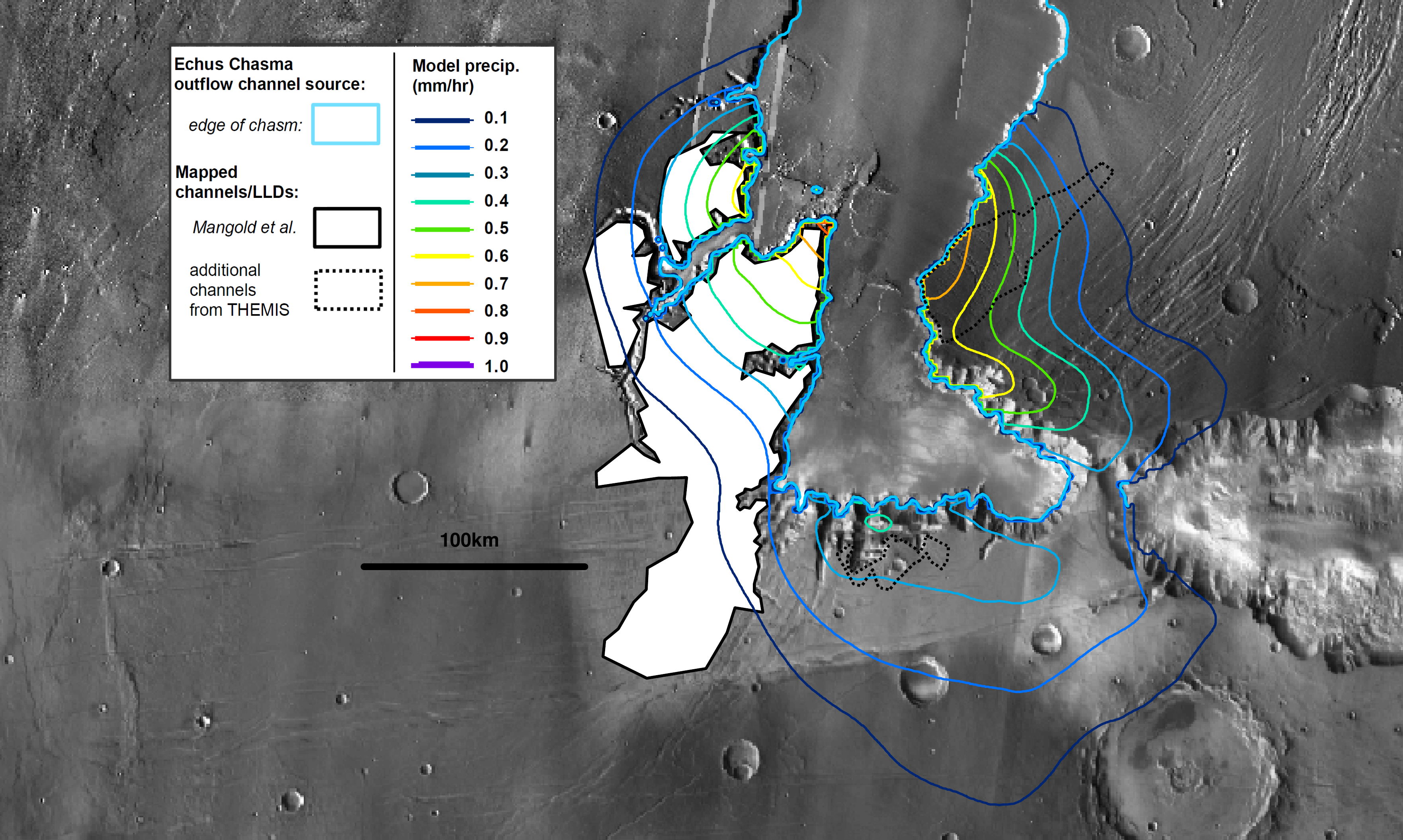}
 \end{figure}

\begin{figure}
 \noindent\includegraphics[width=170mm,clip=true,trim=10mm 0mm 50mm 0mm]{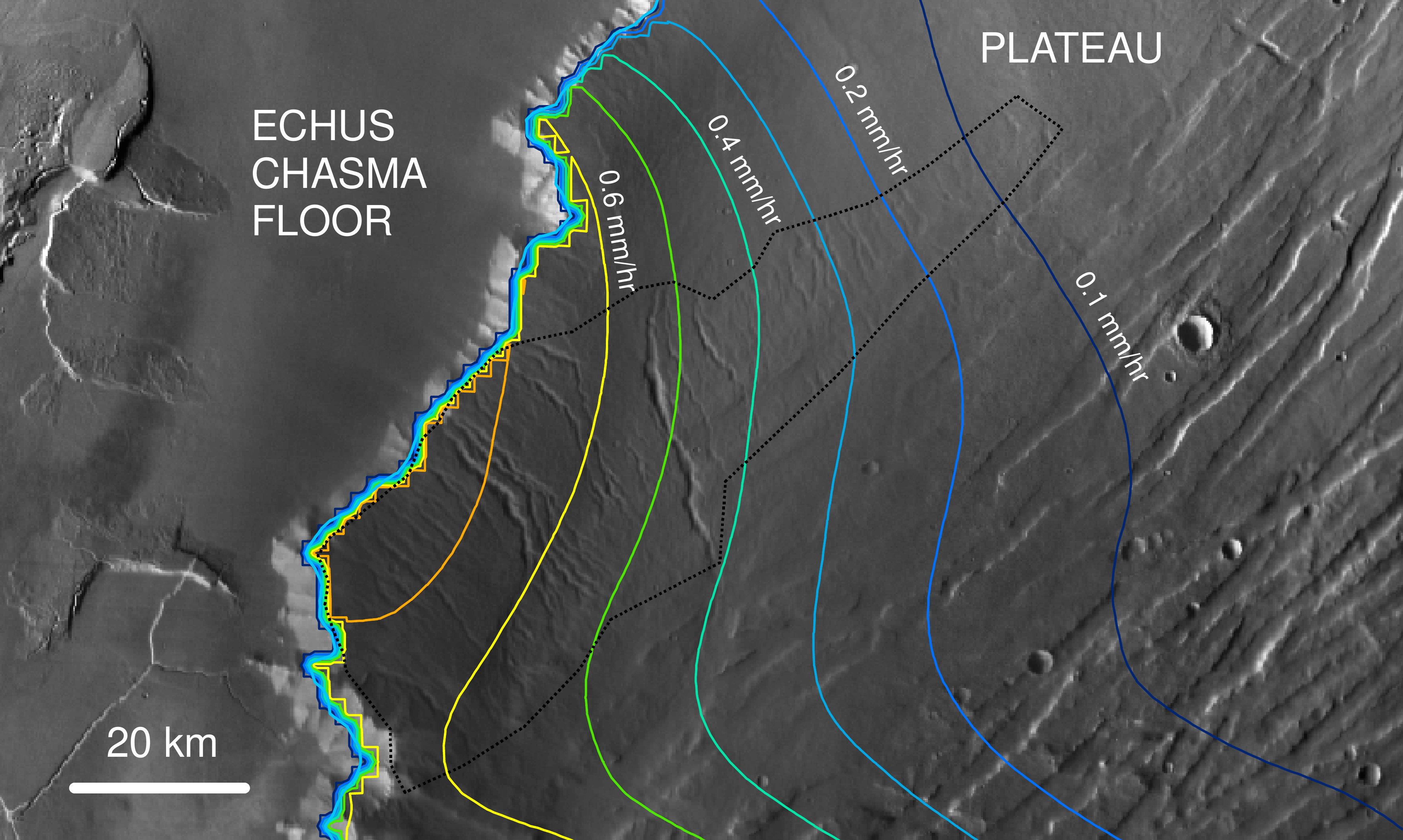}
\caption{\label{ECHUSMANGOLD} a) Modeled precipitation contours overlain on observed geology at Echus. White shading with thick solid black outline is the perimeter of channels reported by \citet{man08}. The dotted black line corresponds to additional channels observed in the THEMIS Day IR 100m per pixel global mosaic. The colored lines are modeled mean precipitation contours at intervals of 0.1 mm/hr water equivalent. Precipitation falling into the lake is not shown. b) Focus on area where channels have not been previously mapped. The dotted line encloses the area of channels visible in Themis IR mosaic. The depth and width of the observed channels decreases away from the chasm edge in step with the decline in modeled precipitation.}
\end{figure}

% \end{document}
%
% \begin{table}
% \caption{}
% \end{table}
%
% ---------------
% TWO-COLUMN figure/table
%
% \begin{figure*}
% \noindent\includegraphics[width=39pc]{samplefigure.eps}
% \caption{Caption text here}
% \end{figure*}
%
% \begin{table*}
% \caption{Caption text here}
% \end{table*}
%
% see below for how to make landscape figures or tables

%%% End the article here:


\begin{thebibliography}{}

\bibitem[{\textit{Andrews-Hanna \& Phillips}(2007)}]{and07}
Andrews-Hanna, J.C., \& R.J. Phillips (2007), Hydrological modeling of outflow channels and chaos regions on Mars, J. Geophys. Res. 112, E08001, doi:10.1029/2006JE002881.

\bibitem[{\textit{Baker et al.}(1991)}]{bak91}
Baker, V.R., et al. (1991), Ancient oceans, ice sheets and the hydrological cycle on Mars, Nature 352, 589-594.

\bibitem[{\textit{Baker}(2001)}]{bak01}
Baker, V.R. (2001), Water and the martian landscape, Nature 412, 228-236, doi:10.1038/35084172

\bibitem[{\textit{Banks et al.}(2009)}]{ban09}
Banks, M.E., et al. (2009), An analysis of sinuous ridges in the southern Argyre Planitia, Mars using HiRISE and CTX images and MOLA data, J. Geophys. Res. 114(E9), E09003, doi:10.1029/2008JE003244.


\bibitem[{\textit{Bargery \& Wilson}(2010)}]{bar10}
Bargery, A.S., \& L. Wilson (2010), Dynamics of the ascent and eruption of water containing dissolved CO$_2$ on Mars, J. Geophys. Res. 115, E05008, doi:10.1029/2009JE003403.


\bibitem[{\textit{Barnhart et al.}(2009)}]{bar09}
Barnhart, C.J., A.D. Howard \& J.M. Moore (2009), Long-term precipitation and late-stage valley network formation: Landform simulations of Parana Basin, Mars, J. Geophys. Res. - Planets 114, E01003, doi:10.1029/2008JE003122.

\bibitem[{\textit{Bishop et al.}(2009)}]{bis09}
Bishop J.L., et al. (2009), Mineralogy of Juventae Chasma: Sulfates in the light-toned mounds, mafic minerals in the bedrock, and hydrated silica and hydroxylated ferric sulfate on the plateau, J. Geophys. Res., 114, E00D09, doi:10.1029/2009JE003352.

\bibitem[{\textit{Box \& Ski}(2007)}]{box07}
Box, J.E., \& K. Ski (2007), Remote sounding of Greenland supraglacial melt lakes: implications for subglacial hydraulics, J. Glaciol 53, 257-265.

\bibitem[{\textit{Burr et al.}(2009)}]{bur09}
Burr, D.M., et al. (2009), Pervasive aqueous paleoflow features in the Aeolis/Zephyria Plana region, Mars, Icarus 200, p. 52-76, doi:10.1016/j.icarus.2008.10.014.

\bibitem[{\textit{Burr et al.}(2010)}]{bur10}
Burr, D. M., R. M. E. Williams, K. D. Wendell, M. Chojnacki, \& J. P. Emery (2010), Inverted fluvial features in the Aeolis/Zephyria Plana region, Mars: Formation mechanism and initial paleodischarge estimates,
J. Geophys. Res. - Planets, doi:10.1029/2009JE003496.

\bibitem[{\textit{Carr \& Malin}(2000)}]{car00}
Carr, M.H., \& M.C. Malin (2000), Meter-scale characteristics of Martian channels and valleys, Icarus 146, 366-386.

\bibitem[{\textit{Carr \& Head}(2003)}]{car03}
Carr, M.H., \& J.W. Head (2003), Basal melting of snow on early Mars: A possible origin of some valley
networks, Geophys. Res. Lett. 30, 2245, doi:10.1029/2003GL018575.

\bibitem[{\textit{Carr \& Head}(2010)}]{car10}
Carr, M.H., \& J.W. Head (2010), Geologic history of Mars,
Earth and Planetary Science Letters 294, 185-203.

\bibitem[{\textit{Catling et al.}(2006)}]{cat06}
Catling, D.C., et al. (2006), Light-toned layered deposits in Juventae Chasma, Mars, Icarus 181, 26-51.

\bibitem[{\textit{Chapman et al.}(2010a)}]{cha09a}
Chapman, M.G., et al. (2010a), Amazonian geologic history of the Echus Chasma and Kasei Valles system on Mars: New data and interpretations, Earth and Planetary Science Letters, Volume 294, 238-255.

\bibitem[{\textit{Chapman et al.}(2010b)}]{cha09b}
Chapman, M.G., et al. (2010b), Noachian-Hesperian geologic history of the Echus Chasma and Kasei Valles system on Mars: New data and interpretations, Earth and Planetary Science Letters, Volume 294, 256-271.

\bibitem[{\textit{Chemtob et al.}(2010)}]{che10}
Chemtob, S.M., et al. (2010), Silica coatings in the Ka'u Desert, Hawaii, a Mars analog terrain: A micromorphological, spectral, chemical, and isotopic study, J. Geophys. Res. 115, E04001, doi:10.1029/2009JE003473.

\bibitem[{\textit{Clancy et al.}(2009)}]{cla10}
Clancy, R.T. et al. (2009), Valles Marineris cloud trails, J. Geophys. Res. - Planets 114, E11002, doi:10.1029/2008JE003323

\bibitem[{\textit{Clow}(1987)}]{clo87}
Clow, G.D. (1987), Generation of liquid water on Mars through the melting of a dusty snowpack, Icarus 72, 95-127.

\bibitem[{\textit{Clow}(1994)}]{clo94}
Clow, G.D. (1994), Minimum Discharge Rates Required for Sustained Water Flow on the Martian Surface, Lunar and Planetary Science Conference 25, p.275.

\bibitem[{\textit{Coleman \& Baker}(2007)}]{col07}
Coleman, N.M., \& V.R. Baker (2007), Evidence that a paleolake overflowed the rim of Juventae Chasma, Mars, LPSC 38, 1338.

\bibitem[{\textit{Conway et al.}(2010)}]{con10}
Conway, S.J., M.P. Lamb, M.R. Balme, M.C. Towner \& J.B. Murray (2010), Enhanced runout and erosion by overland flow at low pressure and subfreezing conditions: experiments and application to Mars, Icarus, in press.

\bibitem[{\textit{Dout\'{e} et al.}(2007)}]{dou07}
Dout\'{e}, S., et al. (2007), South Pole of Mars: Nature and composition of the icy terrains from Mars Express OMEGA observations, Planet. and Space Sci. 55, 113-133.

\bibitem[{\textit{Eichenlaub}(1979)}]{eic79}
Eichenlaub, V. (1979), Weather and climate of the Great Lakes region, University of Notre Dame Press.

\bibitem[{\textit{Edgett}(2005)}]{edg05}
Edgett, K.S. (2005), The sedimentary rocks of Sinus Meridiani: Five key observations from data acquired by the Mars Global Surveyor and Mars Odyssey orbiters, Mars 1, 5-58, doi:10.1555/mars.2005.0002


\bibitem[{\textit{Fassett \& Head}(2008a)}]{fas08b}
Fassett, C.I., \& J.W. Head III (2008), Valley network-fed, open-basin lakes on Mars: Distribution and implications
for Noachian surface and subsurface hydrology, Icarus 198, 37--56.

\bibitem[{\textit{Fassett \& Head}(2008b)}]{fas08}
Fassett, C.I., \& J.W. Head III (2008), The timing of martian valley network activity: Constraints from buffered
crater counting, Icarus 195, 61--89.

\bibitem[{\textit{Fassett et al.}(2010)}]{fas10}
Fassett, C.I., et al. (2010), Supraglacial and proglacial valleys on Amazonian Mars, Icarus 208, 86--100.

\bibitem[{\textit{Forget et al.}(2006)}]{for06}
Forget, F., et al. (2006), Formation of Glaciers on Mars by Atmospheric Precipitation at High Obliquity, Science 368, 311.


\bibitem[{\textit{Gaidos \& Marion}(2003)}]{gai03}
Gaidos E. \& G. Marion (2003), Geological and geochemical legacy of a cold early Mars, J. Geophys. Res. - Planets 108(E6), 5055, doi:10.1029/2002JE002000.

\bibitem[{\textit{Gardner \& Sharp}(2010)}]{gar10}
Gardner, A. S., \& M. J. Sharp (2010),
A review of snow and ice albedo and the development of a new physically based broadband albedo parameterization,
J. Geophys. Res., 115, F01009, doi:10.1029/2009JF001444

\bibitem[{\textit{Garrett}(1992)}]{gar92}
Garrett, J.R. (1992),The Atmospheric Boundary Layer (Cambridge Atmospheric and Space Science Series), Cambridge Univ. Press

\bibitem[{\textit{Golombek et al.}(2006)}]{gol06}
Golombek, M.P., et al. (2006), Erosion rates at the Mars Exploration Rover landing sites and long-term climate change on Mars, J. Geophys. Res. 111, E12S10, doi:10.1029/2006JE002754.

\bibitem[{\textit{Grotzinger et al.}(2006)}]{gro06}
Grotzinger, J., et al. (2006), Sedimentary textures formed by aqueous processes, Erebus crater, Meridiani Planum, Mars, Geology 34(12), 1085-1088; doi:10.1130/G22985A.1

\bibitem[{\textit{Haberle et al.}(1993)}]{hab93}
Haberle, R.M., et al. (1993), Mars atmospheric dynamics as simulated by the NASA Ames General-Circulation Model. 1. The zonal-mean circulation. J. Geophys. Res. - Planets, 98, 3093-3123.

\bibitem[{\textit{Harrison \& Chapman}(2008)}]{har08}
Harrison, K.P., \& M.G. Chapman (2008), Evidence for ponding and catastrophic floods in central Valles Marineris, Mars, Icarus 198, 351-364.

\bibitem[{\textit{Harrison \& Grimm}(2008)}]{har09}
Harrison, K.P., \& R.E. Grimm (2008), Multiple flooding events in Martian outflow channels, J. Geophys. Res. 113, E02002, doi:10.1029/2007JE002951.

\bibitem[{\textit{Hartmann}(2005)}]{harrecent}
Hartmann, W.K. (2005), Martian cratering 8: Isochron refinement and the chronology of Mars, Icarus, 174, 294-320.


\bibitem[{\textit{Halevy et al.}(2007)}]{hal07}
Halevy, I., M.T. Zuber, \& D. Schrag (2007), A Sulfur Dioxide Climate Feedback on Early Mars, Science 318, 1903-1907.

\bibitem[{\textit{Hawyard et al.}(2008)}]{hay06}
Hayward, R.K., et al. (2006), Mars Global Digital Dune Database: MC2-–MC29, U.S. Geological Survey
Open-File Report 2007-1158, version 1.0

\bibitem[{\textit{Hecht}(2002)}]{hec02}
Hecht, M.H. (2002), Metastability of liquid water on Mars, Icarus 156, 373-386.

\bibitem[{\textit{Howard et al.}(2007)}]{howpres}
Howard, A. D.; Moore, J. M.; Irwin, R. P.; Dietrich, W. E. (2007), Boulder Transport Across the Eberswalde Delta, Lunar and Planetary Science Conference 38, abstract no. 1168

\bibitem[{\textit{Hynek \& Phillips}(2003)}]{hyn03}
Hynek, B.M., \& R.J. Phillips (2003), New data reveal mature, integrated drainage systems on Mars indicative of past precipitation, Geology 31, 757–-760

\bibitem[{\textit{Hynek et al.}(2010)}]{hyn10}
Hynek, B. M., M. Beach, \& M. R. T. Hoke (2010),
Updated Global Map of Martian Valley Networks and Implications for Climate and Hydrologic Processes,
J. Geophys. Res.,  doi:10.1029/2009JE003548, in press.

\bibitem[{\textit{Kereszturi et al.}(2011)}]{ker11}
Kereszturi, A., Vincendon, M., \& F. Schmidt, 2011, Water ice in the dark dune spots of Richardson crater on Mars, Planetary and Space Science 59(1), 26-42.

\bibitem[{\textit{Kieffer et al.}(1976)}]{kie76}
Kieffer, H.H., et al., 1976, Martian north pole summer temperatures - Dirty water ice, Science 194, 1341-1344.

\bibitem[{\textit{Kite et al.}(in prep.)}]{kit10b}
Kite, E.S., M. Manga \& I. Halevy (in prep.), Snowmelt model of the formation and distribution of sedimentary rocks on Mars: Thick atmosphere not required?

\bibitem[{\textit{Kite et al.}(in review)}]{kit10}
Kite, E.S., et al., in review, Localized precipitation and runoff on Mars, J. Geophys. Res., available on arXiv astro-ph:EP.
%
\bibitem[{\textit{Kleinhans}(2005)}]{kle05}
Kleinhans, M.G. (2005), Flow discharge and sediment transport models for
estimating a minimum timescale of hydrological
activity and channel and delta formation on Mars, J. Geophys. Res. 110, E12003, doi:10.1029/2005JE002521.

\bibitem[{\textit{Komatsu et al.}(2009)}]{kom09}
Komatsu, G., et al. (2009), Paleolakes, paleofloods, and depressions in Aurorae and Ophir Plana, Mars: Connectivity of surface and subsurface hydrological systems, Icarus  201, 474-491

\bibitem[{\textit{Kraal et al.}(2008a)}]{kraal08alu}
Kraal, E.R., et al. (2008a), Catalogue of large alluvial fans in martian impact craters, Icarus 194(1), 101-110.

\bibitem[{\textit{Kraal et al.}(2008b)}]{kra08}
Kraal, E.R., M. van Dijk, G. Postma, \& M.G. Kleinhans (2008b), Martian stepped-delta formation by rapid water release, Nature, 451, 973-976, doi:10.1038/nature06615.

\bibitem[{\textit{Kuiper et al.}(2008b)}]{kui08}
Kuiper, K.F., et al. (2008), Synchronizing rock clocks of Earth history, Science 320, 500-504.

\bibitem[{\textit{Johnson et al.}(2008)}]{joh08}
Johnson, S.S., Mischna, M.A., Grove, T.L., \& M.T. Zuber (2008), Sulfur-induced greenhouse warming on early Mars, J. Geophys. Res. 113, doi:10.1029/2007JE002962.

\bibitem[{\textit{Laskar et al.}(2004)}]{las04}
Laskar, J. et al. (2004), Long term evolution and chaotic diffusion of the insolation quantities of Mars, Icarus 170, 343-364.

\bibitem[{\textit{Le Deit et al.}(2010)}]{led10}
Le Deit, L., et al. (2010), Morphology, stratigraphy, and mineralogical composition of a layered formation covering the plateaus around Valles Marineris, Mars: Implications for its geological history, Icarus.

\bibitem[{\textit{Langevin et al.}(2005)}]{lan05}
Langevin, Y., et al. (2005), Summer Evolution of the North Polar Cap of Mars as Observed by OMEGA/Mars Express, Science 307, 1581-1584.

\bibitem[{\textit{Le Deit et al.}(2010)}]{led10}
Le Deit, L., et al. (2010), Morphology, stratigraphy, and mineralogical composition of a layered formation covering the plateaus around Valles Marineris, Mars: Implications for its geological history, Icarus, in press.

\bibitem[{\textit{Lewis et al.}(2008)}]{lew08}
Lewis, K.W., et al. (2008), Quasi-periodic bedding in the sedimentary rock record of Mars, Science 322, 1532.

\bibitem[{\textit{Lewis et al.}(2010)}]{lew10}
Lewis, K.W., et al. (2010), Global significance of cyclic sedimentary deposits on Mars, LPSC 41, 2648.

\bibitem[{\textit{Lewis et al.}(1999)}]{lew99}
Lewis, S.R., et al. (1999), A climate database for Mars, J. Geophys. Res. - Planets 104, E10, 24177-24194.

\bibitem[{\textit{L{\"{u}}thje et al.}(2006)}]{lut06}
L{\"{u}}thje, M., et al. (2006). Modelling the evolution of supraglacial lakes on the West Greenland ice-sheet margin, Journal of Glaciology 52(179)

\bibitem[{\textit{McLennan \& Grotzinger}(2008)}]{grotzmclennan}
McLennan, S. M., \& J.P. Grotzinger (2008), The sedimentary rock cycle of Mars, in Bell, J., ed. The Martian Surface - Composition, Mineralogy, and Physical Properties, Cambridge University Press.

\bibitem[{\textit{Madeleine et al.}(2009)}]{mad09}
Madeleine, J.-B., et al. (2009) Amazonian northern mid-latitude glaciation on Mars: A proposed climate scenario, Icarus 203, 390-–405

\bibitem[{\textit{Markowksi \& Richardson}(2010)}]{mar10}
Markowski, P., \& Y. Richardson (2010), Mesoscale meteorology in midlatitudes, John Wiley \& Sons.

\bibitem[{\textit{Malin \& Edgett}(2003)}]{mal03}
Malin \& Edgett (2003), Evidence for persistent flow and aqueous sedimentation on early Mars, Science, 302, 1931-1934

\bibitem[{\textit{Malin et al.}(2010)}]{mal10}
Malin, M.C., et al. (2010), An overview of the 1985-2006 Mars Orbiter Camera science investigation, Mars 5, 1-60, doi:10.1555/mars.2010.0001

\bibitem[{\textit{Mangold et al.}(2004)}]{man04}
Mangold, N. ,et al. (2004), Evidence for Precipitation on Mars from Dendritic Valleys in the Valles Marineris Area, Science 305. 78 - 81
DOI:10.1126/science.1097549

\bibitem[{\textit{Mangold et al.}(2008)}]{man08}
Mangold et al. (2008), Geomorphic study of fluvial landforms on the northern Valles Marineris plateau, Mars, J. Geophys. Res. - Planets 113, E08009, doi:10.1029/2007JE002985

\bibitem[{\textit{McKenzie \& Nimmo}(1999)}]{mck99}
McKenzie, D., \& F. Nimmo (1999), The generation of martian floods by the melting of ground ice above dykes, Nature 397, 231-233.

\bibitem[{\textit{Mellon \& Jakosky}(1995)}]{mel95}
Mellon, M.T., \& B.M. Jakosky (1995), The distribution and behavior of Martian ground ice during past and present epochs, J. Geophys. Res. 100(E6), 11781-11799, doi:10.1029/95JE01027.

\bibitem[{\textit{Mellon et al.}(2000)}]{mel00}
Mellon, M.T., et al. (2000), High-Resolution Thermal Inertia Mapping from the Mars Global
Surveyor Thermal Emission Spectrometer, Icarus 148, 437-455.

\bibitem[{\textit{Mellon et al.}(2009)}]{mel09b}
Mellon, M.T., et al. (2009), Ground ice at the Phoenix Landing Site: Stability state and origin, J. Geophys. Res. 114(E00E07),
doi:10.1029/2009JE003417

\bibitem[{\textit{Metz et al.}(2009)}]{met09}
Metz, J., et al. (2009), Sublacustrine depositional fans in southwest Melas Chasma, J. Geophys. Res. 114, E10002, doi:10.1029/2009JE003365.

\bibitem[{\textit{Michaels et al.}(2006)}]{mic06}
Michaels, T.I., A. Colaprete, \& S.C.R. Rafkin, Significant vertical water transport by mountain-induced circulations on Mars, Geophys. Res. Lett. 33, L16201, doi:10.1029/2006GL026562.

\bibitem[{\textit{Michaels \& Rafkin}(2008b)}]{mic08conf}
Michaels, T.I., \& S.C.R. Rafkin (2008), MRAMS today - One example of current Mars mesoscale modeling capabilities, Third International Workshop on Mars Atmosphere: Modeling and Observations, Williamsburg, Virginia, 9116.

\bibitem[{\textit{Milliken et al.}(2008)}]{mil08}
Milliken, R.E., et al. (2008), Opaline silica in young deposits on Mars, Geology 36, 847–850, doi: 10.1130/G24967A.1.

\bibitem[{\textit{Millour et al.}(2008)}]{dddmcd}
Millour, E., F. Forget, \& S.R. Lewis (2008), Mars Climate Database v4.3 Detailed Design Document, downloaded from http://www-mars.lmd.jussieu.fr/

\bibitem[{\textit{Mischna et al.}(2003)}]{mis03}
Mischna, M.A., et al. (2003), On the orbital forcing of Martian water and CO2 cycles: A general circulation model study with simplified volatile schemes, J. Geophys. Res. - Planets 108(E6), doi:10.1029/2003JE002051.

\bibitem[{\textit{M{\"{o}}hlmann et al.}(2009)}]{moh09}
M{\"{o}}hlmann, D.T.F., et al (2009), Fog phenomena on Mars, Planet. \& Space Sci. 57, 1987-1992.

\bibitem[{\textit{Montgomery \& Gran}(2001)}]{mon01}
Montgomery, D.R., \& K.B. Gran (2001), Downstream variations in the width of bedrock channels, Water Resources Research 37(6), 1841-1846.

\bibitem[{\textit{Murchie et al.}(2009a)}]{mur09}
Murchie, S.L., et al. (2009a), A synthesis of Martian aqueous mineralogy after 1 Mars year of observations from the Mars Reconnaissance Orbiter, J. Geophys. Res. - Planets 114, E00D06, doi:10.1029/2009JE003342

\bibitem[{\textit{Murchie et al.}(2009b)}]{mur09b}
Murchie, S.L., et al. (2009b), Evidence for the origin of layered deposits in Candor Chasma, Mars, from mineral composition and hydrologic modeling, J. Geophys. Res. - Planets 114, E00D06, doi:10.1029/2009JE003343

\bibitem[{\textit{Perron et al.}(2004)}]{lambinfilt}
Perron, J.T., M. Manga, \& M.P. Lamb (2004), Permability, recharge, and runoff generation on Mars, First Workshop on Mars Valley Networks. Downloaded from www.gps.caltech.edu/$\sim$mpl/perronetal\_valleynetworks.pdf on 31 Oct 2010.

\bibitem[{\textit{Perron et al.}(2006)}]{per06}
Perron, J.T., M.P. Lamb, C.D. Koven, I.Y. Fung, \& E. Yager (2006), Valley formation and methane precipitation rates on Titan, J. Geophys. Res. 111, E11001, doi:10.1029/2005JE002602.

\bibitem[{\textit{Pielke}(2002)}]{pie02}
Pielke, R.A. (2002), Mesoscale meteorological modeling, 2nd edition (Intl. Geophysics series vol 78), Academic Press.

\bibitem[{\textit{Pielke et al.}(1992)}]{pie92}
Pielke, R.A., et al. (1992), A comprehensive meteorological modeling system - RAMS, Meteorol. \& Atmos. Phys. 49, 69-91.


\bibitem[{\textit{Pielke \& Mahrer}(1978)}]{pie78}
Pielke, R.A. \& Y. Mahrer (1978), Verification analysis of University-of-Virginia 3-dimensional mesoscale model prediction over South Florida for 1 July 1973, Monthly Weather Rev. 106, 1568-1589.

\bibitem[{\textit{Postberg et al.}(2009)}]{pos09}
Postberg, F., S. Kempf, J. Schmidt, N. Brilliantov, A. Beinsen, B. Abel, U. Buck \& R. Srama (2009), \S 2.2 of supplementary information to: Sodium salts in E-ring ice grains from an ocean below the surface of Enceladus,
Nature 459, 1098-1101, doi:10.1038/nature08046.

\bibitem[{\textit{Quantin et al.}(2005)}]{qua05}
Quantin, C.et al. (2005), Fluvial and lacustrine activity on layered deposits in Melas Chasma, Valles Marineris, Mars, J. Geophys. Res. - Planets, 110, E12S19, doi:10.1029/2005JE002440.

\bibitem[{\textit{Rafkin et al.}(2001)}]{raf01}
Rafkin, S. C. R., Haberle, R. M., and T. I. Michaels (2001), The Mars Regional Atmospheric Modeling System (MRAMS): Model description and selected simulations, Icarus, 151, 228-256.

\bibitem[{\textit{Rafkin \& Michaels}(2003)}]{rafmer}
Rafkin, S. C. R. and T. I. Michaels (2003), Meterological predictions for 2003 Mars Exploration Rover high-priority landing sites. J. Geophys. Res, 108 No. E12,10.1029/2002JE002027

\bibitem[{\textit{Roach et al.}(2010)}]{roa10}
Roach, L.H. et al. (2010), Hydrated mineral stratigraphy of Ius Chasma, Valles Marineris, Icarus 206, 253-268.

\bibitem[{\textit{Scambos et al.}(2003)}]{sca03}
Scambos, T., Hulbe, C., \& M. Fahnestock (2003), Climate-induced ice shelf disintegration in the Antarctic peninsula, Antarct. Res. Ser. 79, 79–92.

\bibitem[{\textit{Schiffman et al.}(2006)}]{sch06}
Schiffman, P. et al., (2006), Acid-fog deposition at Kilauea volcano: A possible mechanism for
the formation of siliceous-sulfate rock coatings on Mars, Geology 34, 921–924; doi:10.1130/G22620A.1.

\bibitem[{\textit{Segura et al.}(2002)}]{seg02}
Segura, T.L., O.B. Toon, T. Colaprete \& K. Zahnle (2002), Environmental Effects of Large Impacts on Mars, Science 298, 1977-1980.

\bibitem[{\textit{Segura et al.}(2008)}]{seg08}
Segura, T.L., O.B. Toon, \& T. Colaprete (2008), Modeling the environmental effects of moderate-sized impacts on Mars, 113, E11007, doi:10.1029/2008JE003147.

\bibitem[{\textit{Smith}(2002)}]{smi02}
Smith, M.D. (2002), The annual cycle of water vapor on Mars as observed by the Thermal Emission Spectrometer, J. Geophys. Res. 107, 5115, doi:10.1029/2001JE001522.

\bibitem[{\textit{Spiga \& Forget}(2009)}]{spigmeso}
Spiga, A., \& F. Forget (2009), A new model to simulate the Martian mesoscale and microscale atmospheric circulation: Validation and first results, J. Geophys Res. 114, E02009.

\bibitem[{\textit{Toon et al.}(2010)}]{too10}
Toon, O.B., T. Segura \& K. Zahnle (2010), The Formation of Martian River Valleys by Impacts, Ann. Rev. Earth \& Planet. Sci. 38, 303-322.


\bibitem[{\textit{USGS}(2009)}]{usgs09}
United States Geological Survey (2009), USGS Astrogeology Science Center Tutorial:
Stereo Processing
using
HiRISE Stereo Imagery
ISIS3
and
SOCET SET, June 2009, downloaded from
http://webgis.wr.usgs.gov/pigwad/tutorials/socetset/ SocetSet4HiRISE.htm

\bibitem[{\textit{van Eaton et al.}(2010)}]{van10}
van Eaton, A., et al., 2010, Microphysical Controls on Ascent of Water-Rich Ash Clouds from Supereruptions, AGU Fall Meeting 2010, abstract \# V13C-2375

\bibitem[{\textit{Vincendon et al.}(2010)}]{vin10}
Vincendon, M., F. Forget, \& J. Mustard (2010),
Water ice at low to mid latitudes on Mars,
J. Geophys. Res. 115, E10001, doi:10.1029/2010JE003584.

\bibitem[{\textit{Warren \& Wiscombe}(1980)}]{war80}
Warren, S.G., \& W.J. Wiscombe (1980), A Model for the Spectral Albedo of Snow. II: Snow Containing Atmospheric Aerosols, J. Atmos. Sciences 37: 2734-2745.

\bibitem[{\textit{Warner et al.}(2010)}]{war10}
Warner, N. et al. (2010), Hesperian equatorial thermokarst lakes in Ares Vallis as evidence for transient warm conditions on Mars, Geology 38, 71-74.

\bibitem[{\textit{Wang et al.}(2005)}]{wan05}
Wang, C.-Y., M. Manga, \& C. Wong (2005), Floods on Mars released from groundwater by impact, Icarus 175, 551–555.

\bibitem[{\textit{Warren}(1984)}]{war84}
Warren, S.G. (1984), Impurities in snow - Effects on albedo and snowmelt, Ann. Glaciol 5, 177-179.

\bibitem[{\textit{Weitz et al.}(2008)}]{wei08}
Weitz, C.M. et al. (2008) Light-toned strata and inverted channels adjacent to Juventae and Ganges chasmata, Mars, Geophys. Res. Lett. 35, L19202,
doi:10.1029/2008GL035317

\bibitem[{\textit{Weitz et al.}(2010)}]{wei10}
Weitz, C.M. et al. (2010) Mars Reconnaissance Orbiter observations of light-toned layered deposits and associated fluvial landforms on the plateaus adjacent to Valles Marineris,
Icarus 205, 73-102

\bibitem[{\textit{Williams}(2007)}]{wil07}
Williams, R.M.E. (2007), Global spatial distribution of raised curvilinear features on Mars, LPSC 38, 1821.

\bibitem[{\textit{Williams et al}(2007)}]{wil07utah}
Williams, R.M.E., Chidsey, T.C., Jr., \& Eby, D.E., (2007), Exhumed paleochannels in central Utah - analogs for raised curvilinear features on Mars, in Willis, G.C., Hylland, M.D., Clark, D.L., and Chidsey, T.C., Jr., editors, Central Utah - diverse geology of a dynamic landscape: Utah Geological Association Publication 36, Salt Lake City, Utah, 220-235.

\bibitem[{\textit{Williams et al.}(2005)}]{wil05}
Williams, R.M.E., M.C. Malin, \& K.S. Edgett (2005), Remnants of the courses of fine-sclae, precipitation-fed runoff streams preserved in the Martian rock record, LPSC 36, 1173.

\bibitem[{\textit{Williams \& Malin}(2008)}]{wil08}
Williams, R.M.E., \& M.C. Malin (2008), Sub-kilometer fans in Mojave Crater, Mars, Icarus 198, 365-383, doi:10.1016/j.icarus.2008.07.013

\bibitem[{\textit{Williams et al.}(2000)}]{wil00}
Williams, R.M.E., et al. (2000), Flow rates and duration within Kasei Valles, Mars: Implications for the formation of a Martian Ocean, Geophys. Res. Lett. 27, 1073--1076.

\bibitem[{\textit{Williams et al.}(2009)}]{wil09}
Williams, R.M.E., et al. (2009), Evaluation of paleohydrologic models for terrestrial inverted channels: Implications for application to martian sinuous ridges, Geomorphology 107, 300-315.

\bibitem[{\textit{Wordsworth et al.}(2010a)}]{wordsworth}
Wordsworth, R., F. Forget, \& V. Eymet (2010a), Infra-red collision-induced and far-line absorption in dense CO2 atmospheres, Icarus, doi:10.1016/j.icarus.2010.06.010, in press.

\bibitem[{\textit{Wordsworth}(2010b)}]{wordexoclimes}
Wordsworth, R., et al. (2010b) Investigating the early Martian climate through three-dimensional atmospheric modelling, Exoclimes: Exploring the diversity of planetary atmospheres, Exeter UK.



%\bibitem[{\textit{Kilby et al.}(2008)}]{jskilbye}
%Kilby, J. S., S. Smith, and R. Jones (2008), Invention of the
%integrated circuit, {\it IEEE Trans. Electron Devices,} \textit{23},
%648--650.

\end{thebibliography}
\end{document}